\documentclass{aa}  
\usepackage{graphicx}
\usepackage{txfonts}
\usepackage{hyperref}
\begin{document}

\title{A$^{3}$COSMOS: A census on the molecular gas mass and extent of main-sequence galaxies across cosmic time}

\subtitle{}

\author{Tsan-Ming Wang
          \inst{1}
          \and
          Benjamin Magnelli
          \inst{1,2}
          \and
          Eva Schinnerer
          \inst{3}
          \and
          Daizhong Liu
          \inst{4}
          \and
          Ziad Aziz Modak
          \inst{1}
          \and
          Eric Faustino Jim{\'e}nez-Andrade
          \inst{5}
          \and
          Christos Karoumpis
          \inst{1}
          \and
          Vasily Kokorev
          \inst{6,7}
          \and
          Frank Bertoldi
          \inst{1}
          }

\institute{Argelander-Institut f\"ur Astronomie, Universit\"at Bonn, Auf dem H\"ugel 71, 53121 Bonn, Germany\\
              \email{twan@uni-bonn.de}
             \and
               AIM, CEA, CNRS, Universit\'e Paris-Saclay, Universit\'e Paris Diderot, Sorbonne Paris Cit\'e, 91191 Gif-sur-Yvette, France
             \and
               Max-Planck-Institut f\"ur Astronomie, K\"onigstuhl 17, 69117 Heidelberg, Germany
             \and
               Max-Planck-Institut f\"ur extraterrestrische Physik, Gie{\ss}enbachstra{\ss}e 1, 85748 Garching b. M\"unchen, Germany
             \and
               National Radio Astronomy Observatory, 520 Edgemont Road, Charlottesville, VA 22903, USA
             \and
               Cosmic Dawn Center (DAWN), Copenhagen, Denmark 
             \and
               Niels Bohr Institute, University of Copenhagen, Lyngbyvej 2, 2100 Copenhagen {\O}, Denmark
             }

\date{Received xxx; accepted xxx}

\abstract
{} 
{To constrain for the first time the mean mass and extent of the molecular gas of a mass-complete sample of normal $>10^{10}\,$M$_{\odot}$ star-forming galaxies at $0.4<z<3.6$.}
{We apply an innovative $uv$-based stacking analysis to a large set of archival Atacama Large Millimeter/submillimeter Array (ALMA) observations using a mass-complete sample of main-sequence (MS) galaxies. This stacking analysis, performed on the Rayleigh-Jeans dust continuum emission, provides accurate measurements of the mean mass and extent of the molecular gas of galaxy populations, which are otherwise individually undetected.}
{The molecular gas mass of MS galaxies evolves with redshift and stellar mass. At all stellar masses, the molecular gas fraction decreases by a factor of $\sim24$ from $z\sim3.2$ to $z\sim0$. At a given redshift, the molecular gas fraction of MS galaxies decreases with stellar mass, at roughly the same rate as their specific star formation rate (${\rm SFR}/M_\star$) decreases. The molecular gas depletion time of MS galaxies remains roughly constant at $z>0.5$ with a value of 300--500\,Myr, but increases by a factor $\sim3$ from $z\sim0.5$ to $z\sim0$. This evolution of the molecular gas depletion time of MS galaxies can be predicted from the evolution of their molecular gas surface density and a seemingly universal MS-only $\Sigma_{M_{\rm mol}}-\Sigma_{\rm SFR}$ relation with an inferred slope of $\sim1.13$, i.e., the so-called Kennicutt-Schmidt (KS) relation. The far-infrared size of MS galaxies shows no significant evolution with redshift or stellar mass, with a mean circularized half-light radius of $\sim$2.2\,kpc. Finally, our mean molecular gas masses are generally lower than previous estimates, likely caused by the fact that literature studies were largely biased towards individually-detected MS galaxies with massive gas reservoirs.} 
{To first order, the molecular gas content of MS galaxies regulates their star formation across cosmic time, while variation of their star formation efficiency plays a secondary role. Despite a large evolution of their gas content and SFRs, MS galaxies evolved along a seemingly universal MS-only KS relation.}

                
\titlerunning{Molecular gas mass and extent of main-sequence galaxies across cosmic time}
\authorrunning{W. Tsan-Ming et al.}         
\maketitle

\section{Introduction}
Understanding galaxy evolution across cosmic time is one of the key topics of modern astronomy. To address this vast and important question, one very successful approach is to assemble and study large and representative samples of galaxies, through multi-wavelength deep extragalactic surveys. Using this approach, much has been learned over the last decades about the global star formation history of the Universe. The cosmic star formation rate density (SFRD) increases from early cosmic times, $z\sim2$, and decreases by a factor of 10 by $z\sim0$ \citep{2014ARA&A..52..415M}. About $80\%$ of this star formation takes place in relatively massive galaxies ($>10^{10}$ M$_{\odot}$) that reside on the so-called main sequence (MS) of star-forming galaxies \citep[e.g.,][]{2007ApJ...660L..43N, 2011ApJ...739L..40R, 2012ApJ...747L..31S}. This MS denotes the tight correlation existing between the stellar mass ($M_{\star}$) and star-formation rate (SFR) of galaxies which is observed up to $z\sim4$ \citep[e.g.,][]{2004MNRAS.351.1151B, 2009ApJ...698L.116P, 2010MNRAS.401.1521M, 2012ApJ...757...54Z, 2013ApJ...777L...8K, 2014ApJ...795..104W, 2014MNRAS.437.3516S, 2014ApJS..214...15S, 2015MNRAS.453.2540J, 2016ApJ...817..118T, 2017MNRAS.467.1360B, 2018A&A...615A.146P, 2019MNRAS.490.5285P, 2020ApJ...899...58L}. The existence of the MS, with its constant scatter of 0.3\,dex and a normalisation that decreases by a factor of 20 from $z\sim2$ to $z\sim0$, suggests that most star-forming galaxies (SFGs) are isolated and secularly evolving with long ($>\,$Gyr) star-forming duty cycles. On the contrary, galaxies above the MS \citep[$\sim5\%$ of the SFG population;][]{2014ApJ...789L..16L} seem to be mostly associated to short, intense starbursts triggered by major mergers and contribute only 10$\%$ to the SFRD at all redshifts \citep[e.g.,][]{2012ApJ...747L..31S}. While the evolution of the MS and SFRD across cosmic time is observationally well established up to $z\sim2$, the mechanisms driving their evolution are yet poorly constrained. At $z>2$, our understanding is even more limited because observations obtained from different rest-frame frequencies (i.e., UV, far-infrared or radio) provide a somewhat discrepant view of the exact evolution of the SFRD \citep[e.g.,][]{2015ApJ...803...34B,2017A&A...602A...5N,2018ApJ...853..172L,2020A&A...643A...8G}.

To shed light on the physical processes that regulate star formation across cosmic time, it is paramount to obtain a precise measurement of the molecular gas content of local and high-redshift galaxies. Indeed, molecular gas fuels star formation, as revealed by the tight correlation between gas mass and star formation rate surface densities, the so-called Kennicutt-Schmidt (KS) relation \citep[][]{1998ApJ...498..541K}. Molecular hydrogen (H$_{2}$) is the most abundant constituent of molecular gas, but it is difficult to observe due to its lack of a dipole moment. For this reason, the carbon monoxide (CO) molecule, which is the most abundant and readily observable constituent of molecular gas, is usually used to trace the molecular gas content of galaxies \citep[see ][for a review]{2013ARA&A..51..207B}. However, even with the Atacama Large Millimeter/submillimeter Array (ALMA), obtaining such measurements for $z>2.0$ MS galaxies with stellar mass of  $\sim$10$^{10}\,$M$_{\odot}$ still requires an hour of observing time per object. Thus, the CO molecule is still poorly suited for the study of large and representative samples of high-redshift galaxies. Therefore, in recent years, an alternative approach focusing on high-redshift galaxies has emerged, which relies on dust mass measurements and a standard gas-to-dust mass ratio calibrated in the local universe. These gas mass measurements, inferred from either multi-wavelength dust spectral energy distribution (SED) fits \citep[e.g.,][]{2012ApJ...760....6M, 2012A&A...548A..22M, 2014A&A...561A..86M, 2014A&A...562A..30S, 2014A&A...569A..98T, 2014A&A...562A..30S, 2015A&A...573A.113B, 2016A&A...587A..73B, 2019A&A...621A..51H} or single Rayleigh–Jeans (RJ) flux density conversion \citep[e.g.,][]{2014ApJ...783...84S, 2015ApJ...799...96G, 2016ApJ...833..112S, 2016ApJ...820...83S, 2019ApJ...880...15K, 2019ApJ...887..235L, 2020ApJ...892...66M, 2020MNRAS.494..293M}, were shown to be surprisingly accurate when compared to state-of-the-art CO measurements \citep[e.g.,][]{2015ApJ...800...20G, 2016ApJ...820...83S, 2017ApJ...837..150S, 2018ApJ...853..179T, 2020ARA&A..58..157T}.

This dust-based approach has since allowed the measurement of the gas content of hundreds of high-redshift SFGs. It was found that the gas fraction (i.e., $M_{\rm gas}/M_{\ast}$) of massive SFGs is relatively constant at $z>2$ but decreases significantly from $z\sim2$ to $z\sim0$ \citep{2013ARA&A..51..105C,  2014ApJ...793...19S, 2016ApJ...833..112S, 2017A&A...606A..17M,  2017ApJ...837..150S,  2018ApJ...853..179T, 2020ARA&A..58..157T, 2019ApJ...875....6G,  2019ApJ...887..235L, 2019ApJ...878...83W, 2020ApJ...891...83C}. This evolution follows that of the normalisation of the MS and implies that the star formation efficiency (SFE$=$ SFR/$M_{\rm gas}$) in these galaxies remains relatively constant across cosmic time. This finding is confirmed by the global evolution of the co-moving gas mass density, which resembles that of the SFRD \citep{2020ApJ...892...66M}. At any redshift, the depletion time ($t_{\rm depl}=$1/SFE) of the gas reservoirs of massive SFGs is found to be relatively short and of the order of $\sim0.5-1\,$Gyr. Without continuous replenishment of their gas reservoirs, star formation in massive MS galaxies would thus cease within $\sim0.5-1\,$Gyr, in tension with the existence of the MS itself, i.e., long star-forming duty cycles. The continuous accretion of fresh gas from the intergalactic or circum-galactic medium would thus be the main parameter regulating star formation across cosmic time, as also suggested by hydro-dynamical simulations \citep[e.g.,][]{2011MNRAS.417.2982F, 2019ApJ...872...13W}.

While all these previous studies provided key information for our understanding of galaxy evolution, they all suffer from a set of limitations. Firstly, all relied on samples of few hundreds to at most thousand galaxies, and thus suffered from small number statistic, especially because these samples were further split into numerous redshift, stellar mass, and $\Delta$MS ($\Delta$MS = log$_{10}$(SFR/SFR$_{\rm MS}$)) bins. Secondly, all these studies were based on subsets of galaxies drawn from a parent sample using underlying complex selection functions. Each sub-sample could thus still fail to provide a complete, and representative view on the gas content of high-redshift galaxies. This likely explains in part why these studies agreed qualitatively but disagree quantitatively on the exact redshift evolution of the gas content of massive galaxies \citep[see][]{2019ApJ...887..235L}. Finally, and most importantly, these studies relied mainly on individually-detected galaxies and were thus limited to the high-mass end ($>10^{10.5}\,$M$_\odot$) of the SFG population. While constraining the gas content of massive galaxies is important, extending our knowledge towards lower stellar masses is crucial because the bulk of the star formation activity of the Universe is known to take place in $10^{10\cdots10.5}\,$M$_{\odot}$ galaxies \citep[e.g.,][]{2011ApJ...730...61K, 2020ApJ...899...58L}. The gas properties of these crucial low-mass high-redshift SFGs remain thus to date largely unknown simply because most are individually-undetected even in deep ALMA observations.

To statistically retrieve the faint emission of this SFG population, one can perform a stacking analysis. Indeed, by grouping galaxies in meaningful ways (e.g., in bins of redshift and stellar mass) and by stacking their observations (e.g., summing or averaging), one effectively increases the observing time toward this galaxy population and can thus infer their average properties. The noise in the stacked image decreases as the root square of the number of stacked galaxies, and thus large samples can lead to robust detection of previously individually-undetected galaxy populations. Such a statistical approach applied to, e.g., \textit{Spitzer}, \textit{Herschel}, or ALMA images, has proven to be extremely powerful and to push measurements well below the conventional instrumental and confusion noise limits of these observatories \citep[e.g.,][]{2006A&A...451..417D,2006ApJ...640..784Z, 2014A&A...561A..86M, 2014ApJ...783...84S, 2015A&A...573A..45M, 2015A&A...575A..74S, 2016MNRAS.462.1192L, 2020ApJ...892...66M}. Although stacking over the entire ALMA archive provides an unique opportunity to study the gas mass content of low-mass high-redshift SFGs, it also presents two challenges when compared to standard stacking analyses performed with \textit{Spitzer}, \textit{Herschel}, or single ALMA projects, as the ALMA archival data is heterogeneous in terms of observed frequencies and spatial resolution. While stacking data obtained at different observing frequencies simply implies a re-scaling of each individual dataset to a common rest-frame luminosity frequency using locally calibrated submillimeter SEDs, stacking data with different spatial resolution is a more uncommon challenge, which has only rarely been tackled in the literature \citep[e.g.,][]{2016MNRAS.462.1192L,  2020ApJ...888...44C}. It can, however, easily be addressed thanks to the very nature of ALMA observations. Indeed, while combining observations with different spatial resolution would involve very uncertain and complex convolutions in the image-domain, combining them in the $uv$-domain is strictly equivalent to performing aperture synthesis on a single object \citep[e.g.,][]{2016MNRAS.462.1192L,  2020ApJ...888...44C}.

In this work, we aim at mitigating most of the limitations affecting current studies on the gas properties of high-redshift SFGs by applying an innovative $uv$-based stacking analysis to a large set of ALMA observations towards a mass-complete sample of $M_{\star}>10^{10}\,$M$_{\odot}$ MS galaxies. This sample is drawn from one of the largest, yet deep, multi-wavelength extragalactic survey, the COSMOS-2015 catalog \citep{2016ApJS..224...24L}. The stellar masses and redshifts of our galaxies were directly taken from the COSMOS-2015 catalog, while their SFRs were estimated from their COSMOS-2015 rest-ultraviolet, mid- and far-infrared photometry following the ladder of SFR indicators of \citet{2011ApJ...738..106W}. From this mass-complete sample of MS galaxies, we only kept those with an ALMA archival band-6 or 7 coverage as assembled by the Automated mining of the ALMA Archive in the COSMOS field (A$^3$COSMOS) project \citep[][]{2019ApJS..244...40L}. This mass-complete sample of MS galaxies was then subdivided into several redshift and stellar mass bins, and a measurement of their mean molecular gas mass and size was performed using a $uv$-based stacking analysis of their ALMA observations. This stacking analysis allows for accurate mean gas mass and size measurements even at low stellar masses where galaxies are too faint to be individually-detected by ALMA. Our results provide for the first time robust RJ-based constraints on the mean cold gas mass of a mass-complete sample of $M_{\star}$> 10$^{10}$ M$_{\odot}$ galaxies up to $z\sim3$. Combined with their mean far-infrared (FIR) size measurements, this yields the first stringent constraint of the KS relation at high-redshift.

The structure of the paper is as follows: in Section~\ref{sec:data}, we introduce the ALMA data used in our study and our mass-complete sample of MS galaxies; in Section~\ref{sec:method}, we describe the method used to estimate the mean gas mass and size of a given galaxy population by stacking their ALMA observations in the $uv$-domain; in Section~\ref{sec:results}, we present our results and discuss them in Section~\ref{sec:discussion}; finally, in Section~\ref{sec:summary}, we summarize our findings and present our conclusions.

Throughout the paper, we assume a flat $\rm \Lambda CDM$ cosmology with $H_{0}$ = 67.8 km s$^{-1}$ Mpc$^{-1}$, $\rm \Omega_{M}$ = 0.308, and $\rm \Omega_{\Lambda}$ = 0.692 \citep{2016A&A...594A..13P}. All stellar masses and SFRs are provided assuming an \citet{2003PASP..115..763C} initial mass function. 

\begin{figure*}
\centering
\includegraphics[width=1.0\textwidth]{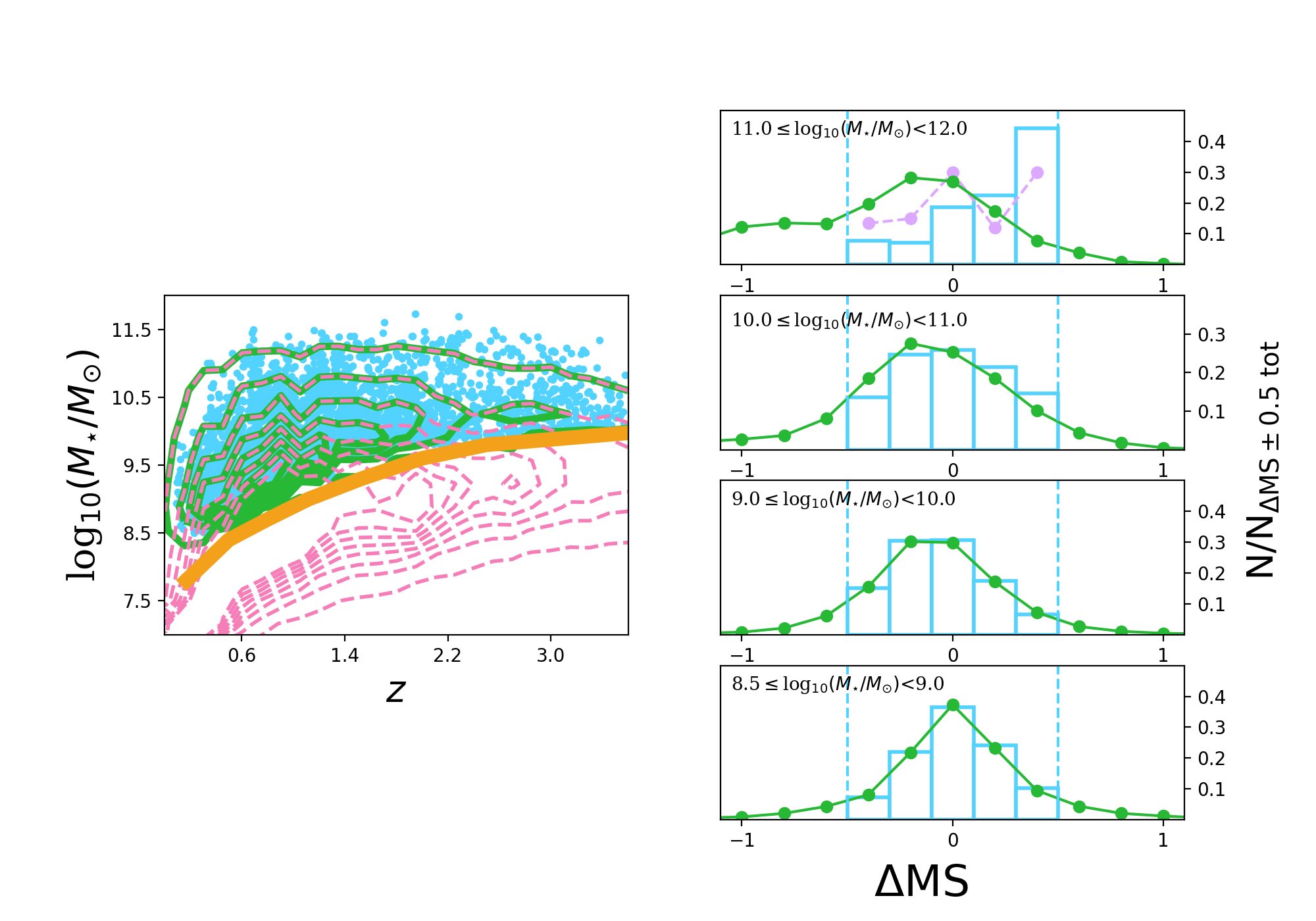}
\caption{(\textit{Left}) Stellar mass and redshift distribution in our final ALMA-covered mass-complete sample of MS galaxies (blue dots). The pink dashed contours displays the number density of SFGs in \citet{2016ApJS..224...24L}, i.e., our parent sample of SFGs. The pink contour levels are in steps of 500 from 200 to 3700 galaxies per $z$--log$_{10} M_{\star}$ bin of size 0.14 and 0.15, respectively. The orange solid line represents the stellar mass completeness limit of SFGs in \citet{2016ApJS..224...24L}. The green contour shows the number density of SFGs in \citet{2016ApJS..224...24L} above this stellar mass completeness limit, i.e., our parent mass-complete sample. The green contour levels are in steps of 500 from 200 to 3700 galaxies per $z$--log$_{10} M_{\star}$ bin of size 0.14 and 0.15, respectively. 
(\textit{Right}) Relative $\Delta$MS distribution in our final ALMA-covered mass-complete sample of MS galaxies (blue histogram) and our mass-complete parent sample of SFGs (green histogram) in different stellar mass bins. In the highest stellar mass bin, the purple dashed line shows the relative $\Delta$MS distribution after having rejected from our final sample all ALMA primary targets, i.e., galaxies at the phase center of the ALMA observation. The vertical blue dashed lines display the $\pm0.5\,$dex interval used to defined MS galaxies. Over this interval, the integral of each histogram is equal to one. This normalization is needed to compare our final and mass-complete parent samples which contain 3,037 and 515,465 galaxies, respectively.} 
\label{fig:completeness}
\end{figure*}

\begin{figure}
\centering
\includegraphics[width=\columnwidth]{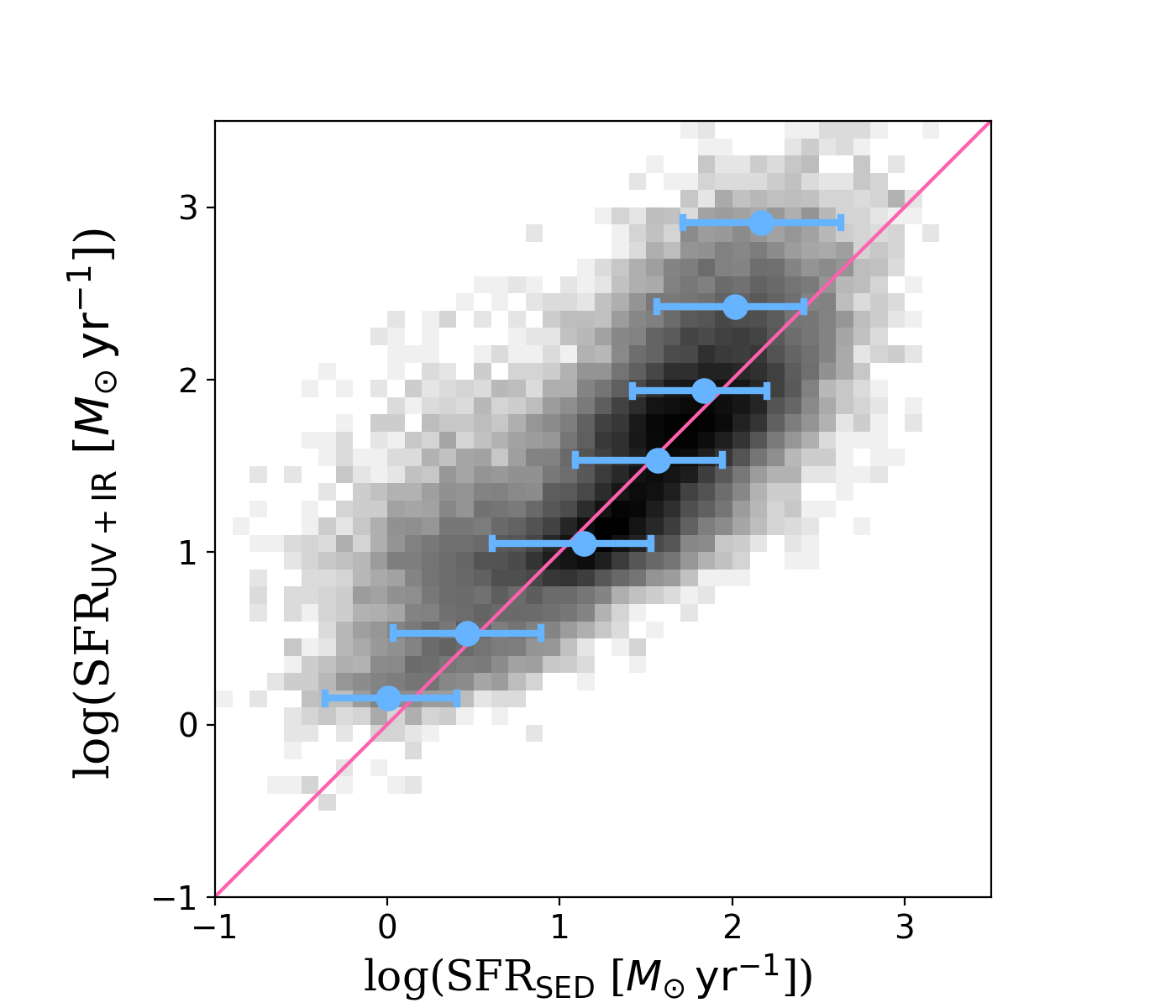}
\caption{Comparison of the SFRs obtained from the COSMOS-2015 catalog, i.e., SFR$_{\rm SED}$, to the SFRs obtained from the ladder of SFR, i.e., SFR$_{\rm UV+IR}$. Number densities are displayed in log-scale. Blue circles represent the median value of log(SFR$_{\rm SED}$) in log(SFR$_{\rm UV+IR}$) bins, starting from -0.25\,dex and with a bin size of 0.5\,dex. Error bars correspond to the 16th and 84th percentiles. The pink line is the one-to-one relation.} 
\label{fig:SFR_compare}
\end{figure}

\section{Data}
\label{sec:data}
\subsection{The A$^3$COSMOS dataset}
\label{subsec:A$^3$COSMOS}
The A$^3$COSMOS project aims at homogeneously processing (i.e., calibration, imaging and source extraction) of all ALMA projects targeting the COSMOS field that are publicly available, and providing these calibrated visibilities, cleaned images, and value-added source catalog via a single access portal \citep[][]{2019ApJS..244...40L}. In our analysis we use the A$^3$COSMOS 20200310 version\footnote{A$^3$COSMOS 20200310 version: \url{https://sites.google.com/view/a3cosmos/data/dataset_v20200310}}, i.e., all ALMA projects publicly available over the COSMOS field by the 10th of March 2020. This database contains 80 independent ALMA projects with band-6 and/or -7 observations. The interferometric calibration was performed by the A$^3$COSMOS project using the Common Astronomy Software Applications package \citep[CASA;][]{CASA} and the calibration scripts provided by the ALMA observatory. During this calibration step, a weight is assigned to each calibrated visibility and this weight is key for the accuracy of our stacking analysis (see Sect.~\ref{subsec:stack}). Unfortunately, the definition of these weights changed between the CASA versions used for the ALMA cycles 0, 1 and 2, and those used for ALMA cycles $>3$. For this reason, we excluded from our analysis all cycle 0, 1, and 2 ALMA projects. Our final database contains 64 ALMA projects, 39 in band-6 and 25 in band-7. These projects include 1893 images (equivalently ALMA pointings), which contain a total of 1002 sources with $>4.35\sigma$ \citep{2019ApJS..244...40L}.

\subsection{Our Sample}
\label{subsec:our_sample}
COSMOS is a deep extragalactic blind survey of two square degrees on the sky centered at R.A. (J2000) = 10$^{\rm h}$00$^{\rm m}$28.6$^{\rm s}$, Dec. = +02$\degr$12$\arcmin$21.0$\arcsec$ \citep{2007ApJS..172....1S}. This survey has been carried out over 46 broad and narrow bands probing the entire electromagnetic spectrum, from X-ray \citep[e.g., XMM-Newton;][]{2009A&A...497..635C}, ultraviolet \citep[e.g., GALEX;][]{2007ApJS..172..468Z}, optical \citep[e.g.,][]{2007ApJS..172..196K, 2007ApJS..172....9T}, infrared \citep[e.g., \textit{Spitzer};][]{2007ApJS..172...86S}, to radio wavelengths \citep[e.g., VLA;][]{2010ApJS..188..384S, 2017A&A...602A...1S}. These observations have triggered numerous spectroscopic follow-up studies, providing nowadays more than 10,000 spectroscopic redshifts for galaxies over this field. From all these photometric and spectroscopic multi-wavelength coverage, \citet{2016ApJS..224...24L} built the reference COSMOS-2015 catalog, providing the photometry, redshift (photometric or spectroscopic), stellar mass, and SFR of more than half a million of galaxies. From their careful analysis, \citet{2016ApJS..224...24L} classified galaxies into quiescent and star-forming based on a standard rest-frame NUV-R-J selection method. The mass-completeness of their SFGs is down to stellar masses of $\sim10^{9.3}\,$M$_{\odot}$ at $z<1.75$ and $\sim10^{9.9}\,$M$_{\odot}$ at $z<3.50$ (see their Table 6). Here we select only SFGs above their mass-completeness limit. Moreover, to avoid contamination from active galactic nuclei (AGN), we exclude from our analysis all galaxies classified as AGNs based on their X-Ray luminosity \citep[$L_{\rm X}$ $\geq$ 10$^{42}$ erg s$^{-1}$;][]{2004ApJS..155..271S} using the latest COSMOS X-ray catalog of \citet{2016ApJ...817...34M}. After the selection of SFGs and exclusion of AGNs, our parent sample is left with 515,465 galaxies (green contours in Fig.~\ref{fig:completeness}). We note that photometric redshifts in the COSMOS-2015 catalog are highly reliable even up to the redshift limit of our study, i.e., $z=3.6$, with a redshift accuracy of $\sigma_{\delta z/(1+z)}$$\sim$0.028 \citep{2016ApJS..224...24L}.

To select from this parent sample galaxies residing within the MS of SFGs, one needs to accurately measured their SFRs. The COSMOS-2015 catalog provides such estimates but those are solely based on optical-to-near-infrared SED fits performed by \citet{2016ApJS..224...24L}. While reliable for stellar masses with $M_{\star}<10^{11}\,$M$_{\odot}$ and moderately star-forming galaxies, observations from the \textit{Herschel Space Observatory} have unambiguously demonstrated that such measurements are inaccurate for starbursting or massive SFGs, in which star formation can be heavily dust-enshrouded \citep[e.g.,][]{2011ApJ...738..106W, 2019MNRAS.485.5733Q}. To accurately measure the SFR of all galaxies in our parent sample, we use thus the approach advocated by \citet{2011ApJ...738..106W}, i.e., applying to each galaxy the best dust-corrected star formation indicator available (the so-called ladder of SFR indicator; see below for details). The SFR of galaxies for which infrared observations were available, were obtained by combining their un-obscured and obscured SFRs, following \citet{1998ARA&A..36..189K} for a \citet{2003PASP..115..763C} initial mass function,
\begin{equation}
{\rm SFR}_{\rm UV+IR}[{\rm M_{\odot}\,yr^{-1}}]=1.09 \times 10^{-10} (L_{\rm IR}[{\rm L_{\odot}}]+3.3 \times L_{\rm UV}[{\rm L_{\odot}}]),
\end{equation}
where the rest-frame $L_{\rm UV}$ at 2300$\AA$ was taken from the COSMOS-2015 catalog, and the rest-frame $L_{\rm IR}$ = $L({\rm 8-1000 \,\mu m})$ was calculated from their mid/far-infrared photometry\footnote{Among the 3037 galaxies of our final sample (see below), 972 (32$\%$) have mid-infrared 24$\,\mu$m photometry and among those 482 (16$\%$) have multiple far-infrared photometry. Among the 1376 galaxies of our final sample with stellar mass $>10^{10}\,$M$_{\odot}$ (those detectable by our stacking analysis; see Sect.~\ref{sec:results}), 852 (62$\%$) have mid-infrared 24$\,\mu$m photometry and among those 461 (33$\%$) have multiple far-infrared photometry.}. For galaxies with multiple far-infrared photometry in the COSMOS-2015 catalog\footnote{The \textit{Herschel} photometry in the COSMOS-2015 catalog is based on the 24$\,\mu$m prior source extraction performed by the PEP \citep{2011A&A...532A..90L} and HerMES \citep{2012MNRAS.424.1614O} consortia.}, we estimated their $L_{\rm IR}$ by fitting their PACS and SPIRE flux densities \citep[][]{2011A&A...532A..90L, 2012MNRAS.424.1614O} with the SED template library of \citet{2001ApJ...556..562C}. For galaxies without a multiple far-infrared photometry but a mid-infrared 24$\mu$m detection in the COSMOS-2015 catalog, we estimated their $L_{\rm IR}$ by scaling the MS SED template of \citet{2011A&A...533A.119E} to their 24$\mu$m flux densities \citep[][]{2009ApJ...703..222L}. This particular MS SED template was chosen because it provides accurate 24$\mu$m-to-$L_{\rm IR}$ conversions over the redshift and stellar mass ranges probed in our study \citep{2011A&A...533A.119E}. For galaxies without any mid- and far-infrared photometry, we used the SFRs measured by \citet{2016ApJS..224...24L} and which were obtained by fitting their optical-to-near-infrared photometry with the \citet{2003MNRAS.344.1000B} SED model. 
We verified that towards intermediate SFRs, i.e., where the fraction of galaxies with a mid/far-infrared detection starts to decrease (i.e., $0<\,$log(SFR$_{\rm IR+UV}$)$\,<1.5$), our (UV$+$IR)-based SFR measurements agree with those solely based on this optical-to-near-infrared SED fits, with a median log(SFR$_{\rm IR+UV}/$SFR$_{\rm SED}$) of $0.09^{+0.39}_{-0.53}$ (Fig.~\ref{fig:SFR_compare}). This agreement ensures a smooth transition between the different steps of our ladder of SFR indicators. Also, among the 269 galaxies of our final sample with stellar mass $>10^{10}\,$M$_{\odot}$ (that is, those detectable by our stacking analysis; see Sect.~\ref{sec:results}) and with SFR>100 M$_{\odot}$ yr$^{-1}$, only 54 have their SFRs solely based on their SED fits and thus potentially underestimated by $\sim0.3 - 0.5\,$dex (see Fig.~\ref{fig:SFR_compare}). Finally, we note that at high SFRs, where a high fraction of galaxies are individually-detected by ALMA, our SFRs agree with those from the A$^3$COSMOS catalog, i.e., inferred with MAGPHYS \citep{2008MNRAS.388.1595D, 2015ApJ...806..110D} SED fitting combining the COSMOS-2015 photometry with super-deblended \textit{Herschel} \citep{2018ApJ...864...56J} and ALMA photometry.

From their redshift, stellar mass, and SFR, we can measure the offset of each of these galaxies from the MS, i.e., $\Delta$MS = ${\rm log(SFR}(z,{\rm SM})/{\rm SFR}_{\rm MS}(z,{\rm SM}))$. To this end, we used the main-sequence calibration of \citet{2020ApJ...899...58L}, as it is also based on the mass-complete COSMOS-2015 catalog:
\begin{equation}
\begin{aligned}
{\rm log(SFR}_{\rm MS}(z,{\rm SM}))= S_{0} - a_{1}t -{\rm log}\left( 1+\left(\frac{10^{M^{'}_{t}}}{10^{M}} \right) \right),\\
M^{'}_{t}=M_{0}-a_{2}t,
\end{aligned}
\end{equation}
where $M$ is log($M_{\star}/M_{\odot}$), $t$ is the age of the universe in Gyr, $S_{0}$=2.97, $M_{0}$=11.06, $a_{1}$=0.22, and $a_{2}$=0.12. Our mass-complete sample of MS galaxies was then constructed by selecting galaxies with $\Delta$MS between $-0.5$ and $0.5$ \citep[e.g.,][]{2014MNRAS.443...19R}. This sample contains 92,739 galaxies.

Finally, from this mass-complete sample of MS galaxies, we selected those with an ALMA band-6 ($\sim\,243$\,GHz) or band-7 ($\sim\,324\,$GHz) coverage in the A$^3$COSMOS database (see Sect.~\ref{subsec:A$^3$COSMOS}). Here, we only consider galaxies well within the ALMA primary beam, i.e., where the primary beam response is higher than 0.5. This conservative primary beam cut was used because uncertainties in the primary beam response far from the phase center can significantly affect our stacking analysis (see Sect.~\ref{subsec:stack}). In addition, to avoid contamination by bright neighbouring sources, we excluded from our analysis galaxy pairs ($<2\farcs0$) with $S_{\rm ALMA}^1/S_{\rm ALMA}^2>2$ or $M_{\star}^1/M_{\star}^2>3$ (for ALMA undetected galaxies, assuming a first-order $M_{\rm gas}-M_{\star}$ correlation). About 8$\%$ of our galaxies are excluded by these criteria. However, we note that most of these excluded galaxy pairs ($\sim95\%$) are due to projection effects ($\Delta z>0.05$). This implies that the exclusion of these galaxies does not introduce any biases into our final ALMA-covered mass-complete sample of MS galaxies. There are 3,037 galaxies in this final sample. The left panel of Fig.~\ref{fig:completeness} shows the stellar mass and redshift distribution of our parent and final samples. Our final sample probes a broad range in redshifts and stellar masses, similar to that probed by our parent sample.
We verified that our parent and final samples have consistent stellar mass, redshift and $L_{\rm IR}$ distributions, with Kolmogorov-Smirvov probabilities of 99$\%$, 99$\%$, and 96$\%$ of being drawn from the same distribution, respectively.

The ALMA archive cannot be treated as a real blind survey and thus our ALMA coverage selection criteria could have introduced a bias in our final ALMA-covered mass-complete MS galaxy sample. As an example (though rather unrealistic), if all ALMA projects in COSMOS would have targeted MS galaxies with $\Delta$MS$=0.3\,$dex, our final sample would naturally be biased toward this population and thus not be representative of the entire MS galaxy population. A simple way to test the presence of such bias is to compare the $\Delta$MS distributions of our final and parent samples for different stellar mass bins (Fig.~\ref{fig:completeness}; right panels). As expected, our parent sample (green histogram) exhibits in all stellar mass bins a Gaussian distribution centered at 0 and with a $0.3\,$dex dispersion. At low stellar masses ($M_{\star}<10^{10.0}\,$M$_{\odot}$), our final sample follows the same distribution, with a Kolmogorov-Smirvov 99$\%$ probability of being drawn from the same sample (this finding remaining true even if we divide further these stellar mass bins into several redshift bins). Indeed, in these low stellar mass bins, only 5$\%$ of our galaxies are located at the phase center of the ALMA image and thus were the primary target of the ALMA observations. In the highest stellar mass bins, we note, however, that the $\Delta$MS distribution of our final sample is significantly skewed towards high $\Delta$MS values (this finding remaining again true if we divide further these stellar mass bins into several redshift bins). In these stellar mass bins, about 63$\%$ of our galaxies are the primary targets of the ALMA observations (i.e., located at the phase center), and thus potentially affected by complex and uncontrollable selection biases. Excluding these primary targets from our galaxy sample yields $\Delta$MS distribution in much better agreement with those of our parent sample. In the rest of our analysis, at high masses, we will show our stacking results before and after excluding these primary-target galaxies. In addition, we will account for these $\Delta$MS distributions while fitting the cosmic and stellar mass evolution of the mean molecular gas content of MS galaxies.

\section{Method}
\label{sec:method}
ALMA has revolutionized the study of high-redshift SFGs at (sub)millimeter wavelengths. Nevertheless, even with its un-parallel sensitivity, ALMA cannot detect within a reasonable observing time MS galaxies with $M_{\star}<10^{10.5}\,$M$_\odot$ at $z>0.5$. Consequently, despite including all individually-detected galaxies within the A$^3$COSMOS images (i.e., primary targets and serendipitous detections), the final sample of \citet{2019ApJ...887..235L} is still mostly restricted to the high-mass end of the SFG population. The emission of such low-mass high-redshift SFGs captured within these images is too faint to be individually-detected, and thus remains unexploited. To statistically retrieve the faint emission of this SFG population, we need to perform a stacking analysis. As already mentioned, stacking over the entire A$^3$COSMOS dataset presents two challenges when compared to standard stacking analysis performed with \textit{Spitzer}, \textit{Herschel}, or individual ALMA projects. Indeed, the A$^3$COSMOS database is heterogeneous in terms of observed frequencies and spatial resolution. The frequency-heterogeneity problem is simply solved by a prior re-scaling of each individual dataset to a common rest-frame luminosity frequency using locally calibrated submillimeter SEDs (Sect.~\ref{subsec:band}), while the spatial resolution-heterogeneity problem is solved by performing our stacking analysis in the $uv$-domain (Sect.~\ref{subsec:stack}).

In the following, we describe in detail the different steps of our stacking analysis, while the validation of this methodology via Monte Carlo simulations is presented in Appendix~\ref{appendix: simu}.
 
\subsection{From observed-frame flux densities to rest-frame luminosities}
\label{subsec:band}
The A$^3$COSMOS observations were performed at different frequencies and the galaxies to be stacked also lie at slightly different redshifts. Therefore, prior to proceeding with our stacking analysis, we needed to convert the ALMA observations of a given galaxy from observed flux density to its rest-frame luminosity at 850\,$\mu$m, i.e., $L_{\rm 850}^{\rm rest}$. To do so, we used the MS SED templates of \citet{2012ApJ...757L..23B}, which accurately capture the monotonic increase of the dust temperature of MS galaxies with redshift \citep[e.g.,][]{2012ApJ...760....6M, 2014A&A...561A..86M}.  First, we computed the SED template luminosity ratio at rest-frame 850$\,\mu$m and the observed rest-frame wavelength of the galaxy of interest,
\begin{equation}
\label{eq:gamma}
\Gamma^{\rm SED}=L^{\rm SED}_{850}\,/\,L^{\rm SED}_{\lambda_{\rm obs} / (1 + z) }.
\end{equation}
The observed ALMA visibility amplitudes toward this galaxy, i.e., $|\,V(u,v,w)\,|_{\rm \lambda_{obs}}$, -- which are in units of flux density -- were then converted into rest-frame 850$\,\mu$m luminosity following:
\begin{equation}
|L(u,v,w)|_{\rm 850}^{\rm rest}=4\,\pi\,D^{2}_{\rm L} \times |\,V(u,v,w)\,|_{\rm \lambda_{obs}} \times \Gamma^{\rm SED} /\,(1+z),
\end{equation}
where $D_{\rm L}$ is the luminosity distance of the galaxy of interest. This re-scaling of the amplitude (and weights) of the ALMA visibilities was performed for each stacked galaxy using the CASA tasks \texttt{gencal} and \texttt{applycal}. 

\subsection{Stacking in the $uv$-domain}
\label{subsec:stack}
Stacking in the $uv$-domain relies on the exact same principle as aperture synthesis. The only difference is that one combines multiple baselines pointing at the same galaxy population instead of multiple baselines pointing at the same galaxy. The tools or tasks needed to perform stacking in the $uv$-domain are thus all readily available in CASA. For each of our stellar mass-redshift bin and each galaxy within these bins, we proceeded as follow. First, we time- and frequency-average their measurement set, producing one averaged visibility per ALMA scan (lasting typically 30$s$ and originally divided into $10\times3s$ integration) and ALMA spectral window (probing typically $2\,$GHz and originally divided into 100s of channels). This step, which was performed using the CASA task \texttt{split}, is crucial to keep the volume of our final stacked measurement sets within current computing capabilities. These averaged visibilities were then re-scaled from observed-frame flux density into rest-frame 850$\,\mu$m luminosity using the CASA tasks \texttt{gencal} and \texttt{applycal} (see Sect.~\ref{subsec:band}). Finally, the phase center of these averaged and re-scaled visibilities were shifted to the coordinate of the stacked galaxy. This step was performed using the CASA package \texttt{STACKER} \citep{2015MNRAS.446.3502L} following, 
\begin{equation}
L_{\rm shifted}(u,v,w)_{\rm 850}^{\rm rest}=L(u,v,w)_{\rm 850}^{\rm rest}\,\frac{1}{A_{N}(\hat{S}_{k})}e^{\frac{2\pi}{\lambda}i B \cdot (\hat{S}_{0}-\hat{S}_{k})}
\end{equation}
where $L(u,v,w)_{\rm 850}^{\rm rest}$ is the averaged and re-scaled visibility, $\hat{S}_{0}$ is a unit vector pointing to the original phase center, $\hat{S}_{k}$ is a unit vector pointing to the position of the stacked galaxy, A$_{N}(\hat{S}_{k})$ is the primary beam attenuation in the direction $\hat{S}_{k}$, $B$ is the baseline of the visibility. The final stacked measurement set of a given stellar mass-redshift bin was then obtained by concatenating the shifted, re-scaled and averaged visibilities, i.e., $L_{\rm shifted}(u,v,w)_{\rm 850}^{\rm rest}$, of all galaxies within this bin using the CASA task \texttt{concat}. Because all these steps were performed in CASA, the original weights of all visibilities (i.e., those accounting for their system temperature, channel width, integration time\dots) were properly re-normalized and could thus be used for the forthcoming $uv$-model fit and image processing.

To measure the stacked rest-frame 850$\,\mu$m luminosity of each of our stellar mass--redshift bins, i.e., $L_{\rm 850}^{\rm stack}$, we used two different approaches. First, we extracted this information from the $uv$-domain by fitting a single component model to the stacked measurement set. This fit was performed using the CASA task \texttt{uvmodelfit}, assuming a single Gaussian component and fixing its position to the stacked phase center. Second, we measured $L_{\rm 850}^{\rm stack}$ from the image-domain. To do so, we imaged the stacked measurement set with the CASA task \texttt{tclean}, using Briggs natural weighting and cleaning the image down to $3\sigma$. Then, we fitted a 2D Gaussian model to the cleaned image using the Python Blob Detector and Source Finder package \citep[\texttt{PyBDSF;}][]{2015ascl.soft02007M}. For all our stellar mass--redshift bins, these two approaches agreed within the uncertainties.

Our $uv$-domain and image-domain fits provide us also with the mean size (or upper limit) of the galaxy population in a given stellar mass--redshift bin. From the intrinsic (i.e., beam-deconvolved) full width at half maximum ($FWHM$) of the major axis outputted by \texttt{uvmodelfit} or \texttt{PyBDSF}, we define the effective --equivalently half-light-- radius ($R_{\rm eff}$) of the stacked population following \citet{2019A&A...625A.114J}, i.e., $R_{\rm eff}\approx FWHM/2.43$. Then, we express these mean size measurements in form of circularized radii, $R^{\rm circ}_{\rm eff}$,
\begin{equation}
R_{\rm eff}^{\rm circ} = R_{\rm eff} \times \sqrt{\frac{b}{a}},
\end{equation}
where $b/a$ is the axis ratio measured with \texttt{uvmodelfit} or \texttt{PyBDSF}.
 
Finally, to infer the uncertainties associated to these stacked rest-frame 850$\,\mu$m luminosity and size measurements, we used a standard re-sampling method. These uncertainties account not only for the instrumental noise in the stacked measurement set (i.e., the detection significance) but also for the intrinsic distribution of $L_{\rm 850}$ and size within the stacked galaxy population. For a stellar mass--redshift bin containing $N$ galaxies, we performed $N$ different realizations of our stacking analysis, removing in each realization one galaxy of the stacked sample. The uncertainties on $L_{\rm 850}^{\rm stack}$ and size are then given by the standard deviation of these quantities measured over these realizations multiplied by $\sqrt{N}$. We note that because there is a possible mismatch of $\sim 0\farcs2$ between the stacked optical-based position and the actual (sub)millimeter position of the sources \citep[e.g.,][]{2018A&A...616A.110E}, the average FIR sizes inferred in our study could be slightly overestimated. This is further discussed in Sect.~\ref{subsec: FIR sizes}.

\begin{figure*}
\centering
\includegraphics[width=0.81\textwidth]{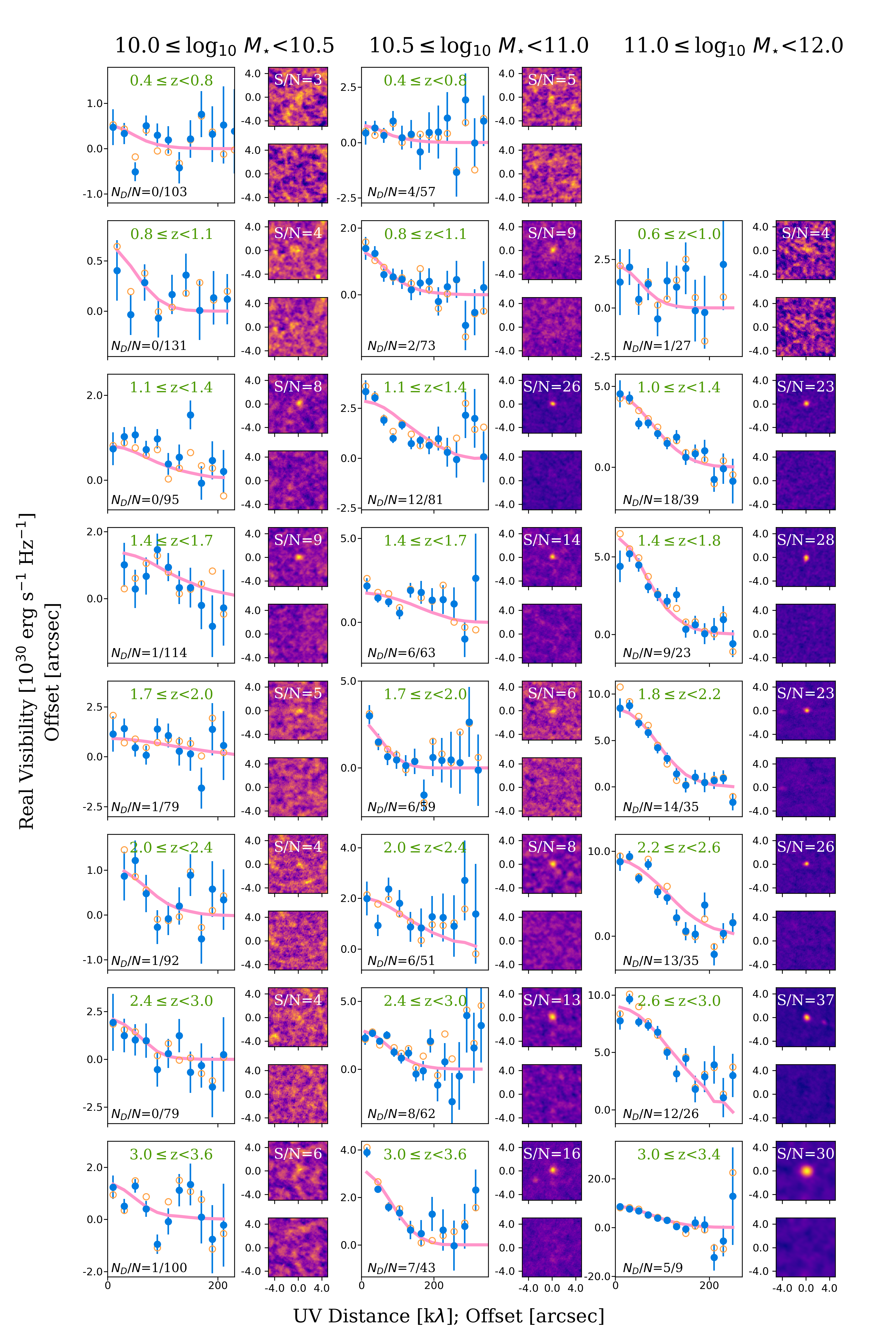}
\caption{Results of our stacking analysis for MS galaxies in the $uv$- and image-domain. For each stellar mass--redshift bin, the left panel shows the single component model (pink solid line) fitted to the (stacked) mean visibility amplitudes (blue filled circles) using the \texttt{CASA} task \texttt{uvmodelfit}. Open orange circles show the median visibility amplitudes, which are consistent, within the uncertainties, with the mean visibility amplitudes. The top-right and bottom-right panels show, respectively, the stacked and residual images, the latter being obtained by subtracting from the former the single 2D Gaussian component fitted by \texttt{PyBDSF}. The number of individually detected galaxies ($N_{\rm D}$) and the number of stacked galaxies ($N$) in each stellar mass--redshift bin is reported in the left panel (i.e., $N_{\rm D}/N$) , while the detection significance, i.e., S$/$N$_{\rm peak}$, is reported in the upper right panel.}
\label{fig:uv_res_median}
\end{figure*}

Note that although some studies have used median stacking to mitigate the contribution of bright outliers to the stacked flux densities \citep[e.g.,][]{2020ApJ...903..138A, 2020A&A...641A.118F, 2020A&A...643A...4F, 2021MNRAS.506.3641G, 2021ApJ...909...73J}, we decided to perform our analysis using a mean stack, i.e., in the $uv$-domain, our models are fitted to the weighted mean visibility amplitudes and our images are created by \texttt{tclean} using weighted mean visibilities. This choice was made for the following reasons: (i) the impact of bright outliers is already mitigated by our $-0.5<\Delta$MS$<0.5$ selection, which by construction excludes gas-rich starbursts; (ii) the impact of bright outliers is accounted for in our uncertainties (i.e., re-sampling method); and finally (iii) \citet{2015A&A...575A..74S} and \citet{2020ApJ...899...58L}, which thoroughly tested mean and median stacking, concluded both that median stacking is biased toward higher values at low S/N because the median is not a linear operation and that the stacked distribution is intrinsically a log-normal distribution skewed toward bright sources. As a result, median stacked fluxes are difficult to interpret and are often not measuring the median nor mean fluxes, but something in between. We note, however, that the median visibility amplitudes of each of our stacked bins are consistent, within the uncertainties, with the mean visibility amplitudes (see open symbols in Fig.~\ref{fig:uv_res_median}).

\subsection{From rest-frame 850$\,\mu$m luminosities to molecular gas masses}
\label{subsec:measure}
The literature contains a plethora of relations linking molecular gas mass of galaxies with their (sub)millimeter luminosities \citep[e.g., ][]{2013MNRAS.436..479B, 2015ApJ...799...96G, 2017ApJ...837..150S, 2018MNRAS.478.1442B, 2018MNRAS.481.3497S, 2019ApJ...880...15K}. All of them rely on an assumed gas-to-dust mass ratio (or a direct 870$\,\mu$m luminosity-to-gas mass ratio) that might or might not depend on the metallicity. \citet{2019ApJ...887..235L} thoroughly studied how these different relations influence our molecular gas mass estimation, using a sample of galaxies down to a stellar mass of $\sim 10^{10.3} M_{\odot}$. They found that metallicity-dependent relations \citep{2014A&A...563A..31R, 2015ApJ...800...20G} and the 850$\,\mu$m luminosity-dependent relation of \cite{2017MNRAS.468L.103H} only differ by $\sim$0.15--0.25\,dex (which is comparable to the observed scatter), and that the relation of \citet{2017MNRAS.468L.103H} provided the best agreement with local observations \citep[e.g.,][]{2018MNRAS.478.1442B, 2018MNRAS.481.3497S}. They concluded that the 850$\,\mu$m luminosity-dependent relation is thus the most preferable relation for galaxies down to a stellar mass of $M_{\star} \sim 10^{10.3} M_{\odot}$ and for which no metallicity measurements are available. Based on their analysis, we decided to use this empirically-calibrated relation of \citet{2017MNRAS.468L.103H}. The mean molecular gas mass of a given galaxy population (i.e., $M_{\rm mol}$) is thus computed from their stacked rest-frame 850$\,\mu$m luminosities following,
\begin{equation}
{\rm log_{10}}\ M_{\rm mol}=(0.93\pm0.01)\cdot {\rm log_{10}}\ L_{850}-(17.74 \pm 0.05),
\end{equation}
where $M_{\rm mol}$ already includes the $1.36$ correction factor to account for helium and assumes a CO-to-$M_{\rm mol}$ conversion factor (i.e., $\alpha_{\rm CO}$) of 6.5\,(K\,km\,s$^{-1}$\,pc$^{2}$)$^{-1}$.

In Section~\ref{subsec:Rj-to-Mol-uncertainties} and Appendix~\ref{appendix: conversion}, we thoroughly present and discuss the impact on our results of using different gas mass calibration relations. In brief, the main conclusions of our paper are not qualitatively affected by this particular choice; the  H17  method  yields  measurements  which  are  bracket by those inferred from other relations; and, finally, measurements obtained using H17 are in good agreement with \citet{2020ARA&A..58..157T} at high stellar masses, i.e., where this latter study can be considered as the reference.

\begin{figure*}
\centering
\includegraphics[width=1.0\textwidth]{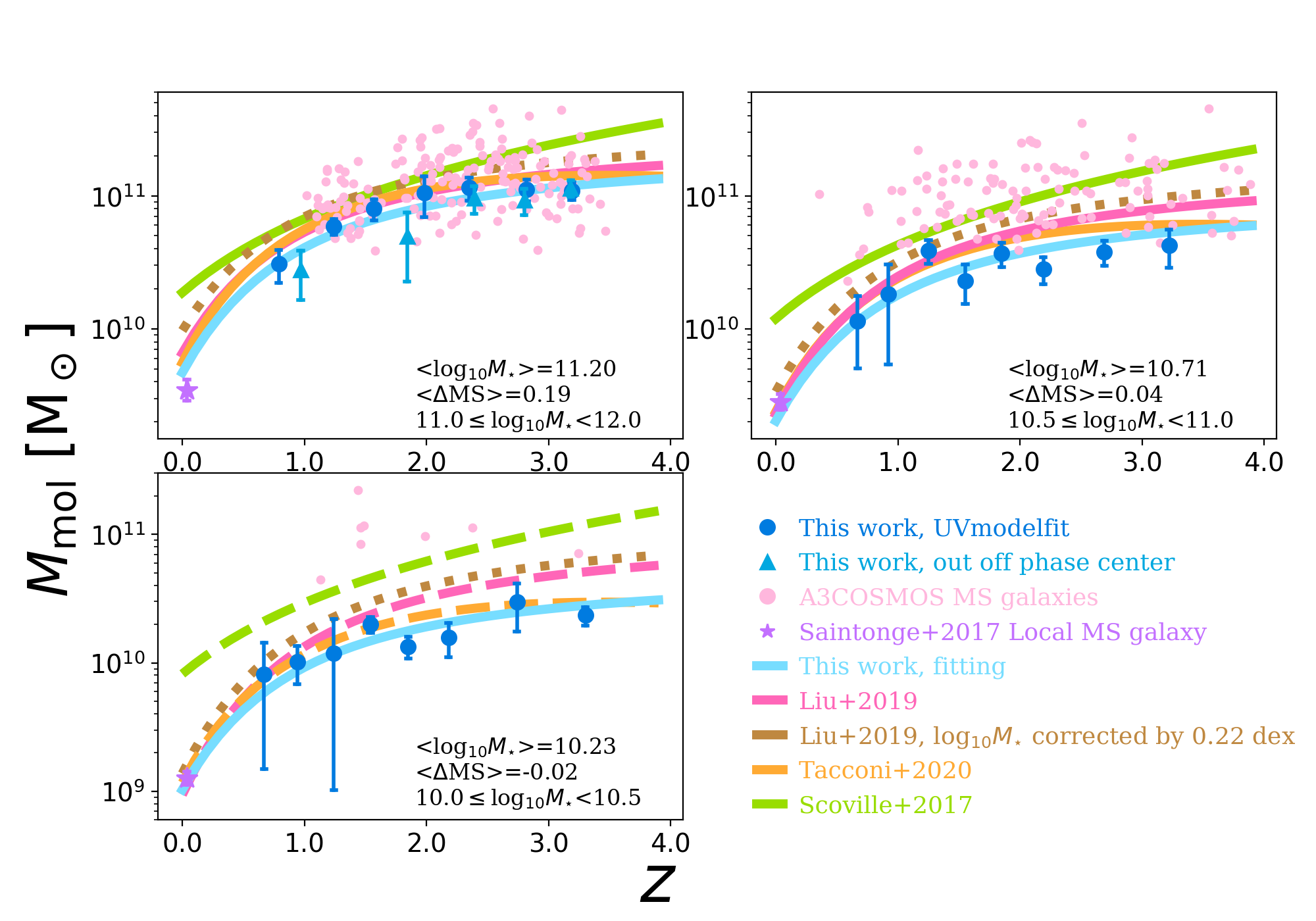}
\caption{Redshift evolution of the mean molecular gas mass of MS galaxies in three stellar mass bins, i.e., $10^{11}\,\leq M_\star/{\rm M}_\odot<10^{12}$, $10^{10.5}\,\leq M_\star/{\rm M}_\odot<10^{11}$, and $10^{10}\,\leq M_\star/{\rm M}_\odot<10^{10.5}$. Blue circles show our $uv$-domain measurements, while in the highest stellar mass bin blue triangles show those obtained after excluding ALMA primary-target galaxies from our stacked sample (see Sect.~\ref{subsec:our_sample}). Pink circles are individually-detected MS galaxies taken from the A$^3$COSMOS catalog \citep[][]{2019ApJ...887..235L}, while purple stars present the local reference taken from \citet[][]{2017ApJS..233...22S}. Lines show the analytical evolution of the gas fraction as inferred from our work (blue lines), from \citet[][green line]{2017ApJ...837..150S}, from \citet[][pink line]{2019ApJ...887..235L}, from \citet[][orange]{2020ARA&A..58..157T}, and finally, from \citet[][brown dotted line]{2019ApJ...887..235L} but this time accounting for the systematic 0.22\,dex offset observed between their and our stellar mass estimates. In our lower stellar mass bin, lines from the literature are dashed as they mostly rely on extrapolations. Note that here and in all following figures, the values of $\langle M_{\star}\rangle$ and $\langle\Delta {\rm MS}\rangle$ given in each panels are simply used to plot the analytical evolution of the gas fraction. These values naturally vary for each stacked measurements and is accounted for by our MCMC analysis. This avoids averaging biases that could arise if one would simply fit our stacked measurements using $\langle t_{\rm cosmic}\rangle$, $\langle M_{\star}\rangle$, and $\langle\Delta {\rm MS}\rangle$.
}
\label{fig:gas}
\end{figure*}
\begin{figure}
\centering
\includegraphics[width=\columnwidth]{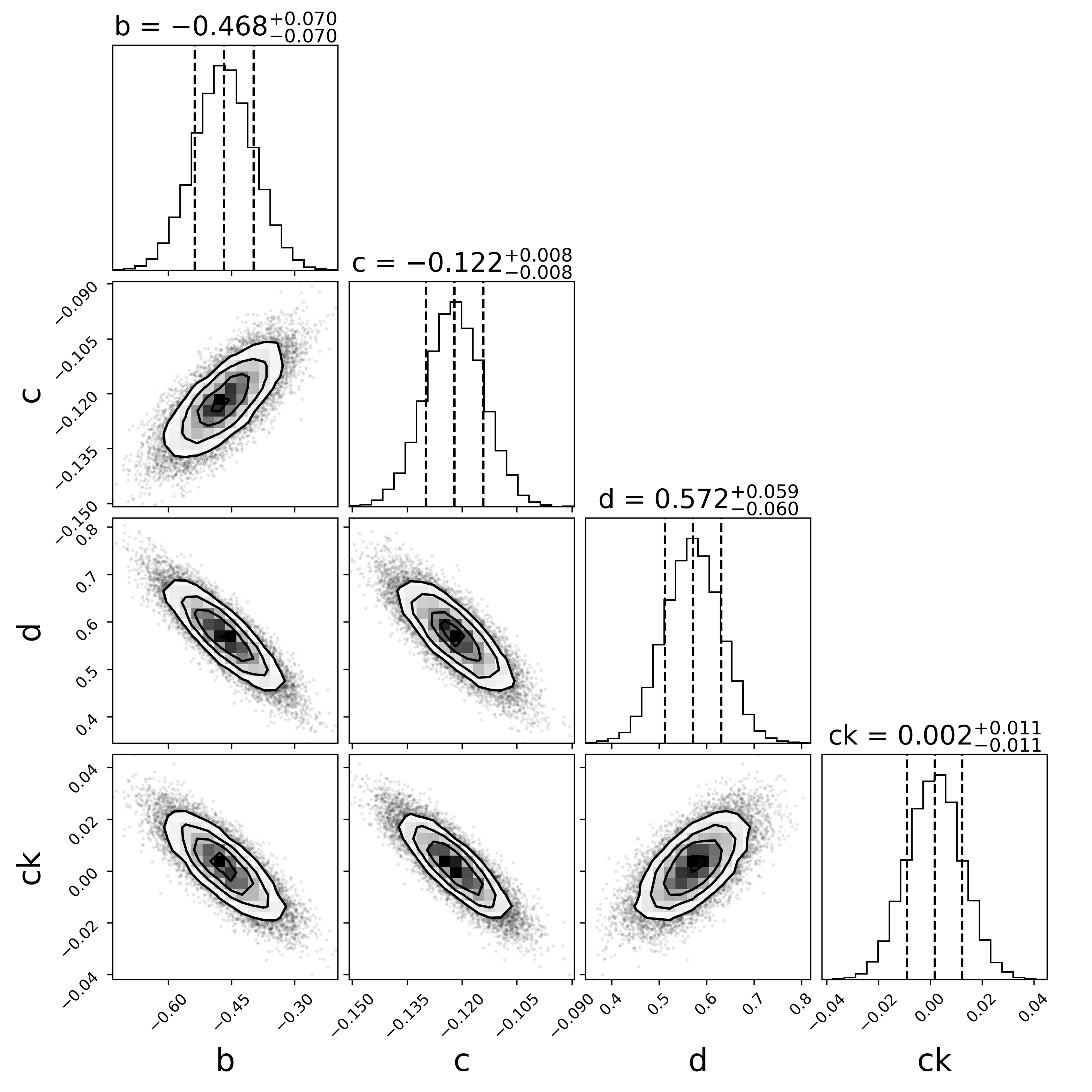}
\caption{Probability distributions of the parameters in Eq.~\ref{eq:gas fraction}, as found by fitting our stacked measurements using a MCMC analysis. The dashed vertical lines show the 16th, 50th, and 84th percentiles of each distribution.}
\label{fig:MCMC}
\end{figure}
\begin{figure}
\centering
\includegraphics[width=\columnwidth]{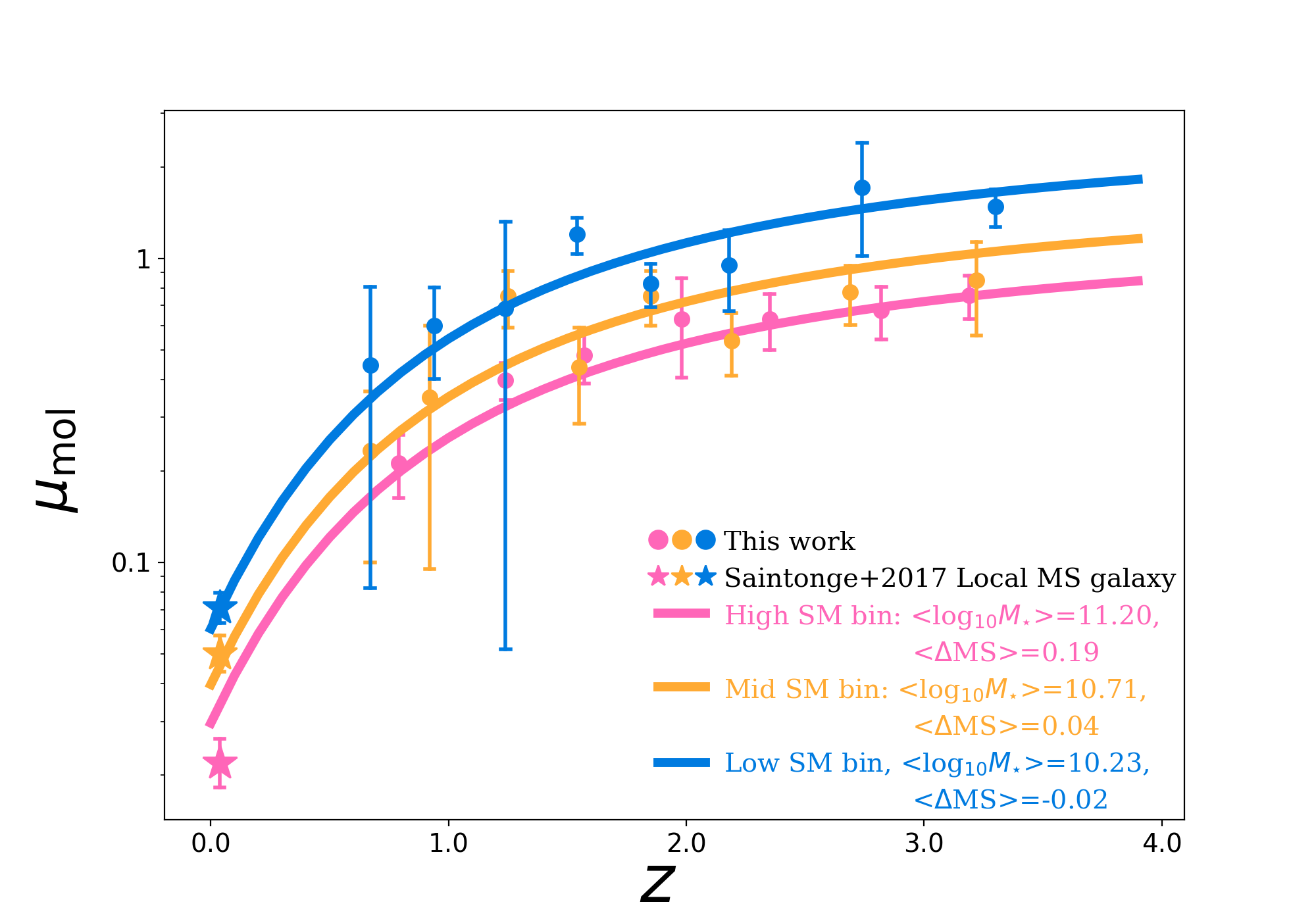}
\caption{Redshift evolution of the mean molecular gas fraction of MS galaxies. Circles show the mean molecular gas fraction from our work. Stars present the local reference taken from \citet[][]{2017ApJS..233...22S}. Lines display the analytical evolution of the molecular gas fraction inferred from our work. Symbols and lines are color-coded by stellar mass, i.e., pink for $10^{11}\,\leq M_\star/{\rm M}_\odot<10^{12}$, orange for $10^{10.5}\,\leq M_\star/{\rm M}_\odot<10^{11}$, and blue for $10^{10}\,\leq M_\star/{\rm M}_\odot<10^{10.5}$.}
\label{fig:frac}
\end{figure}

\begin{sidewaystable*}
\footnotesize
\centering
\caption{Molecular gas mass and size properties of main-sequence galaxies.}
\label{tab:sm}
\begin{tabular}{cccccccccccccccccc}
\hline
$M_\star$ & $z$ & $N$ & $N_{D}$ & $\langle$z$\rangle$ & $\langle M_\star\rangle$ & $\langle \rm SFR \rangle$ & $\langle$$\Delta$MS$\rangle$ & $\langle$$\nu_{\rm obs}$$\rangle$ & $\langle$$L_{850-uv}$$\rangle$ & S/N$_{\rm peak}$ & $M_{\rm mol-uv}$ & $M_{\rm mol-py}$ & $\theta_{\rm beam}^{\rm circ}$ & $\theta_{\rm mol-uv}^{\rm circ}$ & $\theta_{\rm mol-py}^{\rm circ}$ & $R_{\rm \rm eff-uv}^{\rm circ}$ & $R_{\rm \rm eff-py}^{\rm circ}$ \\
 & & & &  & log$_{10}$\,[M$_{\odot}]$ & log$_{10}$\,[M$_{\odot}\,$yr$^{-1}$]  & & GHz & 10$^{30}$\,erg\,s$^{-1}$Hz$^{-1}$ & & log$_{10}$\,[M$_{\odot}]$ & log$_{10}$\,[M$_{\odot}]$ & arcsec & arcsec & arcsec & kpc & kpc\\
 (1) & (2) & (3) & (4) & (5) & (6) & (7) & (8) & (9) & (10) & (11) & (12) & (13) & (14) & (15) & (16) & (17) & (18)\\
\hline
11.0$\leq$ log$_{10}\ M_{\star}$ <12.0 & 0.6$\leq$ z <1.0 & 27 & 1 & 0.79 & 11.16 & 1.58 & 0.09 & 298.02 & 2.3$_{\pm 0.7}$ & 4 & 10.5$_{\pm 0.1}$ & 10.4$_{\pm 0.2}$ & 0.51 & 1.10 & 0.83 & 3.0$_{\pm 1.}$ & 2.1$_{\pm 1.1}$\\
&1.0$\leq$ z <1.4 & 39 & 18 & 1.24 & 11.17 & 2.01 & 0.19 & 313.30 &  4.5$_{\pm 0.7}$ & 23 & 10.8$_{\pm 0.1}$ & 10.8$_{\pm 0.1}$ & 0.50 & 0.75 & 0.75 & 1.9$_{\pm 0.2}$& 1.9$_{\pm 0.2}$\\ 
&1.4$\leq$ z <1.8 & 23 & 9  & 1.57 & 11.22 & 2.21 & 0.20 & 311.27 & 6.3$_{\pm 1.2}$ & 28 & 10.9$_{\pm 0.1}$ & 10.9$_{\pm 0.1}$ & 0.47 & 0.77 &  0.78 & 2.1$_{\pm 0.1}$& 2.2$_{\pm 0.2}$\\
&1.8$\leq$ z <2.2 & 35 & 14 & 1.98 & 11.22 & 2.35 & 0.15 & 312.98 & 8.4$_{\pm 3.0}$ & 23 & 11.0$_{\pm 0.2}$ & 11.1$_{\pm 0.1}$ & 0.42 & 0.65 & 0.67 & 1.7$_{\pm 0.1}$& 1.8$_{\pm 0.2}$\\
&2.2$\leq$ z <2.6 & 35 & 13 & 2.35 & 11.26 & 2.52 & 0.24 & 310.17 & 9.3$_{\pm 1.9}$ & 26 & 11.1$_{\pm 0.1}$ & 11.1$_{\pm 0.1}$ & 0.42 & 0.59 & 0.62 & 1.5$_{\pm 0.1}$& 1.6$_{\pm 0.1}$\\
&2.6$\leq$ z <3.0 & 26 & 12 & 2.82 & 11.22 & 2.57 & 0.22 & 297.10 & 9.0$_{\pm 1.8}$ & 37 & 11.0$_{\pm 0.1}$ & 11.1$_{\pm 0.1}$ & 0.72 & 0.79 & 0.87 & 1.1$_{\pm 0.1}$& 1.5$_{\pm 0.1}$\\
&3.0$\leq$ z <3.4 & 9  & 5  & 3.19 & 11.16 & 2.76 & 0.38 & 271.01 & 8.8$_{\pm 1.5}$ & 30 & 11.0$_{\pm 0.1}$ & 11.0$_{\pm 0.1}$ & 1.39 & 1.60 & 1.60 & 2.6$_{\pm 0.2}$& 2.5$_{\pm 0.2}$\\
\hline
11.0$\leq$ log$_{10}\ M_{\star}$ <12.0 & 0.6$\leq$ z <1.4 & 17 & 0 & 0.97 & 11.17 & 1.62 & -0.03 & 270.21 & 2.0$_{\pm 0.9}$ & 5 & 10.4$_{\pm 0.2}$ & 10.5$_{\pm 0.2}$ & 0.83 & 0.92 & 1.02 & 1.4$_{\pm 0.7}$& 2.0$_{\pm 0.9}$\\
&1.4$\leq$ z <2.2 & 20 & 7 & 1.84 & 11.19 & 2.18 & 0.00 & 301.15 & 3.8$_{\pm 2.1}$ & 13 & 10.7$_{\pm 0.2}$ & 10.7$_{\pm 0.2}$ & 0.69 & 0.81 & 0.87 & 1.5$_{\pm 0.2}$& 1.9$_{\pm 0.2}$\\ 
(off the phase center) &2.2$\leq$ z <2.6 & 13 & 6 & 2.39 & 11.20 & 2.40 & 0.11 & 284.18 & 7.7$_{\pm 1.9}$ & 17 & 11.0$_{\pm 0.1}$ & 11.0$_{\pm 0.1}$ & 0.72 & 0.77 & 0.80 & 0.9$_{\pm 0.1}$& 1.2$_{\pm 0.2}$\\
&2.6$\leq$ z <3.0 & 8 & 3 & 2.80 & 11.20 & 2.55 & 0.19 & 277.93 & 7.4$_{\pm 1.8}$ & 16 & 11.0$_{\pm 0.1}$ & 11.0$_{\pm 0.1}$ & 0.98 & 1.13 & 1.17 & 1.8$_{\pm 0.2}$& 2.1$_{\pm 0.3}$\\
&3.0$\leq$ z <3.4 & 6  & 3  & 3.18 & 11.19 & 2.72 & 0.33 & 259.30 & 9.3$_{\pm 1.4}$ & 15 & 11.1$_{\pm 0.1}$ & 11.1$_{\pm 0.1}$ & 1.29 & 1.57 & 1.57 & 2.9$_{\pm 0.2}$& 2.9$_{\pm 0.3}$\\
\hline
10.5$\leq$ log$_{10}\ M_{\star}$ <11.0 &0.4$\leq$ z <0.8 & 57 & 4  & 0.67 & 10.69 & 1.43 & 0.18 & 316.73 & 0.8$_{\pm 0.5}$ & 5 & 10.1$_{\pm 0.2}$ & 10.1$_{\pm 0.2}$ & 0.60 & 0.82 & 0.84 & 1.6$_{\pm 0.7}$& 1.7$_{\pm 0.8}$\\
&0.8$\leq$ z <1.1  & 73 & 2 & 0.92 & 10.72 & 1.58 & 0.06 & 298.96 & 1.3$_{\pm 0.9}$ & 9 & 10.3$_{\pm 0.3}$ & 10.3$_{\pm 0.3}$ & 0.62 & 0.85 & 0.87 & 1.9$_{\pm 0.2}$& 2.0$_{\pm 0.4}$\\
&1.1$\leq$ z <1.4  & 81 & 12 & 1.25 & 10.71 & 1.83 & 0.08 & 274.75 & 2.9$_{\pm 0.6}$ & 26 & 10.6$_{\pm 0.1}$ & 10.6$_{\pm 0.1}$ & 0.55 & 0.64 & 0.68 & 1.2$_{\pm 0.1}$& 1.4$_{\pm 0.2}$\\
&1.4$\leq$ z <1.7  & 63 & 6 & 1.55 & 10.72 & 1.99 & 0.11 & 299.42 & 1.7$_{\pm 0.6}$ & 14 & 10.4$_{\pm 0.2}$ & 10.4$_{\pm 0.1}$ & 0.68 & 0.73 & 0.73 & 1.0$_{\pm 0.5}$& 1.0$_{\pm 0.2}$\\
&1.7$\leq$ z <2.0  & 59 & 6 & 1.85 & 10.69 & 1.96 & -0.02 & 284.15 & 2.7$_{\pm 0.6}$ & 6 & 10.6$_{\pm 0.1}$ & 10.6$_{\pm 0.1}$ & 0.54 & 1.08 & 1.04 & 3.3$_{\pm 0.4}$& 3.1$_{\pm 0.8}$\\
&2.0$\leq$ z <2.4  & 51 & 6 & 2.19 & 10.72 & 2.07 & -0.04 & 282.95 & 2.0$_{\pm 0.5}$ & 8 & 10.4$_{\pm 0.1}$ & 10.5$_{\pm 0.1}$ & 0.80 & 0.90 & 0.94 & 1.8$_{\pm 0.4}$& 2.6$_{\pm 0.7}$\\
&2.4$\leq$ z <3.0  & 62 & 8 & 2.69 & 10.69 & 2.16 & -0.02 & 279.43 & 2.8$_{\pm 0.6}$ & 13 & 10.6$_{\pm 0.1}$ & 10.5$_{\pm 0.1}$ & 0.81 & 0.98 & 1.01 & 2.2$_{\pm 0.2}$& 2.0$_{\pm 0.4}$\\
&3.0$\leq$ z <3.6  & 43 & 7 & 3.22 & 10.70 & 2.10 & -0.12 & 269.51 & 3.2$_{\pm 1.1}$ & 16 & 10.6$_{\pm 0.1}$ & 10.7$_{\pm 0.1}$ & 0.63 & 0.88 & 0.96 & 2.1$_{\pm 0.2}$& 2.3$_{\pm 0.2}$\\
\hline
10.0$\leq$ log$_{10}\ M_{\star}$ <10.5 &0.4$\leq$ z <0.8& 103 & 0  & 0.67 & 10.26 & 1.24 & 0.11 & 302.70 & 0.5$_{\pm 0.5}$ & 3 & 9.9$_{\pm 0.4}$ & 10.1$_{\pm 0.2}$ & 0.75 & 1.26 & 1.65 & 3.0$_{\pm 1.1}$ & 4.3$_{\pm 2.2}$ \\
&0.8$\leq$ z <1.1 & 131 &0 &  0.94 & 10.23 & 1.34 & 0.02 & 299.47 & 0.7$_{\pm 0.2}$ & 4 & 10.0$_{\pm 0.1}$ & 9.8$_{\pm 0.3}$ & 0.63 & 1.24 & 0.88 & 3.7$_{\pm 0.7}$ & 2.1$_{\pm 1.1}$ \\
&1.1$\leq$ z <1.4 & 95 & 0 & 1.24 & 10.24 & 1.48 & -0.04 & 270.81 & 0.8$_{\pm 0.8}$ & 8 & 10.1$_{\pm 0.4}$ & 10.2$_{\pm 0.3}$ & 0.68 & 0.73 & 1.03 & 0.8$_{\pm 0.7}$ & 2.7$_{\pm 0.7}$ \\
&1.4$\leq$ z <1.7 & 114 & 1 & 1.54 & 10.22 & 1.67 & 0.06 & 289.10 & 1.4$_{\pm 0.2}$ & 9 & 10.3$_{\pm 0.1}$ & 10.4$_{\pm 0.1}$ & 0.71 & 0.82 & 0.94 & 1.4$_{\pm 0.3}$ & 2.2$_{\pm 0.5}$ \\
&1.7$\leq$ z <2.0 & 79 & 1 & 1.85 & 10.21 & 1.67 & -0.04 & 278.14 & 0.9$_{\pm 0.2}$ & 5 & 10.1$_{\pm 0.1}$ & 10.2$_{\pm 0.1}$ & 0.76 & 0.91 & 1.08 & 1.7$_{\pm 0.7}$ & 2.7$_{\pm 1.2}$ \\
&2.0$\leq$ z <2.4 & 92 & 1 & 2.18 & 10.22 & 1.62 & -0.14 & 284.62 & 1.1$_{\pm 0.4}$ & 4 & 10.2$_{\pm 0.1}$ & 10.3$_{\pm 0.1}$ & 0.40 & 0.67 & 0.71 & 1.9$_{\pm 0.4}$ & 2.1$_{\pm 0.7}$ \\
&2.4$\leq$ z <3.0 & 79 & 0 & 2.74 & 10.24 & 1.80 & -0.10 & 269.94 & 2.2$_{\pm 0.9}$ & 4 & 10.5$_{\pm 0.2}$ & 10.4$_{\pm 0.2}$ & 0.58 & 1.20 & 0.91 & 3.5$_{\pm 0.2}$ & 2.3$_{\pm 1.2}$ \\
&3.0$\leq$ z <3.6 & 100 & 1 & 3.30 & 10.20 & 1.79 & -0.11 & 261.39 & 1.7$_{\pm 0.3}$ & 6 & 10.4$_{\pm 0.1}$ & 10.4$_{\pm 0.1}$ & 0.84 & 1.11 & 1.27 & 2.3$_{\pm 0.7}$ & 3.0$_{\pm 1.2}$ \\
\hline
\end{tabular}
\tablefoot{(1) Stellar mass bin, (2) redshift bin, (3) number of stacked galaxies, (4) number of individually-detected stacked galaxies, (5) mean redshift, (6) mean stellar mass, (7) mean SFR, (8) mean $\Delta$MS, (9) mean observed frequency, (10) mean $L_{850}$ inferred from the $uv$-domain, (11) peak S/N on the image, (12) mean gas mass inferred from the $uv$-domain, and (13) from the image-domain, (14) circularized synthesized beam $FWHM$, (15) circularized intrinsic $FWHM$ from the $uv$-domain, (16) circularized intrinsic $FWHM$ from the image-domain, (17) circularized half-light radii from the $uv$-domain, and (18) from the image-domain.}
\end{sidewaystable*}

\section{Results}
\label{sec:results}
The results of our stacking analysis are shown in Fig.~\ref{fig:uv_res_median} and summarized in Tab.~\ref{tab:sm}. In our highest stellar mass bin (i.e., $10^{11}\,\leq M_\star/{\rm M}_\odot<10^{12}$; right-most column), the number of stacked sources per redshift bin varies from 9 to 39, with about $37\%$ of them being individually detected. In this stellar mass bin, our stacking analysis yields high significance detections, with peak signal-to-noise ratios (S/N$_{\rm peak}$) greater than 20, except in our lowest redshift bin with S/N$_{\rm peak}\sim4$. In the $uv$-domain, those high significance detections are characterized by a Gaussian-like decrease of the stacked visibility amplitudes with the $uv$-distance, well fitted by our single component model. These galaxy populations are thus detected and spatially resolved by our stacking analysis. In the image-domain, this translates into bright spatially-resolved phase-center emission -- i.e., with a median synthesised beam $FWHM$ of $0\farcs5$ and an median angular size-to-synthesised beam $FWHM$ ratio of 1.5 -- that are well described by single 2D Gaussian components. In our intermediate stellar mass bin (i.e., $10^{10.5}\leq M_\star/{M}_\odot<10^{11.0}$), the number of stacked sources per redshift bin increases (43--81), while the fraction of them being individually detected decreases to about 10\%. As for our highest stellar mass bin, our stacking analysis yields high significance detections (i.e., S/N$_{\rm peak}>5$) in all of our redshift bins and those are spatially resolved at our median synthesised beam $FWHM$ of $0\farcs6$. Finally, in our lowest stellar mass bin (i.e., $10^{10}\leq M_\star/{M}_\odot<10^{10.5}$), the number of stacked sources per redshift bin increases even further (79--131) and only few of them are individually detected ($1\%$). In this low stellar mass bin, the same patterns are observed, i.e.,  spatially-resolved detections in the uv- and image-domain (with a median synthesize beam $FWHM$ of $0\farcs7$), though at lower significance, i.e.,  $3<\rm{S/N_{peak}}<9$. This implies that the number of stacked galaxies (controlled by the stellar mass function of MS galaxies) does not increase sufficiently to fully counter-balance the decrease of their molecular gas content with respect to the most massive population. Nevertheless, even in this low stellar mass bin, our stacking analysis yields clear detection (S/N$_{\rm peak}\,>3$), especially when considering both the $uv$-domain and image-domain constraints. We note that pushing this stacking analysis to lower stellar masses ($M_{\star}<10^{10}$) did not produce any significant detection. These results are thus not presented here and not discussed further in the paper.

We conclude that our stacking analysis provides robust mean molecular gas mass and FIR size measurements for $M_{\star}>10^{10}\,$M$_\odot$ MS galaxies from $z\sim0.4$ to $3.6$. Considering that in our highest and lowest stellar mass bins only $37\%$ and $\sim1\%$ of these galaxies were individually detected in the A$^3$COSMOS catalog, respectively, our stacking analysis clearly provides the first unbiased ALMA view on the gas content and size of MS galaxies.  

\subsection{The molecular gas content of MS galaxies}
\label{subsec:fgas}
The redshift evolution of the molecular gas mass of MS galaxies inferred from our stacking analysis is shown in Fig.~\ref{fig:gas}. It is compared to analytical predictions from the literature \citep[][]{2017ApJ...837..150S, 2019ApJ...887..235L, 2020ARA&A..58..157T}, individually-detected MS galaxies taken from the A$^3$COSMOS catalog \citep{2019ApJ...887..235L} and a local reference (i.e., $z\sim0.03$) taken from \citet{2017ApJS..233...22S}. In addition, in Fig.~\ref{fig:frac}, we present the evolution of the molecular gas fraction (i.e., $\mu_{\rm mol}=\langle M_{\rm mol}\rangle /\langle M_\star\rangle$) of MS galaxies as a function of redshifts and stellar masses. We note that for our galaxies in common with the A$^3$COSMOS catalog, the stellar masses used here (i.e., those from the COSMOS-2015 catalog) are about 0.22\,dex lower than those reported in the A$^3$COSMOS catalog. This offset, which is also discussed in \citet{2019ApJ...887..235L}, is likely explained by the fact that stellar masses in the A$^3$COSMOS catalog rely on full optical-to-mm energy-balanced SED fits performed with \texttt{MAGPHYS}. While this offset is observed for massive galaxies, it might not be present at $\lesssim\,$10$^{10.5}$M$_{\odot}$, where the number of galaxies available in the A$^3$COSMOS catalog is too scarce to provide meaningful comparison with the COSMOS-2015 catalog. In any case, when comparing our analytical predictions to those from \citet{2019ApJ...887..235L}, we thus show both their original predictions and those inferred by accounting for this systematic 0.22\,dex offset.

Our measurements reveal a significant evolution of the molecular gas mass of MS galaxies with both redshifts and stellar masses. For all stellar mass bins, the molecular gas masses of MS galaxies (equivalently molecular gas fraction) increase by a factor of $\sim$24 from $z\sim0$ to $z\sim3.2$. In addition, at a given redshift, the molecular gas masses of MS galaxies significantly increase with stellar masses. This trend is, however, sub-linear in the log-log space, which implies that the molecular gas fraction of MS galaxies decreases with stellar mass at a given redshift (Fig.~\ref{fig:frac}). To obtain a more quantitative constraint on the stellar mass and redshift dependencies of evolution of the molecular gas fraction of MS galaxies, we fitted our measurements, together with the local reference, following \citet[][]{2019ApJ...887..235L}, i.e.,
\begin{equation}
\label{eq:gas fraction}
\begin{aligned}
{\rm log_{10}\,\mu_{mol}} &=\ (a+ak \times {\rm log_{10}}(M_{\star}/10^{10})) \times \Delta {\rm MS} \\
                                       & + b \times {\rm log_{10}}(M_{\star}/10^{10}) \\
                                       & +(c+ck \times {\rm log_{10}}(M_{\star}/10^{10})) \times t_{\rm cosmic} \\
                                       & +{d},
\end{aligned}
\end{equation}
where $t_{\rm cosmic}$ is the cosmic time in units of Gyr, and $M_{\star}$ is in units of M$_{\odot}$. Because our analysis does not probe a large dynamic range in $\Delta {\rm MS}$, we fixed $a$ and $ak$ to the values reported by \citet[][]{2019ApJ...887..235L}, i.e., $a=0.4195$ and $ak=0.1195$, respectively. To constrain the remaining parameters of Eq.~\ref{eq:gas fraction}, we then performed a standard Bayesian analysis using the python Markov Chain Monte Carlo (MCMC) package \texttt{emcee} \citep{2013PASP..125..306F}. In this analysis, we accounted for the redshift, stellar mass, and $\Delta {\rm MS}$ of each galaxy in a given stacked bin, i.e., in each MCMC step, we compared our stacked measurements, $\langle M_{\rm gas}^i\rangle$, to $\langle f(\,t^i_{\rm cosmic}, M^i_{\star},\Delta {\rm MS}^i\,)\rangle$ and not $f( \langle t^i_{\rm cosmic}\rangle, \langle M^i_{\star}\rangle,\langle\Delta {\rm MS}^i\rangle)$, where $i$ is the $i$th galaxy of our stacked bin and $f$ is the fitted function. This avoids averaging biases that could arise if one would simply fit our stacked measurements using $\langle t^i_{\rm cosmic}\rangle$, $\langle M^i_{\star}\rangle$, and $\langle\Delta {\rm MS}^i\rangle$. Results of this MCMC analysis are shown in Fig.~\ref{fig:MCMC}, with $b=-0.468^{+0.070}_{-0.070}$, $c=-0.122^{+0.008}_{-0.008}$, $d=0.572^{+0.059}_{-0.060}$, and $ck=0.002^{+0.011}_{-0.011}$. These results unambiguously demonstrate that the molecular gas fraction of MS galaxies decreases with stellar masses (i.e., $b<0$) while it increases with redshifts (i.e., $c<0$; see blue solid lines in Fig.~\ref{fig:gas}). Note that repeating this MCMC analysis while fixing $a=0$ and $ak=0$ (i.e., considering that our measurements are for $\Delta$MS\,$=\,0$ galaxies), the likelihood of our fit decreases but the inferred gas fraction evolution remains qualitatively consistent with our previous fit albeit with a somewhat flatter stellar mass dependency, i.e., $b=-0.252^{+0.072}_{-0.072}$, $c=-0.114^{+0.008}_{-0.008}$, $d=0.481^{+0.061}_{-0.061}$, and $ck=-0.016^{+0.011}_{-0.011}$.

By comparing our results with those from the A$^3$COSMOS catalog, one immediately notices that these individually-detected galaxies systematically lie above our measurements. This systematic offset results from an observational bias. First of all, at a given redshift and stellar mass, the A$^3$COSMOS catalog mostly contains galaxies on the upper part of the MS because those galaxies have higher molecular gas mass ($a>0$ in Eq.~\ref{eq:gas fraction}) and are thus more likely to be individually detected. This observational bias was, however, accounted for when fitting the A$^3$COSMOS population using Eq.~\ref{eq:gas fraction}. This `correction' can be seen in Fig.~\ref{fig:gas} by noticing that the A$^3$COSMOS analytical predictions for MS galaxies systematically lies below the A$^3$COSMOS data-points. Nevertheless, at a given redshift, stellar mass, and $\Delta$MS, the A$^3$COSMOS catalog could still be biased toward galaxies with a bright millimeter emission and thus high molecular gas mass \citep[see discussion in][]{2019ApJ...887..235L}. Our measurements, which are not affected by this bias and which lie systematically below those of \citet{2017ApJ...837..150S}, \citet[][]{2019ApJ...887..235L}, and \citet[][]{2020ARA&A..58..157T}, clearly demonstrate the presence of this residual observational bias in these literature studies. By averaging at a given redshift and stellar mass all MS galaxies in the field, our stacking analysis reveals their true mean molecular gas mass. Taken at face value, our findings imply that previous studies might have systematically overestimated by at least $10-40\%$ the gas content of MS galaxies in redshift and stellar mass bins with relatively high detection fraction (i.e., mostly $M_{\star}>10^{11}\,$M$_\odot$), and by $10-60\%$ in bins with low detection fraction. While significant, one should, however, acknowledge that these offsets remain reasonable considering all the selection biases affecting these previous studies. The impact of this finding for galaxy evolution models is discussed in Sect.~\ref{sec:discussion}.

\begin{figure*}
\centering
\includegraphics[width=1.0\textwidth]{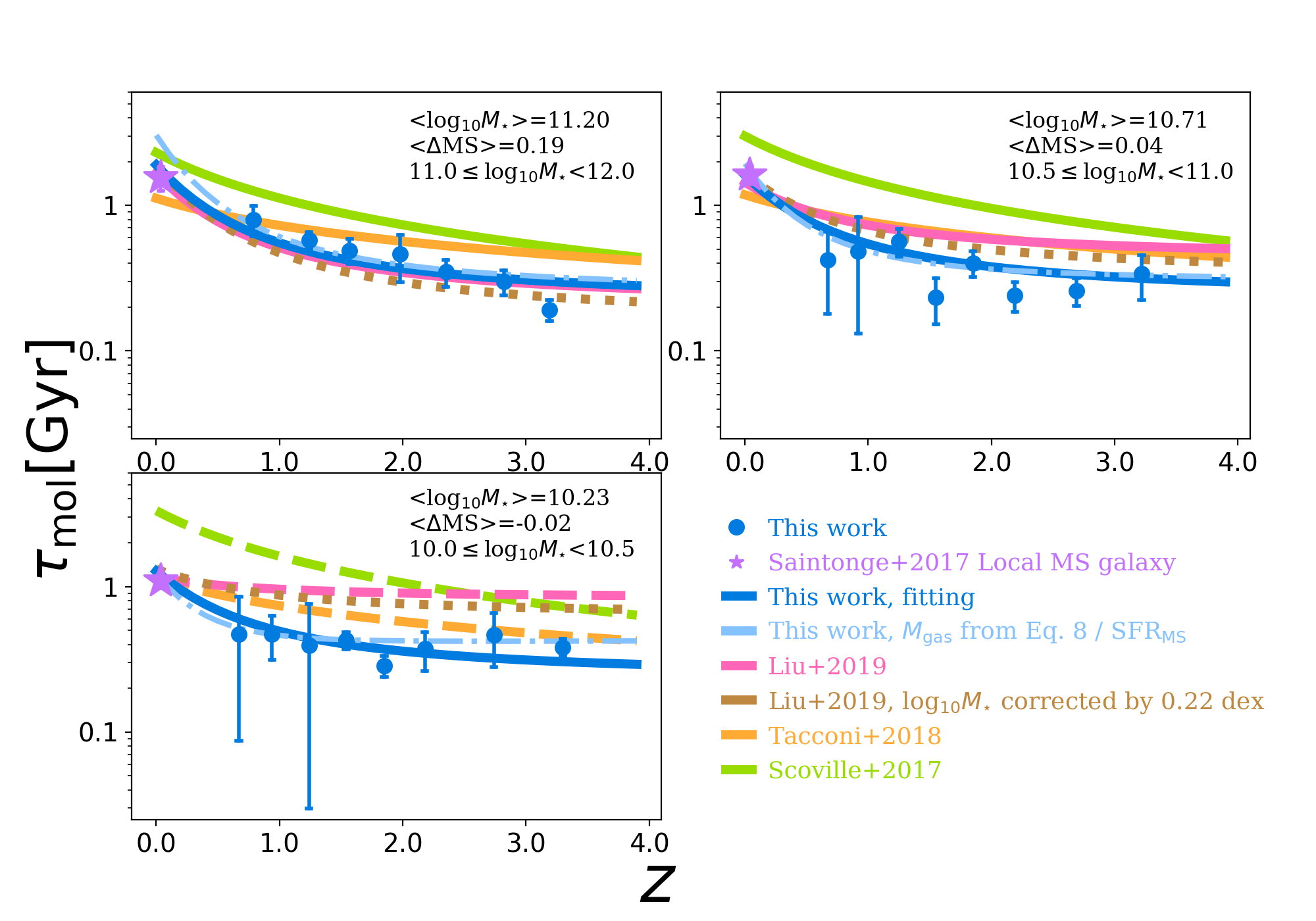}
\caption{The redshift evolution of molecular gas depletion time of MS galaxies in three stellar mass bins, i.e., $10^{11}\,\leq M_\star/{\rm M}_\odot<10^{12}$, $10^{10.5}\,\leq M_\star/{\rm M}_\odot<10^{11}$, and $10^{10}\,\leq M_\star/{\rm M}_\odot<10^{10.5}$. Blue circles show our $uv$-domain molecular gas mass measurements divided by the mean SFR of each of these stacked samples. Purple stars show the local MS reference taken from \citet[][]{2017ApJS..233...22S}. Lines present the analytical evolution of the molecular gas depletion time as inferred from our work (blue lines; see text for details), from \citet[][pink line]{2019ApJ...887..235L}, from \citet[][green line]{2017ApJ...837..150S}, and from \citet[][orange line]{2020ARA&A..58..157T}. In our lower stellar mass bin, lines from the literature are dashed as they mostly rely on extrapolations.}
\label{fig:dep}
\end{figure*}
\begin{figure}
\centering
\includegraphics[width=\columnwidth]{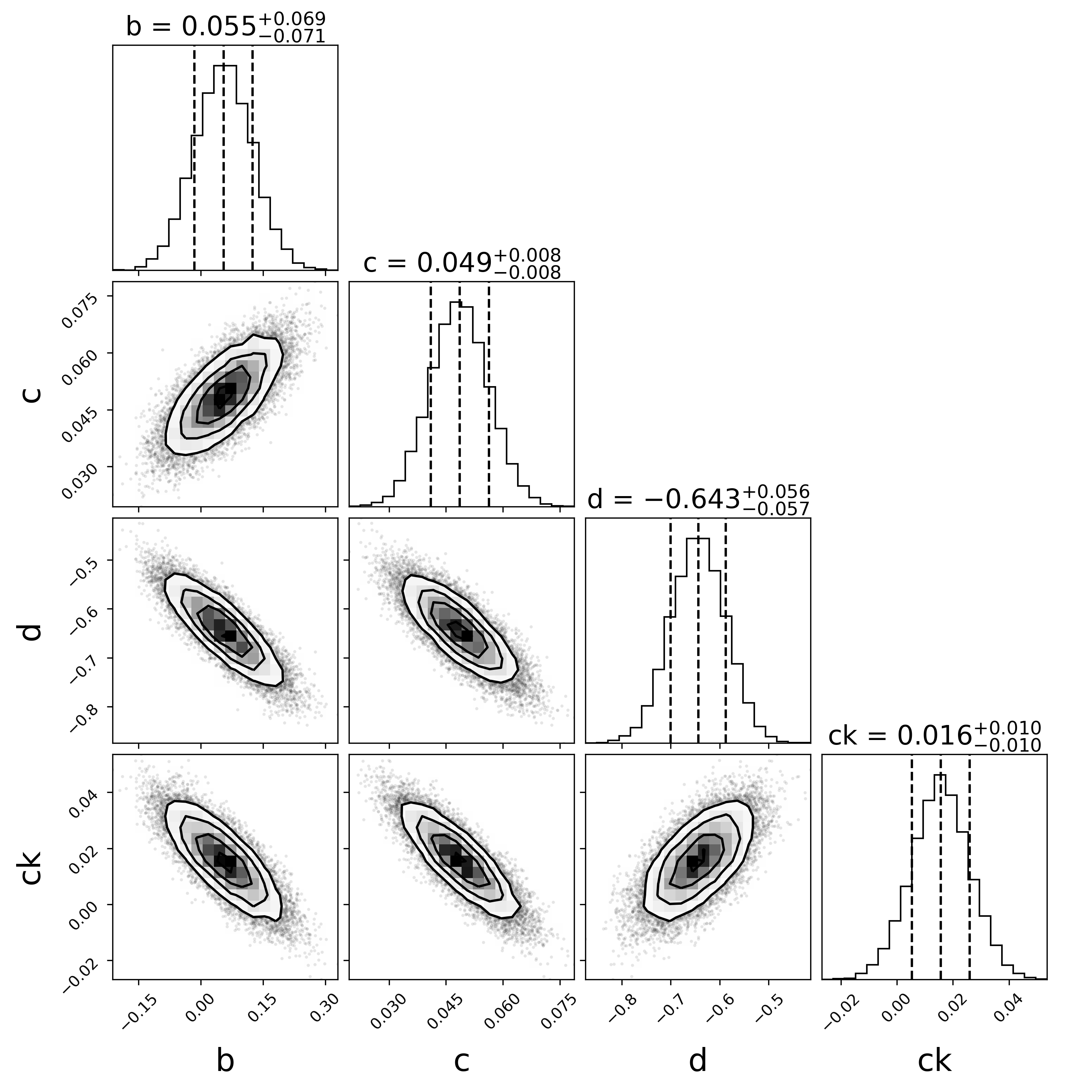}
\caption{Probability distributions of the parameters in Eq.~\ref{eq:depletion}, as found by fitting our stacked measurements using a MCMC analysis. The dashed vertical lines show the 16th, 50th, and 84th percentiles of each distribution.}
\label{fig:MCMC_depletion}
\end{figure}
\begin{figure}
\centering
\includegraphics[width=\columnwidth]{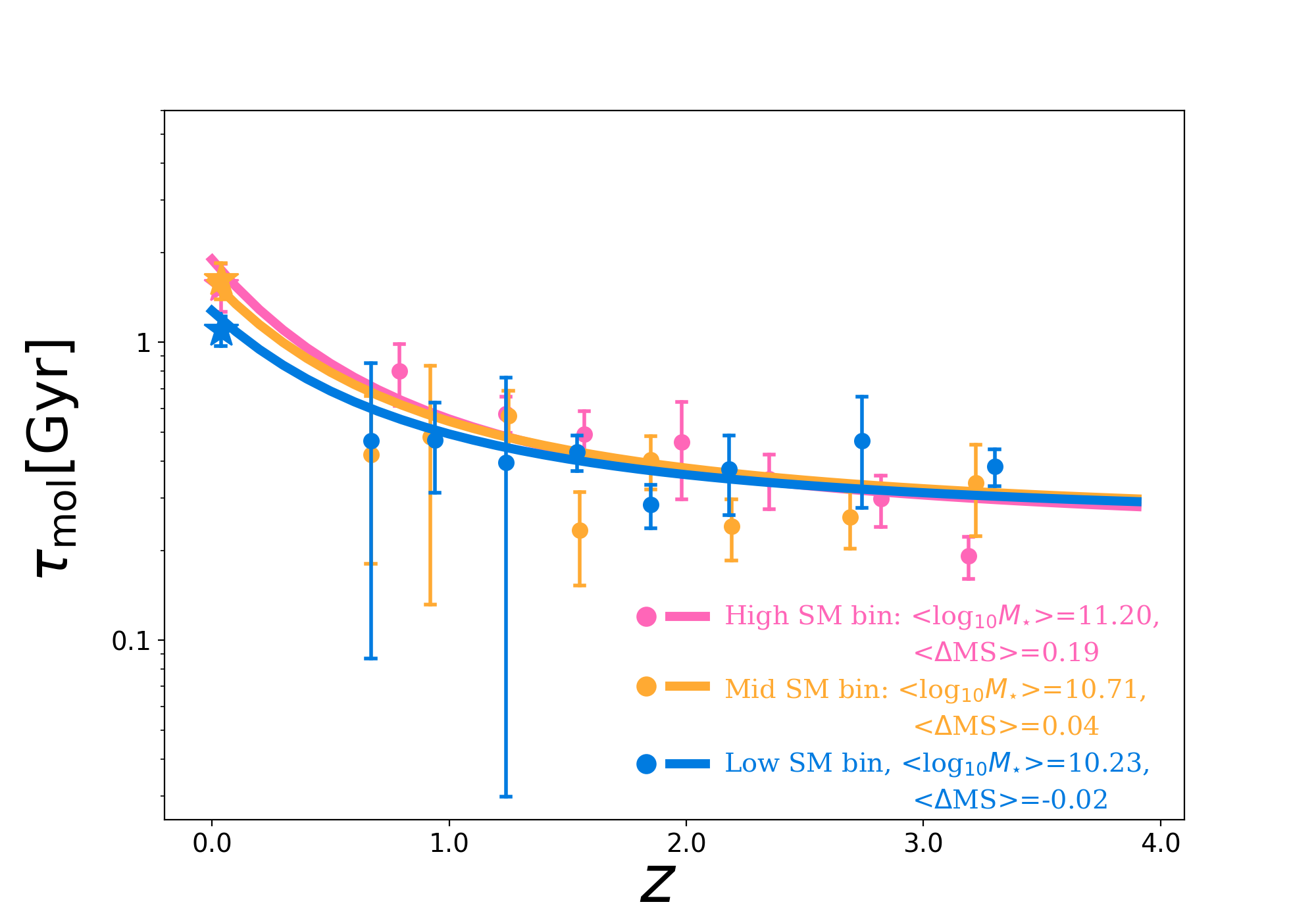}
\caption{Redshift evolution of molecular gas depletion time of MS galaxies. Circles show the mean molecular gas depletion time from our work. Stars show the local MS reference taken from \citet{2017ApJS..233...22S}. Lines display the analytical evolution of the molecular gas fraction inferred from our work. Symbols and lines are color-coded by stellar mass, i.e., pink for $10^{11}\,\leq M_\star/{\rm M}_\odot<10^{12}$, orange for $10^{10.5}\,\leq M_\star/{\rm M}_\odot<10^{11}$, and blue for $10^{10}\,\leq M_\star/{\rm M}_\odot<10^{10.5}$.}
\label{fig:dep_SM}
\end{figure}

\subsection{The molecular gas depletion time of MS galaxies}
\label{subsec:tdepl}
The redshift evolution of the molecular gas depletion time (i.e., ${\rm \tau_{mol}} = M_{\rm mol}/{\rm SFR}$) of MS galaxies as inferred from our stacking analysis is shown in Fig.~\ref{fig:dep}, together with analytical predictions from \citet{2017ApJ...837..150S}, \citet{2019ApJ...887..235L}, and \citet{2020ARA&A..58..157T} as well as the local reference for MS galaxies taken from \citet[][]{2017ApJS..233...22S}. Again, to obtain a more quantitative constraint on the stellar mass and redshift dependencies of the molecular gas depletion time of MS galaxies, we fitted our measurements, together with the local reference, following \citet[][]{2019ApJ...887..235L}, i.e., 
\begin{equation}
\label{eq:depletion}
\begin{aligned}
{\rm log_{10}\,\tau_{mol}} &=\ (a+ak \times {\rm log_{10}}(M_{\star}/10^{10})) \times \Delta {\rm MS} \\
                                       & + b \times {\rm log_{10}}(M_{\star}/10^{10}) \\
                                       & +(c+ck \times {\rm log_{10}}(M_{\star}/10^{10})) \times t_{\rm cosmic} \\
                                       & +{d}.
\end{aligned}
\end{equation}
Our analysis does not probe a large dynamic range in $\Delta {\rm MS}$, we thus fixed $a$ and $ak$ to the values reported by \citet[][]{2019ApJ...887..235L}, i.e., $a=-0.5724$ and $ak=0.1120$. Results of our MCMC analysis are shown in Fig.~\ref{fig:MCMC_depletion}, with $b=0.055^{+0.069}_{-0.071}$, $c=0.049^{+0.008}_{-0.008}$, $d=-0.643^{+0.056}_{-0.057}$, and $ck=0.016^{+0.010}_{-0.010}$. Because the depletion time is the ratio of $M_{\rm mol}$ by ${\rm SFR}$, we also display in Fig.~\ref{fig:dep} the redshift evolution of depletion time as one would infer by dividing $M_{\rm gas}(z,M_\ast,\Delta{\rm MS})$ from Eq.~\ref{eq:gas fraction} by the SFR$_{\rm MS}(z,M_\ast,\Delta{\rm MS})$ from \citet[][dash-dotted light-blue line]{2020ApJ...899...58L}. Finally, in Fig.~\ref{fig:dep_SM}, we compare the redshift evolution of the  molecular gas depletion time as inferred for our three stellar mass bins.

In all our stellar mass bins, the molecular gas depletion time of MS galaxies decreases by a factor of $\sim3-4$ from $z\sim0$ to $z\sim3.2$, with, however, most of this decrease happening at $z\lesssim1.0$. At $z\gtrsim1$, the molecular gas depletion time of MS galaxies remains instead roughly constant with redshifts and stellar masses with a value of $\sim300-500\,$Myr. While such evolution is qualitatively predicted by all literature studies, its amplitude as well as its exact redshift- and stellar mass-dependencies quantitatively disagree (see Fig.~\ref{fig:dep}). For example, our measurements and those from \citet[][]{2019ApJ...887..235L} agree at high stellar masses, but differ by $\sim30-40\%$ in our lower stellar mass bins. These differences are likely explained by the observational biased discussed in Sect.~\ref{subsec:fgas} which implies that the mean molecular gas mass and thus depletion time of MS galaxies inferred by \citet{2019ApJ...887..235L} are slightly overestimated especially at low stellar masses. The same effect likely explains the $\sim20-30\%$ overestimation of the molecular gas depletion time inferred in \citet[][]{2020ARA&A..58..157T} in most redshift--stellar mass bins probed here. 
While our direct analytical fit of the redshift/stellar mass evolution of the molecular gas depletion time (solid blue lines in Fig.~\ref{fig:dep}) matches relatively well the local reference from \citet{2017ApJS..233...22S}, this is not the case of our fit inferred by simply dividing $M_{\rm gas}(z,M_\ast,\Delta{\rm MS})$ from Eq.~\ref{eq:gas fraction} by SFR$_{\rm MS}(z,M_\ast,\Delta{\rm MS})$ from \citet[][dash-dotted light-blue line]{2020ApJ...899...58L}. This disagreement between predictions and observations at $z\sim0$ is entirely attributed to a miss-match in SFR$_{\rm MS}(z,M_\ast,\Delta{\rm MS})$, i.e., at a given stellar mass, the mean SFR of $z\sim0$ MS galaxies as predicted by \citet{2020ApJ...899...58L} does not match that observed by \citet{2017ApJS..233...22S}. This disagreement is, however, not unexpected as the sample used in \citet{2020ApJ...899...58L} was restricted to $z>0.3$ galaxies.

In general, we conclude that our depletion times agree at high stellar masses with \citet{2019ApJ...887..235L}, i.e., where their study relies on a large and robust amount of ALMA-based measurements of MS galaxies; while our depletion time agree better at low stellar masses with \citet{2020ARA&A..58..157T}, i.e., where their study, contrary to that of \citet{2019ApJ...887..235L}, still relies on some observational measurements of MS galaxies thanks to their \textit{Herschel} stacking analysis. Like our measurements, those from \citet[][]{2020ARA&A..58..157T} predict only a minor evolution of the molecular gas depletion time of MS galaxies with stellar masses. This implies that the flattening of the MS at high stellar masses observed in most studies (i.e., log$_{10}\ $SFR$^{\rm MS}=0.7\times\,$log$_{10}\ M^{\rm MS}_\star + {\rm C}$) is not associated/due to lower star-formation efficiencies (SFE; i.e., $1/ \tau_{\rm mol}$) in massive systems but rather lower molecular gas fraction (see Sect.~\ref{subsec:fgas}). This is further discussed in Sect.~\ref{sec:discussion}. In addition, we note that extrapolating our molecular gas depletion time predictions to $z\sim5$, i.e., $\tau_{\rm mol}^{\rm pred}=\,$250\,Myr (from Eq.~\ref{eq:depletion}) or $\tau_{\rm mol}^{\rm pred}=\,$620\,Myr (from Eq.~\ref{eq:gas fraction}/specific SFR$_{\rm MS}$), our prediction qualitatively agrees with the latest observational constraints from the ALPINE [C\,{\footnotesize II}] ALMA large project, i.e., $\tau_{\rm mol}^{\rm obs}=$520$\pm$70\,Myr \citep[][]{2020A&A...643A...5D}. 

Finally, we note that accuracy of the depletion times relies not only on accurate gas masses but also accurate SFRs. In our study, the latter were estimated using the so-called ladder of SFR indicators, i.e., by applying to each galaxy the best dust-corrected star formation indicator available (Sect.~\ref{subsec:our_sample}). In particular, among the 1376 galaxies in our final sample with stellar mass >$10^{10}\,$M$_{\odot}$, 852 (62\%) have very robust dust-corrected SFRs based on the combination of IR and UV measurements. Of the remaining 524 galaxies whose SFRs are solely based on their UV-to-optical fits, most (470) should also have robust SFRs, as they falls below the $\sim100\,$M$_{\odot}$\,yr$^{-1}$ limit above which SFR$_{\rm SED}$ starts to be systematically underestimated (Fig.~\ref{fig:SFR_compare}). We verified that our results remain unchanged (within the uncertainties) when excluding from our stacking analysis these 54 galaxies with SFR$_{\rm SED}>100\,$M$_{\odot}$\,yr$^{-1}$ and without IR detection.

\subsection{The FIR sizes of MS galaxies}
\label{subsec: FIR sizes}
\begin{figure*}
\centering
\includegraphics[width=0.8\textwidth]{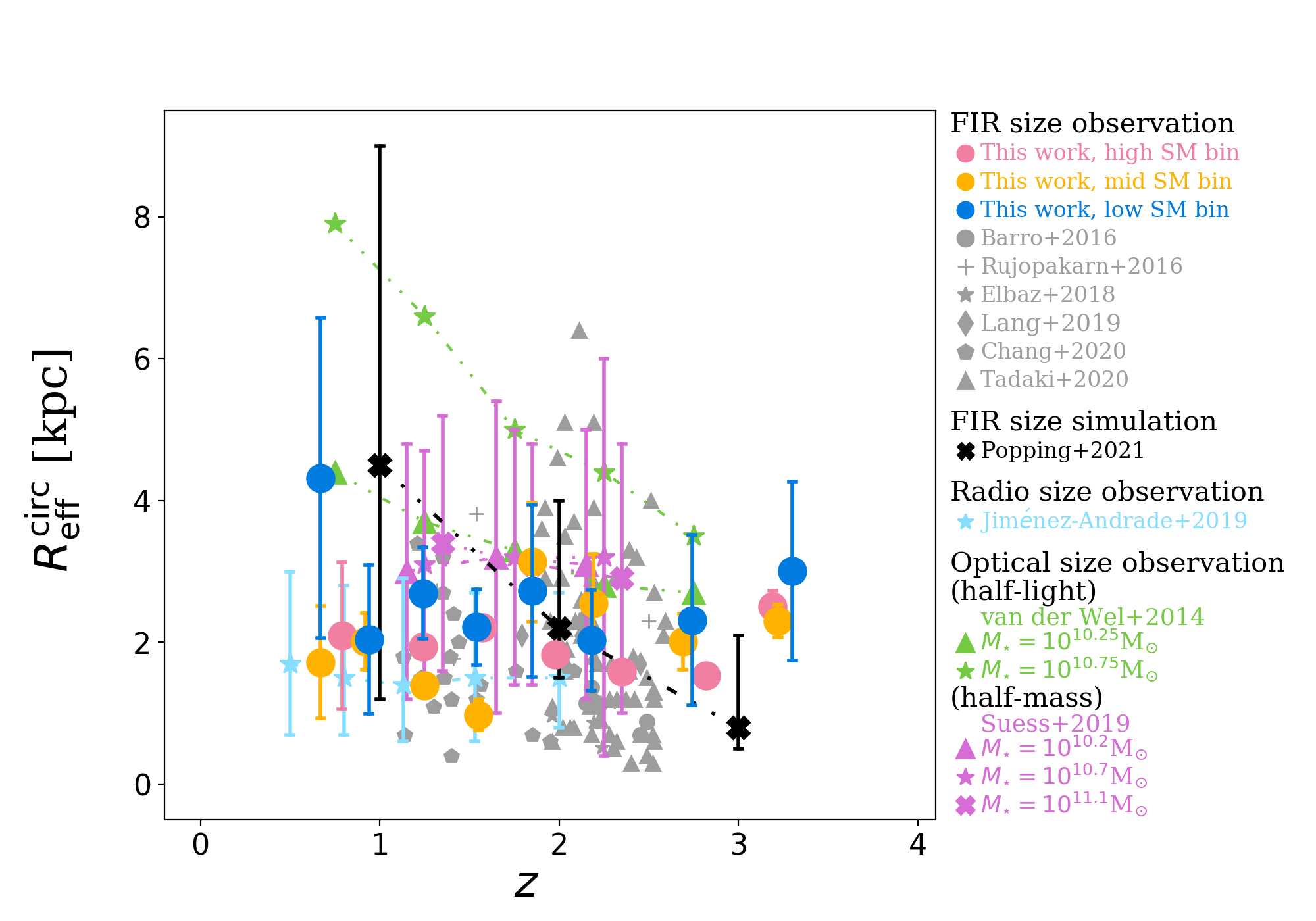}
\caption{Redshift evolution of the half-light (or half-mass) radius of MS galaxies. Pink, orange, and blue circles present our stacking results for our high, mid, and low stellar mass bins, i.e., $10^{11}\,\leq M_\star/{\rm M}_\odot<10^{12}$, $10^{10.5}\,\leq M_\star/{\rm M}_\odot<10^{11}$, and $10^{10}\,\leq M_\star/{\rm M}_\odot<10^{10.5}$, respectively. The gray data points are the FIR sizes of MS galaxies from \citet[][dots]{2016ApJ...827L..32B}, \citet[][pluses]{2016ApJ...833...12R}, \citet[][stars]{2018A&A...616A.110E}, \citet[][diamonds]{2019ApJ...879...54L}, \citet[][pentagons]{2020ApJ...888...44C}, and \citet[][triangles]{2020ApJ...901...74T}; while light blue stars are radio sizes of MS galaxies from \citet[][]{2019A&A...625A.114J}. The black crosses are FIR sizes of $M_{\star}=10^{10.5}M_{\odot}$ MS galaxies from the simulations of \citet{2022MNRAS.510.3321P}. The green triangles and stars are optical half-light sizes of $M_{\star}=10^{10.25}M_{\odot}$ and $M_{\star}=10^{11.25}M_{\odot}$ MS galaxies from \citet{2014ApJ...788...28V}. Finally, the purple triangles, stars and crosses are optical half-mass sizes of $M_{\star}=10^{10.25}M_{\odot}$, $M_{\star}=10^{10.75}M_{\odot}$ and $M_{\star}=10^{11.25}M_{\odot}$ MS galaxies from \citet{2019ApJ...877..103S}. Because most of these literature studies relied on image-plane fits, the stacked FIR sizes displayed here are those from our 2D Gaussian image-plane fits using \texttt{PyBSDF}.}
\label{fig:size}
\end{figure*}

Our stacking analysis provides the first measurements of the mean FIR size of MS galaxies across cosmic time. These mean FIR (at the observed-frame 850--1300\,$\mu$m) sizes of MS galaxies are presented in Fig.~\ref{fig:size}, and compared to optical, FIR and radio sizes measurements from \citet{2014ApJ...788...28V}, \citet{2016ApJ...827L..32B}, \citet{2016ApJ...833...12R}, \citet{2018A&A...616A.110E}, \citet{2019A&A...625A.114J}, \citet{2019ApJ...877..103S}, \citet{2020ApJ...888...44C}, and \citet{2020ApJ...901...74T}. Because most of the FIR and radio size measurements from the literature were made on the image-plane, we displayed in Fig.~\ref{fig:size} our 2D Gaussian image-plane measurements inferred using \texttt{PyBDSF}. Displaying instead our $uv$-plane size measurements would, however, not change any of our conclusions, as both agree within their uncertainties with no apparent systematic offset between them.

These FIR sizes do not seem to evolve significantly with redshift or stellar mass, with a mean circularized effective -- equivalently half-light -- radius of 2.2\,kpc ( i.e., corresponding to a median angular size-to-synthesised beam $FWHM$ ratio of 1.5). Because there is a possible mismatch between our stacked position and the actual millimeter position of the sources \citep[e.g.,][]{2018A&A...616A.110E}, these average FIR sizes could, however, be slightly overestimated. Nevertheless, such bias does not seem to be significant as our measurements agree qualitatively and quantitatively with the mean star-forming size of massive ($M_{\star}\sim 10^{10.7-11.7}$M$_{\odot}$) MS galaxies inferred by the most recent literature studies. In contrast, the half-light stellar size of massive MS galaxies is typically larger than these FIR extents by a factor 2 and 4 at $z\sim3$ and 1, respectively \citep[see Fig.~\ref{fig:size};][]{2014ApJ...788...28V}. In lower mass MS galaxies (i.e., $M_{\star}\sim 10^{10.3}$M$_{\odot}$), larger half-light stellar size than FIR extents are also observed but mostly at low redshifts. As discussed in Sect.~\ref{sec:discussion}, this apparent discrepancy between optical and FIR sizes of MS galaxies does not, however, necessarily translate into stellar half-mass radius discrepancy, as complex obscuration biases need to be accounted for when converting half-light stellar radius into half-mass stellar radius \citep[e.g.,][]{2019ApJ...879...54L, 2019ApJ...877..103S, 2022MNRAS.510.3321P}. For example, the FIR sizes inferred in our study agree quantitatively with the mean redshift-independent half-mass stellar radius of SFGs measured by \citet{2019ApJ...877..103S}.

\subsection{The Kennicutt-Schmidt relation}
\begin{figure*}
\centering
\includegraphics[width=0.8\textwidth]{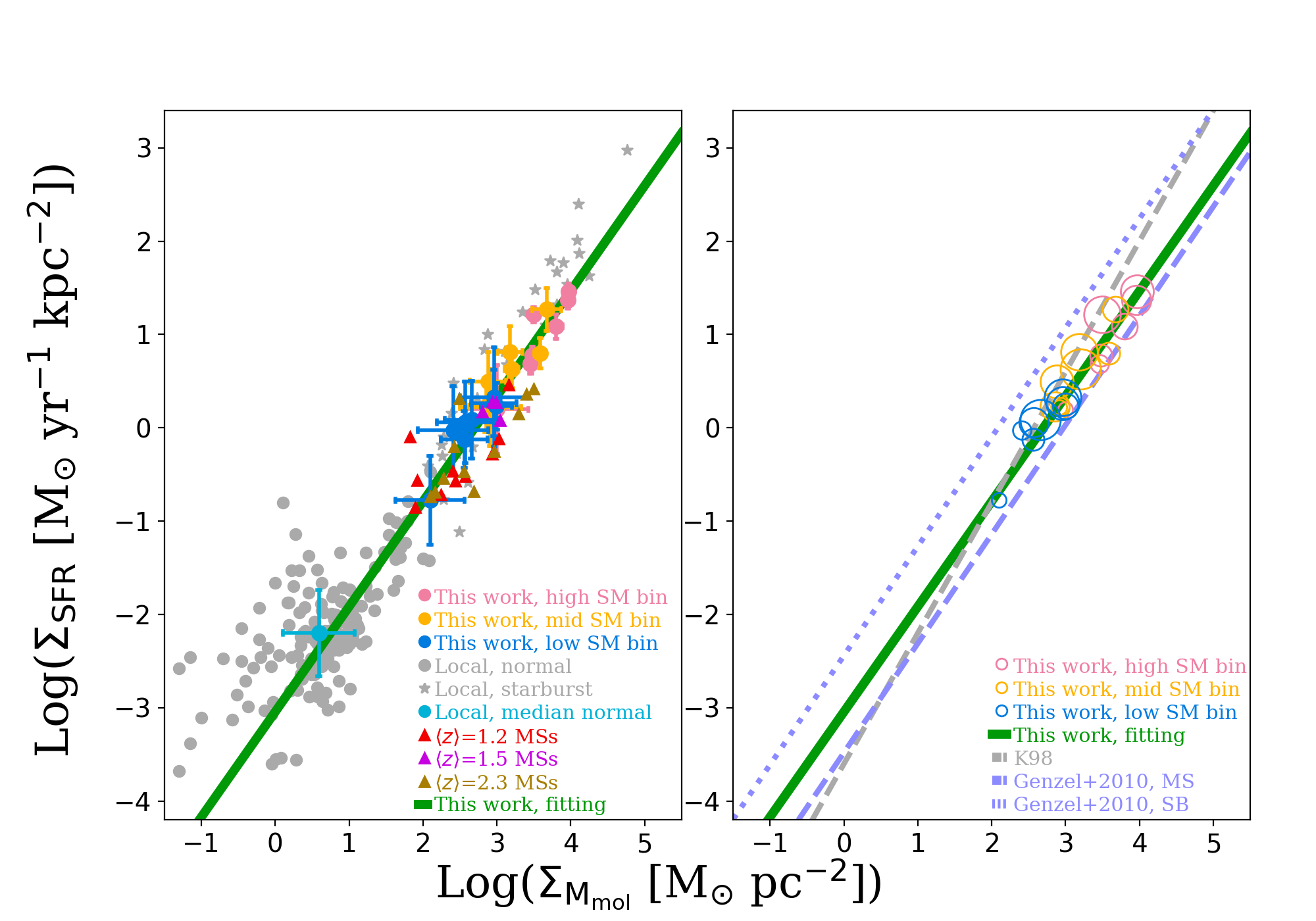}
\caption{Relation between the SFR and gas mass densities of SFGs, i.e., the so-called KS relation. (\textit{Left}) Pink, orange, and blue circles present our stacking results for our high, mid, and low stellar mass bins, i.e., $10^{11}\,\leq M_\star/{\rm M}_\odot<10^{12}$, $10^{10.5}\,\leq M_\star/{\rm M}_\odot<10^{11}$, and $10^{10}\,\leq M_\star/{\rm M}_\odot<10^{10.5}$, respectively. Gray circles and stars are normal and starburst local galaxies from K98 and \citet[][taking only their molecular gas mass estimates, i.e., excluding the atomic phase]{2019ApJ...872...16D}. Red, purple, and brown triangles are $\langle z\rangle=1.2$ \citep{2010ApJ...714L.118D}, $\langle z\rangle=1.5$ \citep[][]{2007ApJ...660L...1D, 2007ApJ...660L..43N, 2010Natur.463..781T}, and $\langle z\rangle=2.3$ \citep{2006ApJ...647..128E} MS galaxies, respectively. The green line is a fit to the KS relation considering only MS galaxies, i.e., our measurements together with the K98 normal local galaxies average (turquoise dot). (\textit{Right}) Comparison of our MS-only KS relation to the global fit of K98, the MS-only (blue long dashed line) and SB-only (blue dotted line) fits of \citet{2010MNRAS.407.2091G}. Open circles show our measurements, with symbols size increasing with redshifts and colour-coded by stellar masses.}
\label{fig:ks}
\end{figure*}

Combining our half-light FIR radii (from our image-plane fits), molecular gas mass, and SFR measurements, we study in Fig.~\ref{fig:ks} the relation between the SFR and gas mass surface densities of MS galaxies (i.e., $\Sigma_{\rm SFR} = {\rm SFR} / ( 2\,\pi\,{R_{\rm \rm eff-py}^{\rm circ}}^2)$ vs. $\,\Sigma_{M_{\rm mol}} = M_{\rm mol} / ( 2\,\pi\,{R_{\rm \rm eff-py}^{\rm circ}}^2)$; the so-called KS relation). We compare our estimates with results from the literature: local normal and starburst galaxies from \citet[][hereafter K98]{1998ApJ...498..541K} and \citet{2019ApJ...872...16D} (taking only their molecular gas phase measurements and thus excluding contribution from the atomic gas phase) as well as the global fit of the KS relation from K98; $\langle z\rangle=1.2$ MS galaxies from \citet{2010ApJ...714L.118D} as well as their MS-only  galaxies fit of the KS relation; $\langle z\rangle=1.5$ MS galaxies from \citet[][]{2007ApJ...660L...1D}, \citet{2007ApJ...660L..43N}, and \citet{2010Natur.463..781T}; $\langle z\rangle=2.3$ MS galaxies from \citet{2006ApJ...647..128E}; and finally the MS-only galaxies fit of the KS relation from \citet{2010MNRAS.407.2091G}. Note that we take here the FIR size of galaxies as a proxy of their SFR and gas mass distributions (under the hypothesis that the dust and gas are co-spatial). This assumption is justified by recent simulations in which the FIR half-light radius of galaxies is found to be consistent with the radius containing half their star formation and to be only slightly more compact than the radius containing half their molecular gas mass, at least in $z\lesssim2$ galaxies \citep{2022MNRAS.510.3321P}. 

There exists a tight correlation between the $\Sigma_{M_{\rm mol}}$ and $\Sigma_{\rm SFR}$ of MS galaxies, with no significant dependencies of this relation on stellar mass or redshift, i.e., at a given $\Sigma_{M_{\rm mol}}$, measurements from different stellar mass or redshift bins agree within their uncertainties. Our measurements are consistent with previous individually-detected MS galaxy estimates while they fall below those from individually-detected starbursts. In general, at a given redshift, MS galaxies with higher stellar masses are located at the higher end of the $\Sigma_{\rm SFR}-\Sigma_{M_{\rm mol}}$ relation due to the increase of their molecular gas content and the absence of significant size evolution with stellar mass, which translates into an overall increase of their $\Sigma_{M_{\rm mol}}$. Similarly, at a given stellar mass, MS galaxies at higher redshifts are mostly located at the higher end of the $\Sigma_{\rm SFR}-\Sigma_{M_{\rm mol}}$ relation due to the increase of their molecular gas content and the absence of significant size evolution with redshifts, which also translates into an overall increase of their $\Sigma_{M_{\rm mol}}$.

We performed a linear fit of the KS relation of MS-only galaxies in log-log space, combining our high-redshift MS galaxies measurements with those from the local universe obtained by K98 and \citet{2019ApJ...872...16D},
\begin{equation}
{\rm log_{10}\,}\Sigma_{\rm SFR} = (1.13\pm0.09) \cdot {\rm log_{10}\,}\Sigma_{\rm M_{mol}}\ -\ (3.06\pm0.33). 
\end{equation}
The inferred power index of the MS-only KS relation (i.e., $\alpha$=1.13) is smaller than that found by \citet{2010ApJ...714L.118D} considering MS-only galaxies ($\alpha$=1.42), but similar to that found by \citet{2010MNRAS.407.2091G} for MS-only galaxies ($\alpha$=1.17). We note that previous high-redshift investigations \citep[i.e.,][]{2010ApJ...714L.118D, 2010MNRAS.407.2091G} were only based on relatively small samples of massive high-redshift SFGs (i.e., $N<50$, $z>1$, and $M_\star>10^{11}\,M_\odot$), and are thus likely limited by selection biases. A power law index for the MS-only KS relation that is greater than unity implies that the depletion time (i.e., $\tau_{\rm mol}$) -- equivalently, SFE (i.e., $1/ \tau_{\rm mol}$) -- of MS galaxies is controlled by their $\Sigma_{M_{\rm mol}}$. In other words, the evolution of the depletion time with redshift and stellar mass seen in Fig.~\ref{fig:dep} can be predicted from their $\Sigma_{M_{\rm mol}}$ and this universal redshift-independent MS-only KS relation. The KS of MS-only galaxies remains thus one of the most fundamental relation to understand the stellar mass growth of the universe over the last 10\,Gyr.

Finally, as already pointed out by, e.g., \citet[][]{2010ApJ...714L.118D} and \citet[][]{2010MNRAS.407.2091G}, we found that MS galaxies seems to follow a KS relation that at high $\Sigma_{M_{\rm mol}}$ falls below the relation followed by starburst galaxies. In this high $\Sigma_{M_{\rm mol}}$ regime, starbursts exhibit $2-3$ times higher SFE. \\

Note that in this analysis, we implicitly assume that the dust and gas are co-spatial. However, this assumption might not be always verified as suggested by some ALMA high-resolution observations of submillimeter-selected galaxies \citep[e.g.,][]{2017ApJ...846..108C,2018ApJ...863...56C}, which revealed $\sim$$\times2$ more compact dust continuum emission than gas CO emission. Increasing by a factor two our FIR sizes would shift our data points toward lower surface densities along the one-to-one line in the log-log space but would not change significantly the slope of the inferred KS relation. However, such large offset/discrepancy in spatial distribution would also translate into very uncertain dust-based gas mass measurements and would thus impact in a more complex way the inferred KS relation. 
Regardless, submillimeter-selected galaxies are extreme object located far above the MS \citep[e.g.,][]{2012A&A...539A.155M, 2014PhR...541...45C} and MS galaxies do not seem to exhibit any significant discrepancies between their gas and dust sizes \citep{2019ApJ...877L..23P}.

\subsection{Limitations and uncertainties}

Naturally, our analysis suffers from a number of limitations and uncertainties. Those can be mostly divided into two categories: those inherent to all studies measuring molecular gas masses from single RJ dust continuum flux densities; and those specifically associated to our stacking analysis that are related to the averaged nature of our stacked measurements. In the following, we try to exhaustively list these limitations and uncertainties, and discuss their impact on the main conclusions of our analysis. 

\subsubsection{From observed-frame flux densities to molecular gas masses}
\label{subsec:Rj-to-Mol-uncertainties}
To convert observed-frame flux densities into molecular gas masses, we applied a two steps approach, i.e., first converting observed-frame flux densities into rest-frame 850$\,\mu$m luminosities using a standard SED template (the so-called $k$-correction) and then converting these rest-frame luminosities into molecular gas masses using a standard $L_{850}$-to-$M_{\rm mol}$ relation. To study how these particular choices of SED templates and $L_{850}$-to-$M_{\rm mol}$ relations influence our results, we repeated our analysis using alternatives commonly adopted in the literature. 

Instead of using the SED template of \citet{2012ApJ...757L..23B} to perform our $k$-corrections, we repeated our analysis using the SED template of \citet{2018A&A...609A..30S} or a single grey-body emission with $T_{\rm dust} = 25\,$K and $\beta=1.8$ \citep[as it is assumed in, e.g.,][]{2016ApJ...820...83S}. These two $k$-correction methods yield molecular gas masses which are, respectively, 12\% higher and 16\% lower at $z\sim$0.6 than our original calculation and 5\% higher and 5\% lower at $z\sim$3.2 than our original calculation. Because our original $k$-corrections are bracket by these alternatives and because the inferred offsets are in any cases well within the uncertainties of our original constraints, we conclude that the specific choice of this SED template has no significant impact on our results.

As extensively discussed in \citet{2019ApJ...887..235L}, systematic offsets are found between all different metallicity-dependent or -independent $L_{850}$-to-$M_{\rm mol}$ relations. We evaluate the impact of these relations on our results in Appendix~\ref{appendix: conversion} by repeating our analysis using instead of the metallicy-independent $L_{850}$-to-$M_{\rm mol}$ relation of \citet[][hereafter H17]{2017MNRAS.468L.103H}, (i) the H17 relation inferred assuming a $\alpha_{\rm CO}$ of 4.35 $M_{\odot}\,$(K\,km\,s$^{-1}$\,pc$^{2}$)$^{-1}$ (hereafter H17$_{\alpha_{\rm CO}=4.35}$) instead of 6.5 $M_{\odot}\,$(K\,km\,s$^{-1}$\,pc$^{2}$)$^{-1}$; (ii) the metallicity-independent $L_{850}$-to-$M_{\rm mol}$ relation of \citet[][hereafter S16]{2016ApJ...820...83S}, (iii) a $L_{850}$-to-$M_{\rm dust}$ relation assuming $T_{\rm dust} = 25\,$K and $\beta=1.8$ combined to the metallicity-dependent $M_{\rm dust}$-to-$M_{\rm mol}$ relation of \citet[][hereafter B18]{2018MNRAS.478.1442B} and finally (iv) a $L_{850}$-to-$M_{\rm dust}$ relation assuming $T_{\rm dust} = 25\,$K and $\beta=1.8$ combined to the metallicity-dependent $M_{\rm dust}$-to-$M_{\rm gas}$ relation of \citet[][hereafter $\delta_{\rm GDR}$]{2011ApJ...737...12L}. As for our $k$-corrections, our original calculation (i.e., H17) yields estimates which are bracket by these alternatives : H17$_{\alpha_{\rm CO}=4.35}$ produces estimates which are systematically lower than ours by $\sim0.17\,$dex, B18 gives values which are systematically lower than ours by $\sim0.17\,$dex; S16 yields values which are consistent with those reported here within $\sim0.04\,$dex; while the $\delta_{\rm GDR}$ method produces estimates which are systematically higher than ours by $\sim0.13\,$dex. In addition to this global offsets, the metallicity-dependent methods (i.e., $\delta_{\rm GDR}$ and B18) introduces redshift- and stellar mass-dependent trends, which are due to the fact that lower mass and higher redshift galaxies have increasingly lower metallicities. Consequently, the offsets between our measurements and those inferred with the $\delta_{\rm GDR}$ method increase towards lower masses and higher redshifts, while the offsets with B18 decrease towards lower masses and higher redshifts. Even if present, these stellar mass- and redshift-dependent trends do not change qualitatively the main conclusions of our papers, i.e., irrespective of the assumed methods (i) the molecular gas fraction of MS galaxies still increases with redshifts and decreases with stellar masses; (ii) their depletion time still remains mostly dependent on their redshifts and not their stellar masses and finally, (iii) MS galaxies still evolve along a seemingly universal MS-only KS relation. These evolutions are, however, quantitatively changed, with, e.g., the molecular gas fraction of MS galaxies increasing by a factor of $\sim$15, $\sim$45, $\sim$33, and $\sim$17 from $z\sim$0 to $z\sim$3.2 for the H17$_{\alpha_{\rm CO}=4.35}$, $\delta_{\rm GDR}$, S16, and B18 gas mass calibrations, respectively, as compared to the factor of 24 found for the H17 method; and their molecular gas depletion time being 200--300, 400--600, 400--700, and 200--500\,Myr for the H17$_{\alpha_{\rm CO}=4.35}$, $\delta_{\rm GDR}$, S16, and B18 gas mass calibrations, respectively, instead of 300--500\,Myr for our original calculation. Again, our original constraints are roughly bracket by these alternatives and values are consistent within 1-2$\sigma$. 

We note that combining the local measurements from \citet[][]{2017ApJS..233...22S} with our high-redshift H17 estimates yields a redshift evolution of the molecular gas content of SFGs that is in very good agreement with that of \citet{2020ARA&A..58..157T} at high stellar masses (Fig.~\ref{fig:gas}), i.e., where this latter can be considered as the reference as it is based on a fairly complete sample of massive SFGs across cosmic time and a thorough cross-calibration of the CO- and dust-based methods. This agreement could seem surprising as the $L_{850}$-to-$M_{\rm mol}$ relation of H17 was calibrated using $\alpha_{\rm CO}$=6.5, while our local reference, i.e., \citet[][]{2017ApJS..233...22S}, converted their CO measurements into molecular gas masses using $\alpha_{\rm CO}\sim$4 (at the high stellar masses of our study). This agreement between our high-redshift H17 measurements and those from \citet{2020ARA&A..58..157T} is due to the fact that using $\alpha_{\rm CO}$=6.5 instead of 4.3 to calibrate the local $L_{850}$-to-$M_{\rm mol}$ relation corrects indirectly (and to first order) for the fact that at a given stellar mass, high-redshift galaxies have lower metallicities than local galaxies, and thus have a higher gas-to-dust ratio \citep[e.g.,][]{2011ApJ...737...12L} and consequently should have a lower $L_{850}$-to-$M_{\rm mol}$ ratio.

Finally, we note that the particular choice of a $L_{850}$-to-$M_{\rm mol}$ relation cannot explain the differences observed between the molecular gas masses of MS galaxies at a given stellar mass and redshift inferred by our study and that from \citet{2019ApJ...887..235L} and \citet{2016ApJ...820...83S}. Indeed, \citet{2019ApJ...887..235L} also used as us H17 to infer their molecular gas mass estimates, while the method used in \citet{2017ApJ...837..150S}, i.e., S16, provides consistent results with H17 (within $\sim0.04\,$dex). In both cases, differences between our and their measurements are likely caused by the fact that these literature studies were largely biased towards individually-detected MS galaxies with massive gas reservoirs.

\subsubsection{Limitation and uncertainties associated with stacking}
Stacking in the $uv$-domain is a difficult task, which could be subject to a series of potential downfalls when applied to the heterogeneous A$^3$COSMOS database. To test the reliability of our stacking analysis, we used realistic simulations, in which mock sources with different flux densities and sizes were introduced in an A$^3$COSMOS-like interferometric database and subsequently stacked using the same procedure as the real sources. The results of these simulations, which are shown in Appendix~\ref{appendix: simu}, unambiguously demonstrate the reliability of our stacking analysis to accurately retrieve the intrinsic flux densities and sizes of a stacked population. As a reminder, performing such stacking analysis on images with drastically different spatial resolutions would be virtually impossible or very uncertain. 

While our simulations demonstrated that we were able to accurately measure the mean flux density of a galaxy population, one still has to remember that these stacked measurements are averaged values for galaxy populations with intricate stellar mass, SFR, and redshift distributions. How these mean molecular gas measurements, $\langle M_{\rm mol}\rangle$, can be related to $\langle t_{\rm cosmic}\rangle$, $\langle M_{\star}\rangle$, and $\langle\Delta {\rm MS}\rangle$ to infer $\mu_{\rm gas}(M_\ast,t_{\rm cosmic},\Delta {\rm MS})$ and $\tau_{\rm mol}(M_\ast,t_{\rm cosmic},\Delta {\rm MS})$ is an none trivial question and depends on the intrinsic stellar mass, SFR, and redshift distributions of each stacked populations. To address this issue, we applied a Bayesian analysis in which using the true distributions of these stacked populations we estimated the likelihood to measure their stacked $\langle M_{\rm mol}\rangle$ for a given analytical functions of $\mu_{\rm gas}(M_\ast,t_{\rm cosmic},\Delta {\rm MS})$ and $\tau_{\rm mol}(M_\ast,t_{\rm cosmic},\Delta {\rm MS})$. Without such an approach, i.e.,  simply inferring $\mu_{\rm gas}(M_\ast,t_{\rm cosmic},\Delta {\rm MS})$ and $\tau_{\rm mol}(M_\ast,t_{\rm cosmic},\Delta {\rm MS})$ by fitting $\langle M_{\rm mol}\rangle$, $\langle t_{\rm cosmic}\rangle$, $\langle M_{\star}\rangle$, and $\langle\Delta {\rm MS}\rangle$, those analytical functions would differ from our original calculation by up to 25\%. While this approach is currently the best way to deal with this averaging issue, only future high-sensitivity ALMA observations individually-detecting all our MS galaxies would be able to definitively constrain $\mu_{\rm gas}(M_\ast,t_{\rm cosmic},\Delta {\rm MS})$ and $\tau_{\rm mol}(M_\ast,t_{\rm cosmic},\Delta {\rm MS})$.

Recent findings in the literature suggests that there might exist an mean offset of $0\farcs2$ between the optical position and the actual (sub)millimeter position of our sources \citep[e.g.,][]{2018A&A...616A.110E}. While we verified via simulations that such an offset does not affect significantly our stacked flux density and size measurements, one should at worst consider our molecular gas sizes as upper limits or more realistically keep in mind that those should be corrected from this extra convolution kernel. We decided, however, not to correct our size measurements from this effect in Tab.~\ref{tab:sm} because this optical-to-FIR position offset still needs to be confirmed and because deconvolving those intrinsic sizes (i.e., $\theta_{\rm mol-uv}^{\rm circ}$ or $\theta_{\rm mol-py}^{\rm circ}$) by an Gaussian kernel with a $FWHM$ of $0\farcs2$ would only have lowered them by $\sim$6\%, leaving all our results unchanged.

\section{Discussion}
\label{sec:discussion}
Our analysis reveals that (i) the molecular gas fraction of MS galaxies increases with redshift and decreases with stellar mass; (ii) the depletion time of MS galaxies does not depend on their stellar masses but mostly on their redshift, increasing from 0.4\,Gyr at $z\sim3.6$ to 1.3\,Gyr at $z\sim0$; (iii) the FIR size of MS galaxies does not evolve with redshifts nor stellar masses, with a mean half-light radius of 2.2\,kpc; and finally, (iv) MS galaxies evolve along a seemingly universal MS-only KS relation with a slope of $\sim1.13$.

In the following, we discuss some of these results in light of recent observational findings and galaxy evolution scenarios.

\subsection{An universal Kennicutt-Schmidt relation}
It is crucial to accurately measure the relation between the SFR and gas (surface) densities of galaxies because it provides theoretical models with key information about the mechanisms and efficiency with which these galaxies turn their gas into stars. The pioneering work of \citet{1959ApJ...129..243S} suggests that in the Galactic plane the SFR volume density ($\rho_{\rm SFR}$) is proportional to the gas volume density ($\rho_{\rm gas}$) with a power law index of $\sim2$. This power law index directly reflects the physical conditions for star formation and can be studied, assuming a constant gas scale height, via the observationally more convenient $\Sigma_{\rm gas}$--$\Sigma_{\rm SFR}$ relation (i.e., the so-called KS relation). At sub-kpc scales, this power law index is found to vary from $\alpha\sim$0.75 to 2, rendering difficult any theoretical interpretation of this relation at small scales \citep[e.g.,][]{2002ApJ...569..157W, 2007A&A...461..143S, 2008AJ....136.2846B, 2013AJ....146...19L, 2013ApJ...772L..13M, 2014ApJ...788..167M, 2014MNRAS.437L..61S, 2019ApJ...882....5W, 2020A&A...641A..24W, 2021MNRAS.501.4777E, 2021A&A...650A.134P, 2021MNRAS.503.1615S}. For example, a power law index of $\sim$0.75 is expected if giant molecular clouds (GMCs) convert all their gas into stars over a free-fall time \citep[e.g.,][]{2005ApJ...630..250K}, while a power index of $\sim$2.0 is expected if star formation is mostly induced by collisions of small clouds of gas \citep[e.g.,][]{1986ApJ...311L..41W, 2002ApJ...569..157W}. 

K98 provides the first accurate measurement of the $\Sigma_{\rm gas}$--$\Sigma_{\rm SFR}$ relation at global scales by combining data from both normal and starburst galaxies. K98 find $\alpha=1.4$, i.e., near the expected value of $1.5$ for self-gravitating disks if the SFR scales as the ratio of gas volume density ($\rho_{\rm gas}$) to the free-fall timescale ($\rho_{\rm gas}^{-0.5}$). However, \citet{2008AJ....136.2846B} argue that the conditions for star formation in starbursts are too different to be combined with normal galaxies \citep[e.g.,][]{2004ApJ...606..271G, 2005ApJ...623..826R} and find $\alpha=1.0$ when considering only kpc-scale star-forming regions of nearby spirals. They conclude that stars are forming in GMCs with relatively uniform properties and that at supra-kpc scales, star formation remains unresolved and $\Sigma_{\rm SFR}$ becomes thus more a measure of the filling fraction of GMCs than changes in conditions for star-formation. Recent studies at high redshift and global scales support this conclusion \citep[e.g.,][]{2010MNRAS.407.2091G, 2017A&A...602L...9M}. In particular, \citet{2010MNRAS.407.2091G} argue that normal and starburst galaxies seem to follow two different KS relations both with a near-unity power law index but with different normalization, the latter galaxy population being more efficient in turning gas into stars. Our analysis also supports the existence of an universal KS relation for MS galaxies with a near-unity power law index. Our results are the first to extend this finding up to $z\sim3.6$ and using a mass-complete sample of $>10^{10}\,M_{\odot}$ MS galaxies. We note in particular that the extension of our analysis to very high redshift is crucial because high-redshift MS galaxies have sufficiently high $\Sigma_{\rm gas}$ to provide adequate leverage to accurately constrain the power law index of the KS relation on global scales. The high $\Sigma_{\rm gas}$ of these high-redshift MS galaxies also allow us to compare their star formation efficiency with that of local starbursts which have similarly high $\Sigma_{\rm gas}$ \citep[see local starbursts on Fig.~\ref{fig:ks} from K98 and][]{2019ApJ...872...16D}.

While the power law index of the KS relation for MS galaxies found in our analysis is consistent with that of \citet{2010MNRAS.407.2091G}, our normalisation differs by about $0.2\,$dex (see right panel of Fig.~\ref{fig:ks}), with our findings predicting shorter depletion times for MS galaxies than theirs. This difference can most likely be explained by the same limitation than that affecting \citet{2019ApJ...887..235L} and \citet{2020ARA&A..58..157T}, i.e., literature studies on the KS relation are based on CO individually-detected galaxy observations and thus likely biased towards gas-rich galaxies at fix $\Sigma_{\rm SFR}$. Naturally, one cannot also exclude that part of this offset could be due to some remaining offset between CO-based and dust-based molecular gas mass estimates \citep[e.g., ][]{2020ARA&A..58..157T}.

\begin{figure*}
\centering
\includegraphics[width=1.0\textwidth]{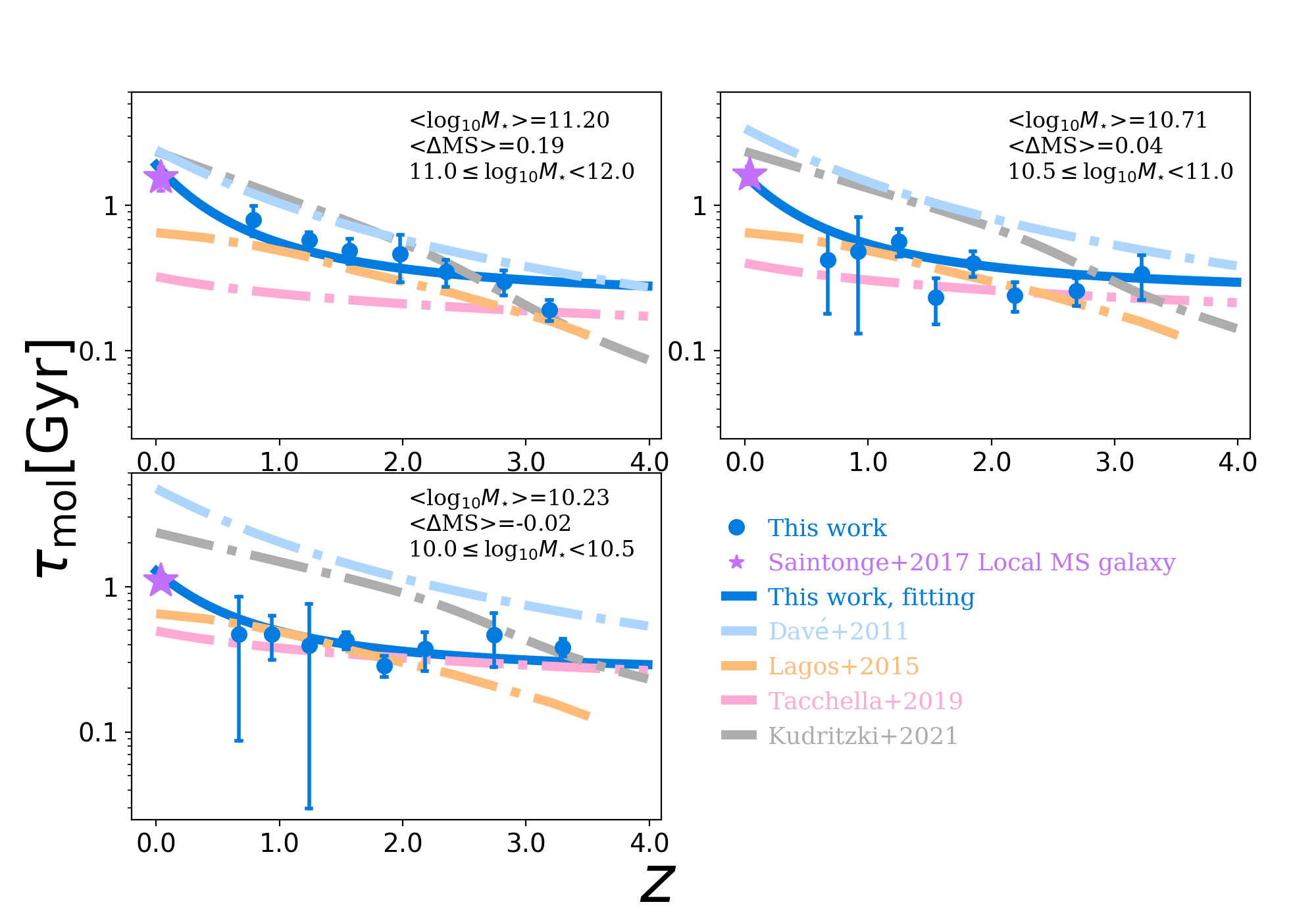}
\caption{Same as Fig.~\ref{fig:dep}, but comparing our results to predictions from the cosmological hydrodynamic simulations of \citet[][dash-dotted light-blue line]{2011MNRAS.416.1354D}, \citet[][dash-dotted orange line]{2015MNRAS.452.3815L}, \citet[][dash-dotted pink line]{2016MNRAS.457.2790T}, and \citet[][dash-dotted grey line]{2021ApJ...910...87K}.}
\label{fig:dep_simulation}
\end{figure*}
\subsection{Molecular gas depletion time}
The molecular gas depletion time is defined as the time that a galaxy would need to deplete its molecular gas reservoir through star formation, provided a constant SFR and no gas accretion, i.e., $\tau_{\rm mol}$ = $M_{\rm mol}$/SFR. It can theoretically be written as $\tau_{\rm mol}$ = $M_{\rm mol}$/SFR = $t_{\rm ff}$/$\varepsilon_{\rm ff}$, where $t_{\rm ff}$ is the free-fall time and $\varepsilon_{\rm ff}$ is a dimensionless measure of the SFR efficiency, linking the mass of gas available for star formation and that effectively turning into stars \citep{2012ApJ...745...69K}. Theoretically, $\varepsilon_{\rm ff}$ is supposed to be roughly constant ($\approx0.01$), rendering any variations in depletion time mostly due to variations in $t_{\rm ff}$ \citep{2012ApJ...745...69K}.

Observationally, the molecular depletion time of SFGs was found to follow tight scaling relations with their $\Delta$MS, redshifts, and stellar masses, providing thereby key information for models of galaxy evolution \citep[e.g.,][]{2016MNRAS.462.1749S, 2018ApJ...853..179T, 2020ARA&A..58..157T, 2019ApJ...887..235L, 2020A&A...643A.180H, 2020MNRAS.496.2531P}. For example, the depletion time of galaxies situated well above the MS (i.e., $\times 4$) was found to be significantly shorter ($\sim 0.1$Gyr) than that of MS galaxies, suggesting a different star formation mode for this galaxy population, likely triggered by the major merger of two gas-rich galaxies \citep[e.g.,][]{2010MNRAS.407.2091G, 2011ApJ...743..159H, 2012MNRAS.424.2232A, 2013Natur.496..329R}. In addition, the depletion time of MS galaxies was found to slightly decrease with redshift up to $z\sim3$ \citep[e.g.,][]{2016ApJ...833..112S, 2016MNRAS.457.2790T, 2018ApJ...853..179T, 2020ARA&A..58..157T, 2020MNRAS.496.1124P}. Our analysis, which confirms this finding, reveals that this apparent evolution is, however, not associated per se to a redshift evolution of the star formation mode of MS galaxies but rather to the increase of their gas content with redshift, a relatively constant star-forming extent, and a seemingly universal KS relation with a power law index of $\sim1.13$: at a given stellar mass, the gas content of MS galaxies increases with redshift while their star-forming size remains roughly constant; their $\Sigma_{\rm M_{\rm mol}}$ increases thus smoothly with redshift, shifting towards higher $\Sigma_{\rm SFR}$ to $\Sigma_{\rm M_{\rm mol}}$ ratios. The molecular gas depletion time of MS galaxies was finally found to slightly decrease with stellar masses \citep[e.g.,][]{2016MNRAS.462.1749S, 2018ApJ...853..179T, 2020ARA&A..58..157T, 2020MNRAS.496.1124P}. On the contrary, our analysis finds that the evolution of their depletion time is mostly independent from their stellar masses. We note that this stellar mass-independent evolution of the depletion time suggests that the flattening of the MS at high stellar masses ($>10^{10.5}\,$M$_{\odot}$) and $z\lesssim2.5$ \citep{2020ApJ...899...58L} is mostly due to their lower gas content rather than lower star formation efficiencies (see Sect.~\ref{subsec:flattening MS}).

The scaling relations between the molecular gas depletion time of MS galaxies and their redshifts and stellar masses provide stringent constraints to hydrodynamic simulations performed in a cosmological context. In Fig.~\ref{fig:dep_simulation}, we compare our findings to predictions from the simulations of \citet[]{2011MNRAS.416.1354D}, \citet[]{2015MNRAS.452.3815L}, \citet[]{2016MNRAS.457.2790T}, and \citet[]{2021ApJ...910...87K}. Overall, all these simulations predict a decrease of the molecular gas depletion time of MS galaxies with redshift and stellar mass, in qualitative agreement with our observations. However, the slope and overall normalisation of these scaling relations vary by at least a factor three between all these simulations and none can accurately reproduce the observed relations. As discussed in \citet{2021ApJ...910...87K}, predictions of the molecular depletion time are indeed strongly affected by the exact star formation, accretion and feedback models implemented in these simulations. The large disagreement between simulations and with the observations demonstrates that our understanding of these complex mechanisms across cosmic time is far from being complete and it also demonstrates the power of simple scaling relations to constrain models of galaxy evolution.

Finally, irrespective of the exact slope of these various scaling relations, all observations point towards relatively short molecular gas depletion times ($\sim0.5-1\,$Gyr) for MS galaxies of any stellar masses and redshifts. Without a constant replenishment of their gas reservoirs, the population of MS galaxies observed at, for example, $z\sim2$, would thus have fully disappeared by $z=1.5$ \citep[see, e.g.,][]{2020ApJ...902..111W}. These findings strongly support the so-called gas regulator models \citep[e.g.,][]{2008ApJ...674..151E, 2010ApJ...718.1001B, 2012MNRAS.421...98D, 2013ApJ...772..119L, 2014MNRAS.443.3643P, 2016MNRAS.458.3168R}, in which galaxy growth is mostly driven by a continuous supply of fresh gas from the cosmic web \citep[][]{2009Natur.457..451D}.

\subsection{Compact star forming extent ?}
Our analysis as well as numerous recent studies have revealed that the star-forming half-light radius of MS galaxies is relatively compact, i.e., 1--3\,kpc, and does not evolve significantly with redshift nor stellar mass \citep[e.g.,][]{2016ApJ...827L..32B,2016ApJ...833...12R,2018A&A...616A.110E,2019ApJ...879...54L,2019A&A...625A.114J,2020ApJ...888...44C,2020ApJ...901...74T}. In contrast, \citet{2017ApJ...850...83F} found that the FIR size of MS galaxies evolves slightly with redshift. However, as stressed in their study, their individually-detected ALMA sample is SFR-selected and could therefore be biased at high redshift towards compact star-forming galaxies, i.e., galaxies with high surface brightness. In any cases, the optical half-light radius of late-type galaxies of similar masses and at $z\lesssim3$ is found to be about two times larger ($\sim3-8\,$kpc) than their star-forming extent \citep[Sect.~\ref{subsec: FIR sizes}][]{2014ApJ...788...28V, 2017ApJ...850...83F, 2018A&A...616A.110E, 2019ApJ...879...54L, 2019A&A...625A.114J, 2021ApJ...910..106J, 2020A&A...635A.119C, 2020ApJ...901...74T}. This centrally enhanced star formation is usually interpreted in the literature as a sign that the cold gas accreted by MS galaxies falls preferentially onto their central region and triggers the formation of their bulge \citep[e.g.,][]{2006ApJ...642L..17F, 2016ApJ...827...28G, 2016MNRAS.459.4109T}. These bulges would therefore grow from inside out and quench in the latest evolutionary stage of MS galaxies, leaving solely the outer disk with star formation activities \citep[e.g.,][]{2015Sci...348..314T, 2018MNRAS.474.2039E, 2018MNRAS.480.2544R, 2020A&A...644A..97C}. While possible for massive star-forming galaxies, which are known to have massive central bulges \citep[e.g.,][]{2020ApJ...899...58L}, such interpretation seems less likely for less massive MS galaxies (i.e., $M_{\star} < 10^{11.0}\,$M$_\odot$) that are explored for the first time here. Using high-resolution ALMA and \textit{Hubble Space Telescope} observations of 20 submillimeter-selected galaxies, \citet{2019ApJ...879...54L} argues instead that the discrepancy between FIR and optical sizes is mostly due to observational biases in which important radial color gradients yield very discrepant half-light and half-mass radii. This observational finding has recently been supported by TNG50 simulations coupled with state-of-the-art radiative transfer code to study the FIR, optical, and half-mass radius of thousands high-redshift $10^{9}-10^{11}\,$M$_\odot$ MS galaxies \citep[][]{2022MNRAS.510.3321P}. Indeed, in these simulations it is found that while the FIR half-light radius correlates with the radius containing half the star formation in galaxies, strong and un-corrected obscuration of the stellar light toward the galaxy centre increases significantly the apparent extent of the disk sizes in the optical. \citet{2022MNRAS.510.3321P} conclude that the compact dust-continuum emission of MS galaxies with respect to the optical size is not necessarily evidence of the buildup of a dense central stellar component. Future high-resolution near-infrared observations performed by the \textit{James Webb Space Telescope} will certainly play a key role in validating or invalidating these later findings.

\subsection{The flattening of the MS relation at high masses}
\label{subsec:flattening MS}
It is now relatively well established that the slope of the MS of star-forming galaxies flattens at high stellar masses, with this flattening becoming more and more prominent at $z$ $\lesssim2.5$ \citep[e.g.,][]{2015A&A...575A..74S, 2019MNRAS.490.5285P, 2020ApJ...899...58L}. Such flattening of the MS is usually associated to the so-called mass-quenching model \citep[e.g.,][]{2010ApJ...721..193P, 2018MNRAS.477..450N, 2019MNRAS.487.3740W}. In this model, massive galaxies with their high gravitational potential hold a large accretion rate, growing their core rapidly in a few Gyr. As the core keeps growing, however, the ever larger gravitational potential could shock heat and/or an AGN could heat the new infalling gas, slowing down the gas accretion rate and reducing thereby the specific SFR of massive galaxies \citep[e.g.,][]{2006MNRAS.368....2D, 2019MNRAS.490.2139R, 2021MNRAS.500.4004D}. On the other hand, less massive galaxies could take more than a Hubble time to trigger such feedback and could thus efficiently accrete fresh cold gas even at low redshift.  Our analysis reveals that the molecular gas fraction of our MS galaxies decreases with stellar mass at a rate mirroring than that of the flattening of the MS, yielding almost constant depletion time (equivalently star formation efficiency) with stellar mass. This implies that the slow downfall of the star formation in massive MS galaxies is principally due to an decrease of their molecular gas content rather than a decrease of their star formation efficiency, in agreement with recent observations of low-redshift galaxies \citep{2020A&A...644A..97C}. Our findings support thus an interpretation in which the flattening of the MS at high masses is primarily controlled by the ability of galaxies to efficiently accrete or not accrete gas from the IGM.

Finally, we note that \citet{2020ApJ...899...58L} found that the flattening of the MS must be linked with changes in the morphological composition of galaxies: bulge-dominated late-type galaxies, which dominate the SFG population at high stellar masses, show a flattening of the MS, while disk-dominated late-type galaxies have align on a SFR-$M_\ast$ sequence with a slightly higher normalization and with a power law index in the log-log space close to unity. Although this result seems to favor a scenario in which the gas in bulge-dominated MS galaxies is stabilized against fragmentation \citep[the so the-called morphological quenching;][]{2013MNRAS.432.1914M}, our findings suggests that the gas is instead simply not present to form stars in these galaxies, i.e., either it has been removed by feedback or cold gas is no longer able to be accreted efficiently onto bulge-dominated late-type galaxies.

\section{Summary}
\label{sec:summary}
We investigate the evolution of the molecular gas content of MS galaxies from $z\sim3.6$ to $z\sim0.4$. We apply an innovative $uv$-based stacking analysis to a large set of ALMA observations towards a mass-complete sample of $>\,10^{10}\,$M$_{\odot}$ MS galaxies. This $uv$-based stacking analysis, performed on the Rayleigh-Jeans dust continuum emission of these galaxies, provides an accurate measurement of their mean molecular gas content \citep{2017MNRAS.468L.103H}. With this unique dataset and innovative approach, we constrain the redshift and stellar mass evolution of the mean molecular gas mass, molecular gas fraction, molecular gas depletion time, and molecular gas size of MS galaxies down to $10^{10}\,$M$_{\odot}$ and up to $z\sim3.4$. Finally, we also study for the first time -- using a mass-complete sample of MS galaxies -- the KS relation at high-redshift. Our main findings are:
\begin{enumerate}
    \item The mean molecular gas mass of MS galaxies evolves significantly with redshift and depends on the stellar mass. At all stellar masses, the molecular gas fraction (i.e., $\mu_{\rm gas}=M_{\rm mol}/M_\star$) decreases by a factor $\sim24$ from $z\sim3.2$ to $z\sim0$. In addition, at a given redshift, $\mu_{\rm gas}$ decreases with stellar mass, at roughly the same rate than the decrease of the specific SFR (i.e., ${\rm SSFR}={\rm SFR}/M_\star$) of MS galaxies.
    \item Our mean molecular gas mass measurements are generally lower ($\sim$10--60$\%$) than literature estimates \citep[e.g.,][]{2017ApJ...837..150S, 2019ApJ...887..235L, 2020ARA&A..58..157T}, especially at low stellar masses. Literature measurements, which mostly relied on individually-detected galaxies, were likely biased toward gas-rich galaxies.
    \item The molecular gas depletion time (i.e., $\tau_{\rm mol}=M_{\rm mol}/{\rm SFR}$) of MS galaxies remains mostly constant at $z>0.5$ with a value of 300--500\,Myr, but increases by a factor of $\sim3$ by $z\sim0$. 
    \item The mean FIR size MS galaxies does not seem to evolve significantly with redshift nor stellar mass, with a mean circularized half-light radius of $\sim$2.2\,kpc. This result agrees qualitatively and quantitatively with the star-forming extent of MS galaxies measured in \citet[][]{2019A&A...625A.114J} using high-resolution radio observations of the COSMOS field (i.e., $R_{\rm radio}^{\rm circ}$ $\sim$ 1.5$^{+1.5}_{-0.8}$ kpc).
    \item The redshift evolution of $\tau_{\rm mol}$ can be accurately predicted from the redshift evolution of the molecular gas surface density (i.e., $\Sigma_{M_{\rm mol}}$) of MS galaxies and a seemingly universal MS-only $\Sigma_{M_{\rm mol}}-\Sigma_{M_{\rm mol}}$ relation with a slope of $\sim1.13$, i.e., the so-called Kennicutt-Schmidt relation.
\end{enumerate}
Our findings provide key constraints for galaxy evolution models, as $>10^{10}\,$M$_{\odot}$ MS galaxies are known to be responsible for the bulk of the star-forming activity of the universe over the last 10\,Gyr. To first order, it seems that the molecular gas content of MS galaxies regulates the evolution of their star formation activity across cosmic time, while variation of their star formation efficiency (i.e., $1/\tau_{\rm mol}$) plays only a secondary role. The short depletion time of the molecular gas reservoir of MS galaxies ($<1\,$Gyr) contrasts with the long duty cycle inherent to the existence of the MS itself. This suggests that the continuous replenishment of the molecular gas reservoir of MS galaxies plays a fundamental role in regulating star formation across cosmic time. Finally, despite large variations of the gas content and star formation rate of MS galaxies over the last 10\,Gyr, their star formation seems to take place in their inner 2\,kpc radius and to follow a seemingly universal MS-only $\Sigma_{M_{\rm mol}}-\Sigma_{\rm SFR}$ relation.

\begin{acknowledgements}
This research was carried out within the Collaborative Research Centre 956, sub-project A1, funded by the Deutsche Forschungsgemeinschaft (DFG) – project ID 184018867. ES and DL acknowledge funding from the European Research Council (ERC) under the European Union's Horizon 2020 research 
and innovation programme (grant agreement No. 694343). ALMA is a partnership of ESO (representing its member states), NSF (USA) and NINS (Japan), together with NRC (Canada), MOST and ASIAA (Taiwan), and KASI (Republic of Korea), in cooperation with the Republic of Chile. The Joint ALMA Observatory is operated by ESO, AUI/NRAO and NAOJ. This paper makes use of the following ALMA data: ADS/JAO.ALMA\#2015.1.00026.S, ADS/JAO.ALMA\#2015.1.00055.S, ADS/JAO.ALMA\#2015.1.00122.S, ADS/JAO.ALMA\#2015.1.00137.S, ADS/JAO.ALMA$\#$2015.1.00260.S, ADS/JAO.ALMA$\#$2015.1.00379.S, ADS/JAO.ALMA$\#$2015.1.00388.S, ADS/JAO.ALMA$\#$2015.1.00540.S, ADS/JAO.ALMA$\#$2015.1.00568.S, ADS/JAO.ALMA$\#$2015.1.00664.S, ADS/JAO.ALMA$\#$2015.1.00695.S, ADS/JAO.ALMA$\#$2015.1.00704.S, ADS/JAO.ALMA$\#$2015.1.00928.S, ADS/JAO.ALMA$\#$2015.1.01074.S, ADS/JAO.ALMA$\#$2015.1.01105.S, ADS/JAO.ALMA$\#$2015.1.01111.S, ADS/JAO.ALMA$\#$2015.1.01171.S, ADS/JAO.ALMA$\#$2015.1.01212.S, ADS/JAO.ALMA$\#$2015.1.01345.S, ADS/JAO.ALMA$\#$2015.1.01495.S, ADS/JAO.ALMA$\#$2016.1.00279.S, ADS/JAO.ALMA$\#$2016.1.00463.S, ADS/JAO.ALMA$\#$2016.1.00478.S, ADS/JAO.ALMA$\#$2016.1.00624.S, ADS/JAO.ALMA$\#$2016.1.00646.S, ADS/JAO.ALMA$\#$2016.1.00778.S, ADS/JAO.ALMA$\#$2016.1.00804.S, ADS/JAO.ALMA$\#$2016.1.01012.S, ADS/JAO.ALMA$\#$2016.1.01040.S, ADS/JAO.ALMA$\#$2016.1.01184.S, ADS/JAO.ALMA$\#$2016.1.01208.S, ADS/JAO.ALMA$\#$2016.1.01240.S, ADS/JAO.ALMA$\#$2016.1.01355.S, ADS/JAO.ALMA$\#$2016.1.01426.S, ADS/JAO.ALMA$\#$2016.1.01559.S, ADS/JAO.ALMA$\#$2016.1.01604.S, ADS/JAO.ALMA$\#$2017.1.00326.S, ADS/JAO.ALMA$\#$2017.1.00413.S, ADS/JAO.ALMA$\#$2017.1.00428.L, ADS/JAO.ALMA$\#$2017.1.01020.S, ADS/JAO.ALMA$\#$2017.1.01176.S, ADS/JAO.ALMA$\#$2017.1.01276.S, ADS/JAO.ALMA$\#$2017.1.01358.S, ADS/JAO.ALMA$\#$2017.1.01451.S, ADS/JAO.ALMA$\#$2017.1.01618.S, ADS/JAO.ALMA$\#$2017.1.00046.S,
ADS/JAO.ALMA$\#$2017.1.01217.S, ADS/JAO.ALMA$\#$2017.1.01259.S,
ADS/JAO.ALMA$\#$2018.1.00085.S, ADS/JAO.ALMA$\#$2018.1.00216.S, ADS/JAO.ALMA$\#$2018.1.00681.S, ADS/JAO.ALMA$\#$2018.1.00992.S,
ADS/JAO.ALMA$\#$2018.1.01044.S, ADS/JAO.ALMA$\#$2018.1.01359.S.
\end{acknowledgements}

\begin{appendix}
\section{Reliability of the stacked flux density and size measurements}
\label{appendix: simu}
To test the reliability of the flux densities and sizes measured by stacking in the $uv$- and image-domain, we use simulations tailored to reproduce the peculiar properties of the A$^3$COSMOS archive, i.e., with a heterogeneous frequency, depth and spatial resolution. The logic of our approach is to (i) simulate 100 realistic ALMA observations of a galaxy population with a given intrinsic flux density and size properties, (ii) stack these 100 mock observations, and (iii) finally compare the measured stacked flux density and size to the intrinsic ones. For a given set of flux density and size properties, we judge the reliability of our stacking analysis by computing $(S_{\rm in}-S_{\rm out})/S_{\rm in}$ and $(R_{\rm in}^{\rm circ}-R_{\rm out}^{\rm circ})/R_{\rm in}^{\rm circ}$, i.e., the error on the retrieved stacked flux density and size, respectively. 

Our mock ALMA observations were generated using the CASA task \texttt{simobserve}. To realistically reproduce the heterogeneity of the A$^3$COSMOS archive, the observing properties of each simulation -- i.e., frequency resolution, integration time, and antenna positions -- were randomly picked from one of the A$^3$COSMOS dataset. Then, the stacked position of the mock galaxies were randomly selected within a radius of 3\arcsec from the simulated phase center. Finally, to account for possible mismatches between the stacked position and the actual mm position of the stacked sources, we randomly placed these mock galaxies around their stacked position, following a 2D Gaussian distribution with a dispersion of $0\farcs2$ \citep[e.g.,][]{2018A&A...616A.110E}. Note that we do not test here the effect of the observed-frame flux density to rest-frame luminosities conversion discussed in Sect.~\ref{subsec:band}. Indeed, in a technical point of view this conversion is strictly equivalent to a simple multiplication and thus an increase or decrease of the noise in the initial to-be-stacked dataset. Our simulation already test the effect of stacking dataset with different noise, re-distributing this noise distribution is thus unnecessary. 

\begin{figure*}
\centering
\includegraphics[width=0.75\textwidth]{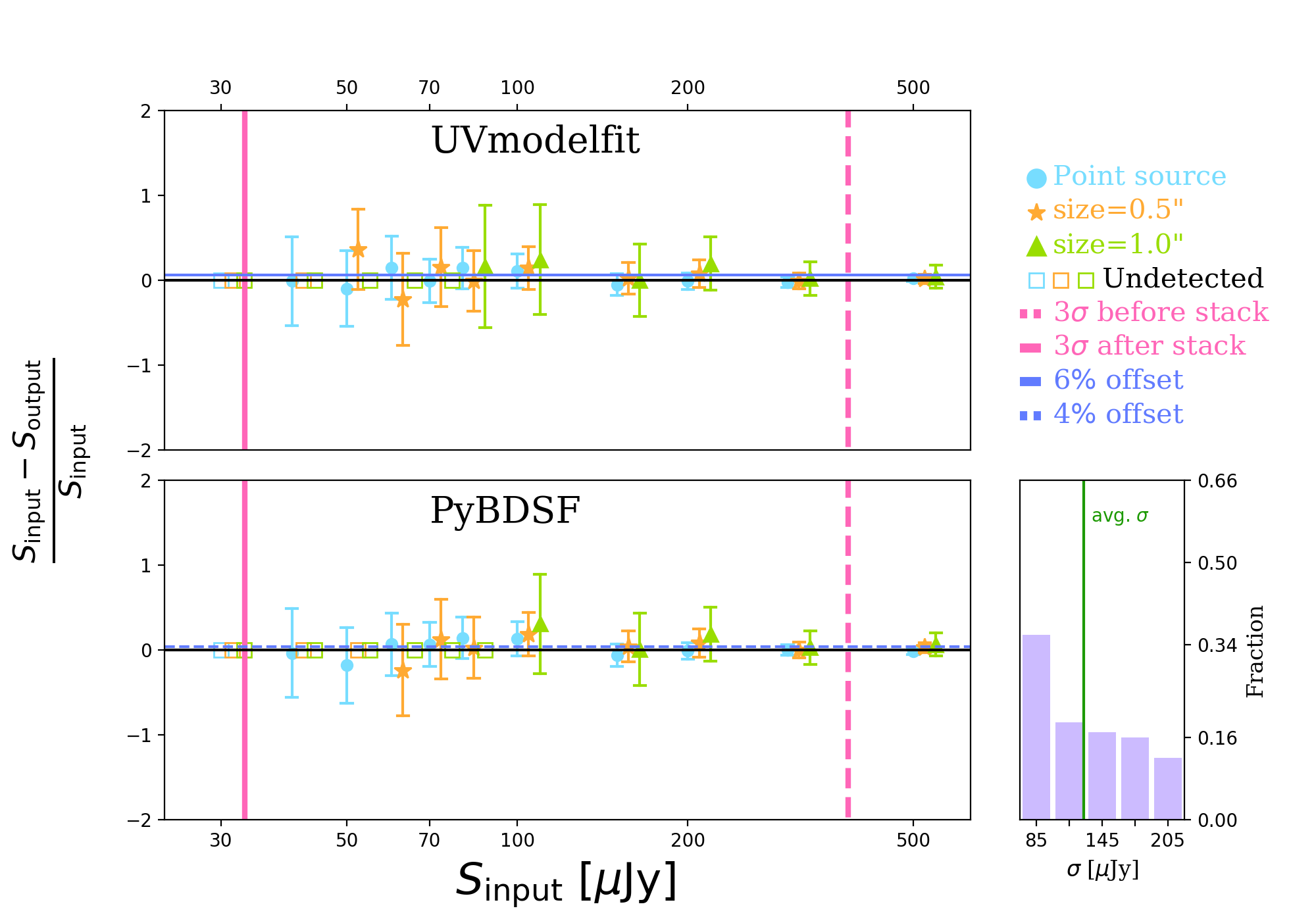}
\caption{Uncertainty on the $uv$-domain (upper panel) and image-domain (lower panel) stacked flux density measurements of 100 simulated galaxies as a function of their intrinsic flux densities and sizes. The vertical pink dashed line shows the average `point source' $3\sigma$ detection threshold of the measurement sets to be stacked, while the lower right sub-panel displays their $1\sigma$ distribution. The vertical solid pink line presents the `point source' $3\sigma$ detection threshold in the stacked measurement set. Our stacking analysis allows accurate mean flux density measurements (i.e., $(S_{\rm in}-S_{\rm out})/S_{\rm in}\sim 0$) for galaxy populations which are otherwise individually undetected, i.e., with intrinsic flux densities lower than the vertical pink dashed line. Stacked flux densities are in average underestimated by 6$\%$ and 4$\%$ in the $uv$-domain and image-domain, respectively, as illustrated by the dark blue solid and dashed horizontal lines, respectively.}
\label{fig:append1}
\end{figure*}

\begin{figure*}
\centering
\includegraphics[width=0.75\textwidth]{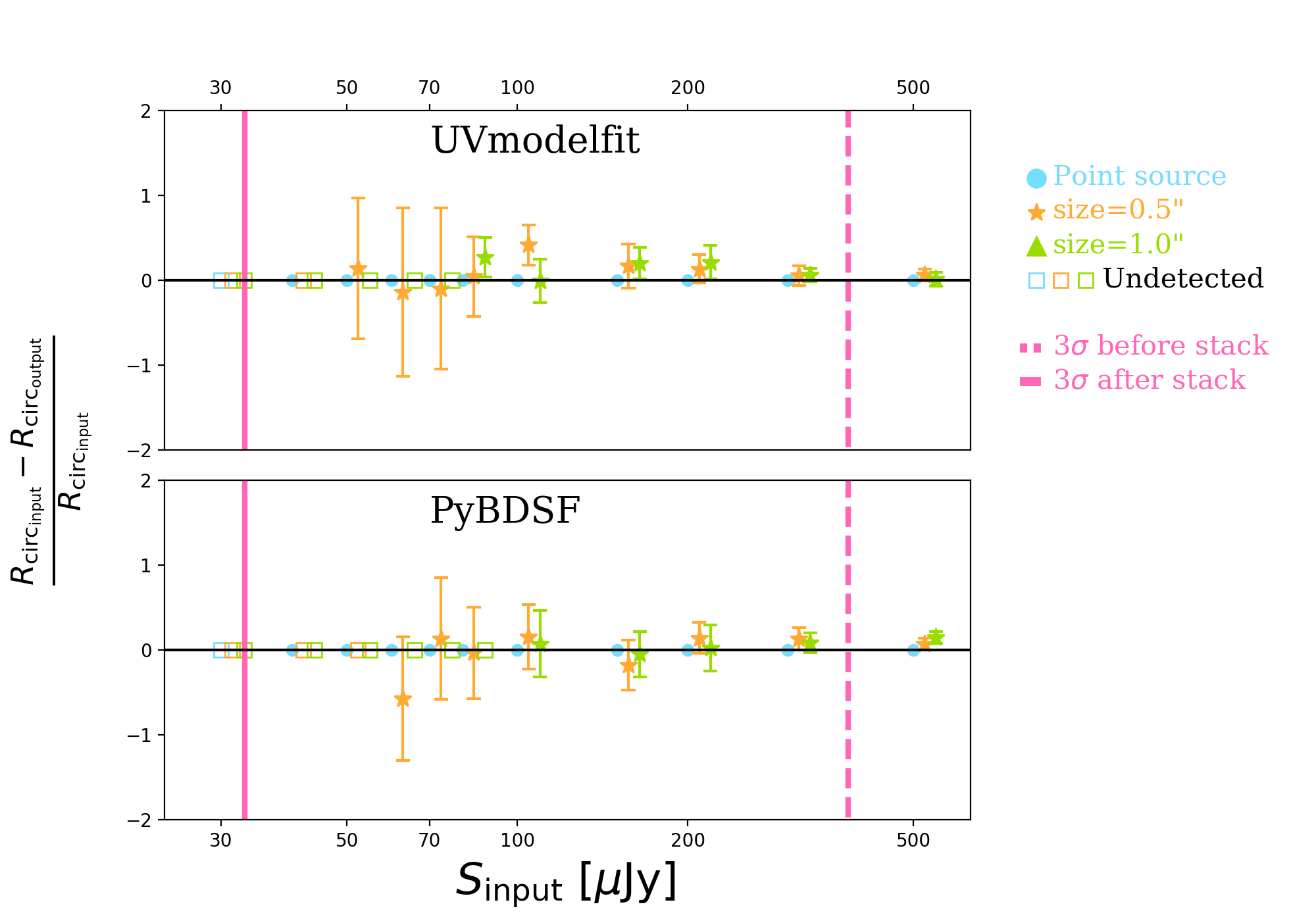}
\caption{Uncertainty on the $uv$-domain (upper panel) and image-domain (lower panel) stacked size measurements of 100 simulated galaxies as a function of their intrinsic flux densities and sizes. Lines and symbols are the same as in Fig.~\ref{fig:append1}. Our stacking analysis allows accurate mean size measurements (i.e., $(R_{\rm in}^{\rm circ}-R_{\rm out}^{\rm circ})/R_{\rm in}^{\rm circ}\sim 0$) for galaxy populations which are otherwise individually undetected, i.e., with intrinsic flux densities lower than the vertical pink dashed line.}
\label{fig:append2}
\end{figure*}

Results of these Monte Carlo simulations for three realistic angular sizes (i.e., point source, $FWHM=0\farcs5$ and $FWHM=1\farcs0$) and twelves flux densities combinations, are shown in Fig.~\ref{fig:append1} and \ref{fig:append2}. These results demonstrate the reliability of our stacked flux density and size measurements. Indeed, all size-flux density combinations which yield a detection have $(S_{\rm in}-S_{\rm out})/S_{\rm in}$ and $(R_{\rm in}^{\rm circ}-R_{\rm out}^{\rm circ})/R_{\rm in}^{\rm circ}$ values consistent with 0 within the measured uncertainties. We only notice a slight systematic overestimation of our stacked flux densities by $6\%$ and $4\%$ when inferred from the $uv$- and image-domain, respectively. Such an offset is virtually insignificant with respect to the measured uncertainties. Lastly we note that galaxies with larger intrinsic sizes (i.e., $1\farcs0$) do not yield a detection down to our lowest flux density combination. This is not inherent to our stacking analysis but simply to the spreading of the flux densities of these galaxies over several synthesized beams.   

\section{Different gas mass calibration}
\label{appendix: conversion}

\begin{figure*}
\centering
\includegraphics[width=\textwidth]{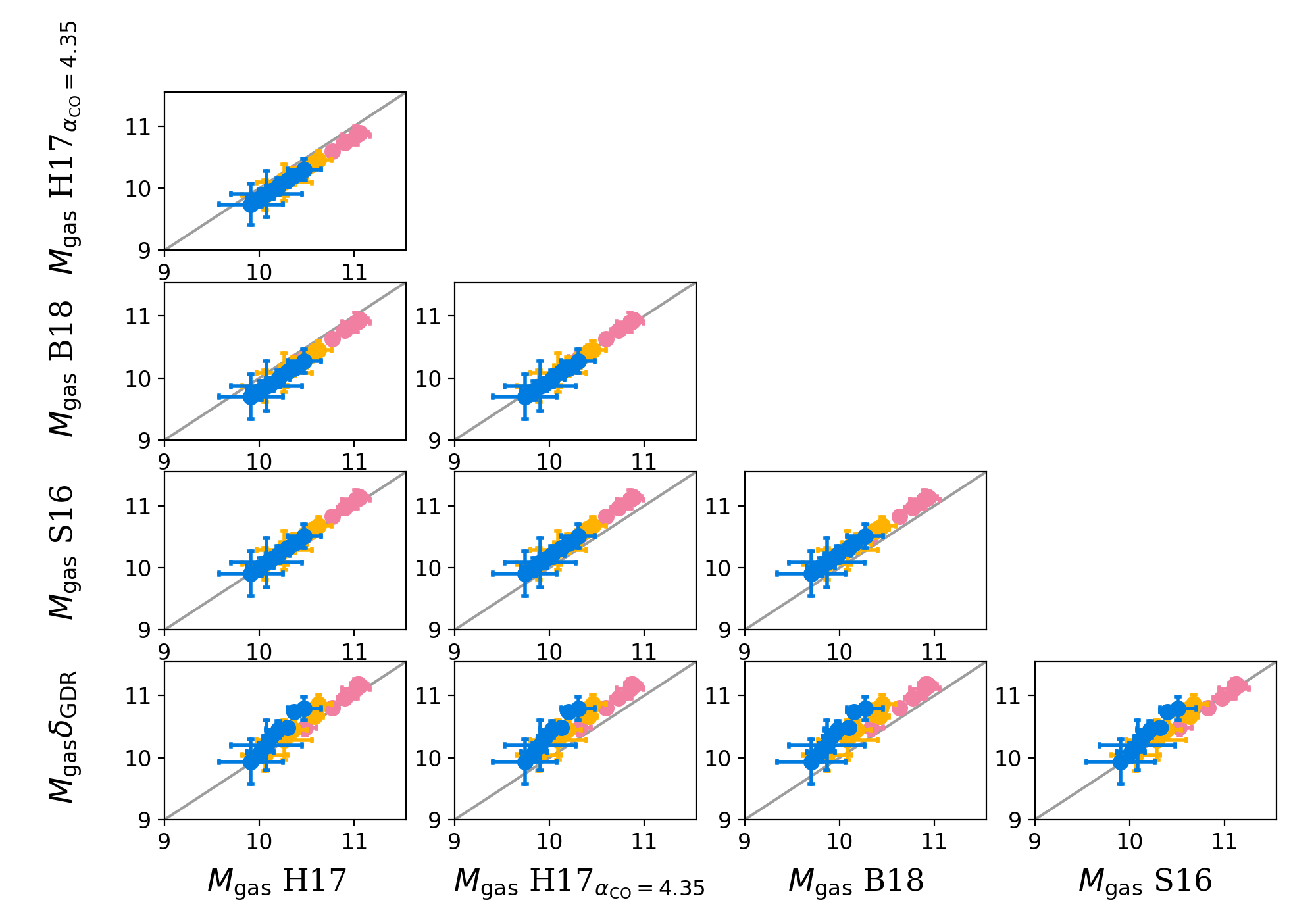}
\caption{Comparisons between four molecular gas mass calibration, i.e., from \citet[][]{2017MNRAS.468L.103H} assuming $\alpha_{\rm CO}=6.5$ (H17; used throughout our paper), \citet[][]{2017MNRAS.468L.103H} assuming $\alpha_{\rm CO}=4.35$ (H17$_{\alpha_{\rm CO}=4.35}$), \citet[][S16]{2016ApJ...820...83S}, \citet[][B18]{2018MNRAS.478.1442B}, and \citet[][$\delta_{\rm GDR}$]{2011ApJ...737...12L}. In each panel, the gray solid line is the one-to-one relation. Color of symbols are the same as in Fig.~\ref{fig:ks}.}
\label{fig:append_B_lum_met}
\end{figure*}


\begin{figure}
\centering
\includegraphics[width=\columnwidth]{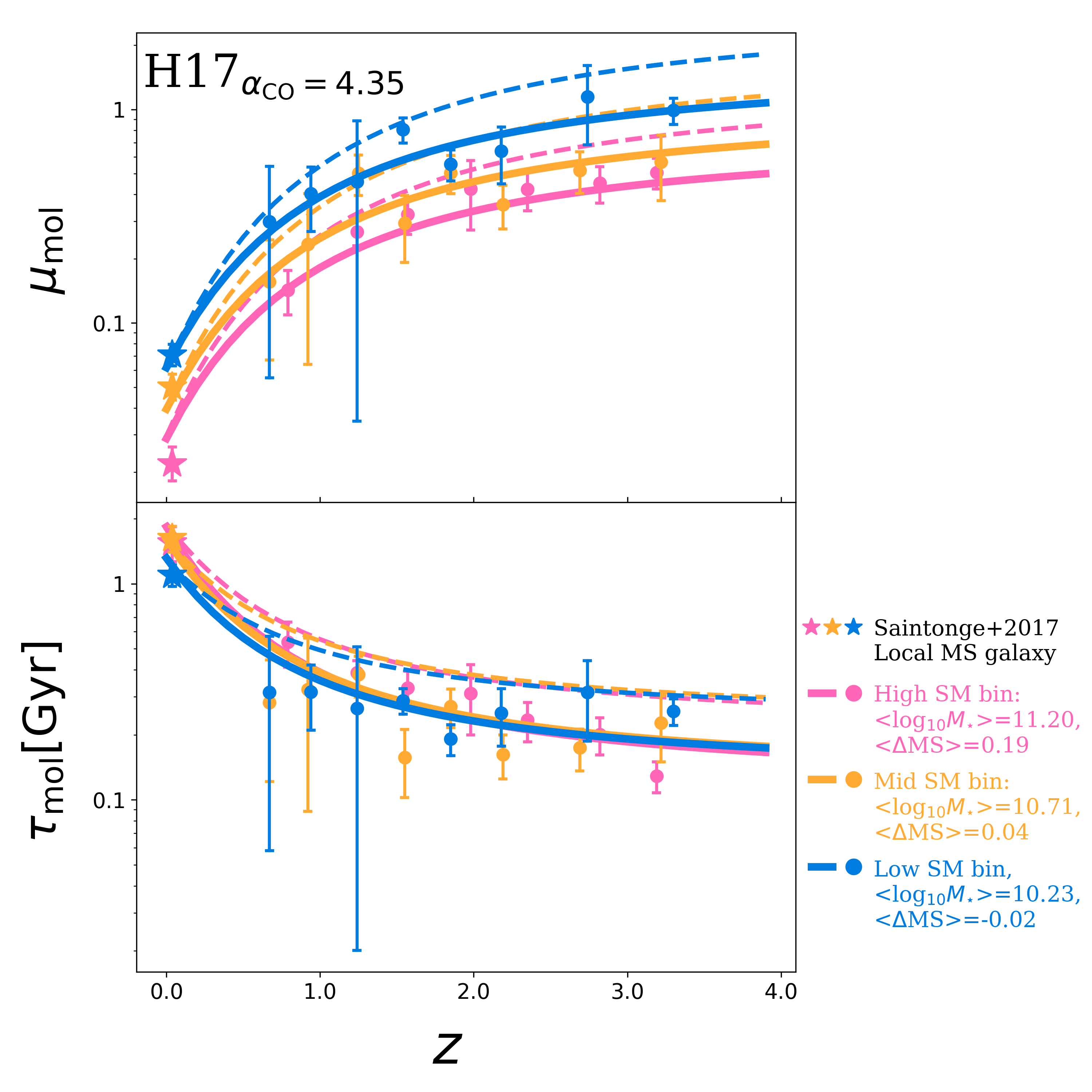}
\caption{Redshift evolution of the mean molecular gas fraction and molecular gas depletion time of MS galaxies, with the $M_{\rm gas}$ from the H17$_{\alpha_{\rm CO}=4.35}$ gas mass calibration. Dots show the mean molecular gas fraction and molecular gas depletion time from our work. Stars represent the local reference taken from \citet[][]{2017ApJS..233...22S}. Solid lines display the analytical evolution of the molecular gas fraction and molecular gas depletion time inferred with the H17$_{\alpha_{\rm CO}=4.35}$ gas mass calibration. Dashed lines show the analytical evolution of the molecular gas fraction and molecular gas depletion time inferred with the H17 gas mass calibration (used throughout the paper). Symbols and lines are color-coded by stellar mass.}
\label{fig:append_B_H17_4p3}
\end{figure}

\begin{figure}
\centering
\includegraphics[width=\columnwidth]{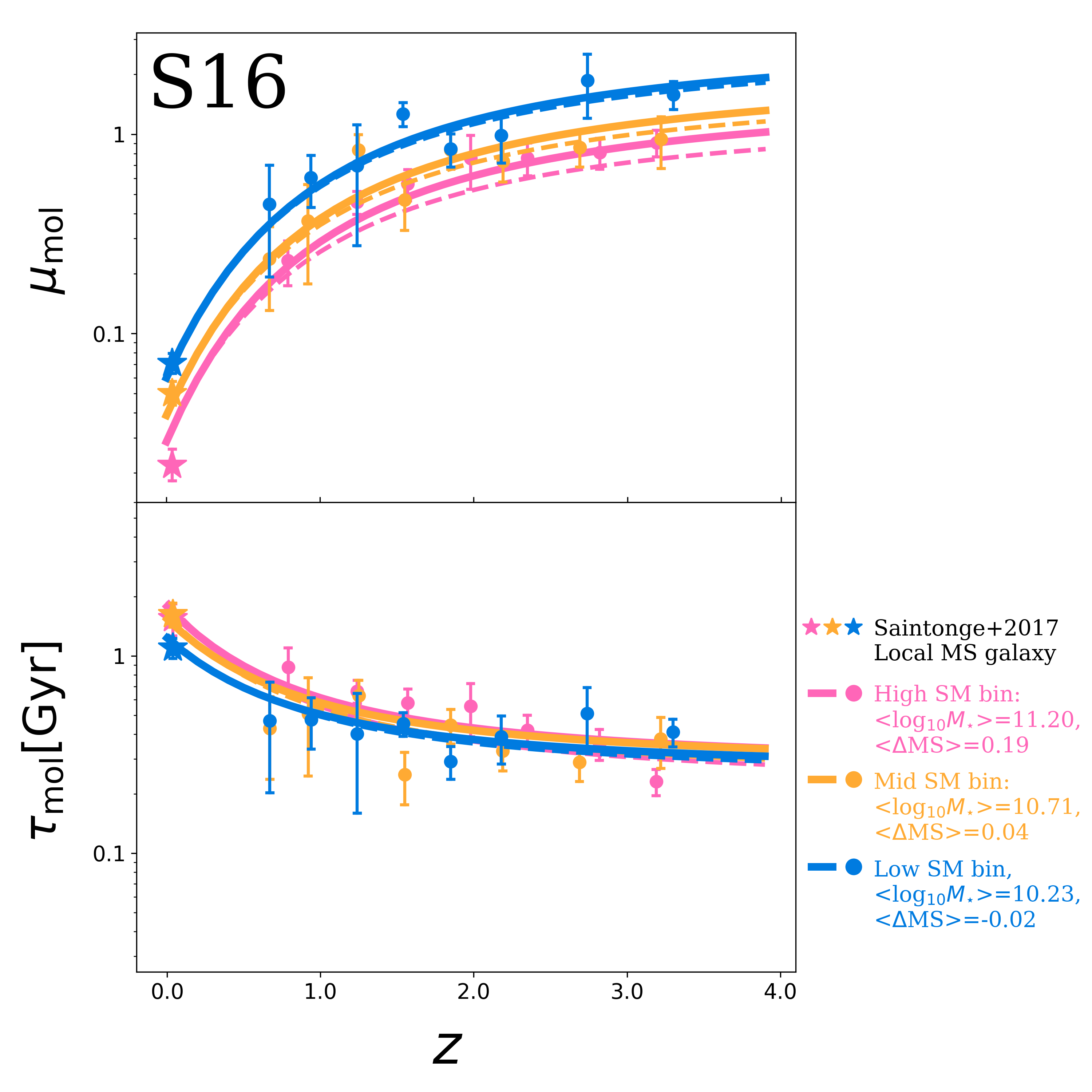}
\caption{Same as Fig.~\ref{fig:append_B_H17_4p3} but for the gas mass calibration from S16.}
\label{fig:append_B_S16}
\end{figure}

\begin{figure}
\centering
\includegraphics[width=\columnwidth]{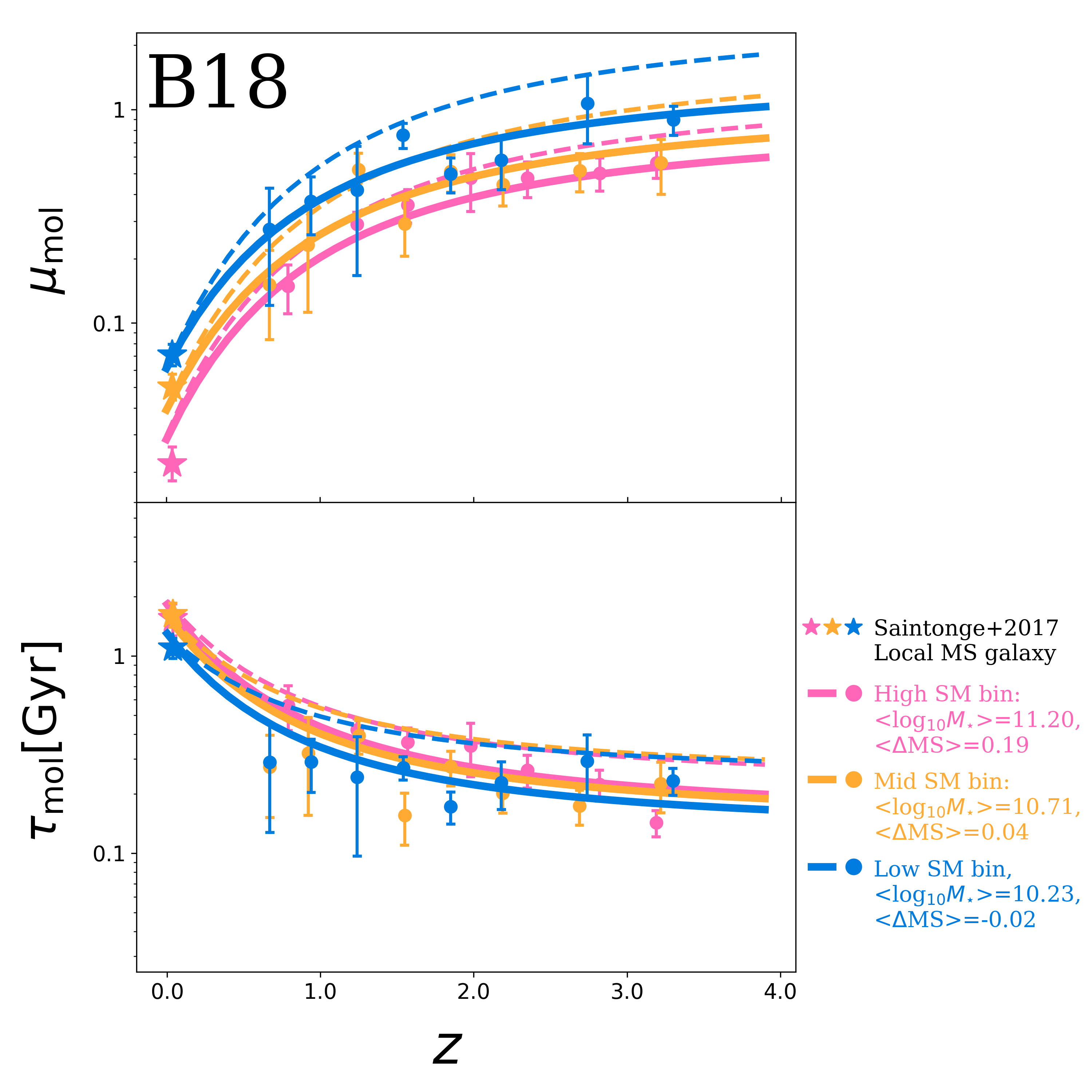}
\caption{Same as Fig.~\ref{fig:append_B_H17_4p3} but for the gas mass calibration from B18.}
\label{fig:append_B_B18}
\end{figure}

\begin{figure}
\centering
\includegraphics[width=\columnwidth]{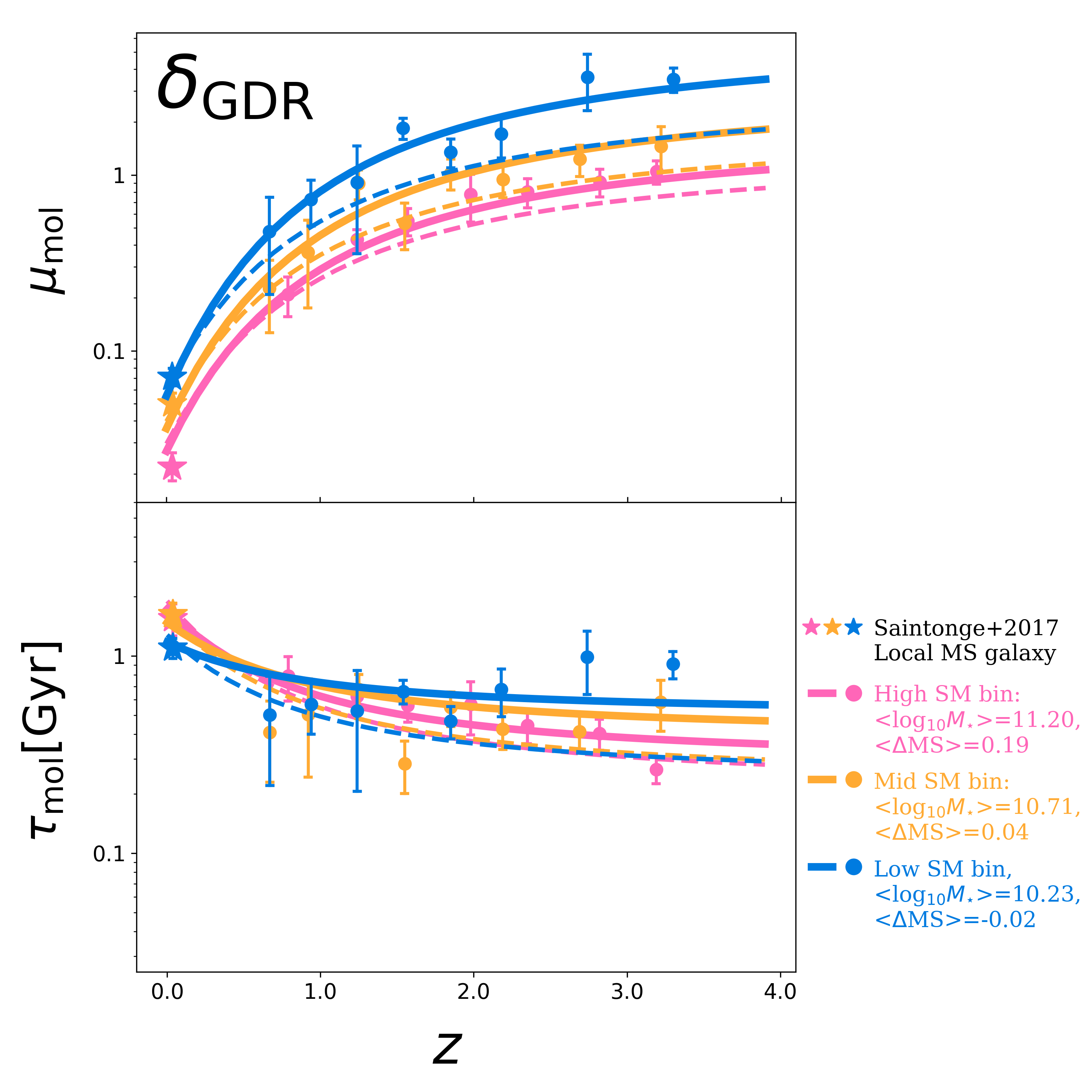}
\caption{Same as Fig.~\ref{fig:append_B_H17_4p3} but for the gas mass calibration from \citet[][$\delta_{\rm GDR}$]{2011ApJ...737...12L}.}
\label{fig:append_B_L11}
\end{figure}


The results of our study could naturally be influenced by the particular assumptions made here to convert stacked rest-frame $850\,\mu$m luminosities into molecular gas masses \citep[see, e.g., discussion in][]{2019ApJ...887..235L}. To evaluate the impact of these assumption on our results, we compare molecular gas masses obtained here using the metallicity-independent $L_{850}$-to-$M_{\rm mol}$ relation of H17 with those obtained using others relations commonly applied in the literature (Fig.~\ref{fig:append_B_lum_met}).

H17 inferred two $L_{850}$-to-$M_{\rm mol}$ relations: the one used in the core of our paper was calibrated using a CO-to-$M_{\rm mol}$ conversion factor (i.e., $\alpha_{\rm CO}$) that assumes that the density and properties in the gas reservoirs of high-redshift galaxies are similar to Milky Way giant molecular clouds, i.e., $\alpha_{\rm CO}$=6.5 $M_{\odot}\,$(K\,km\,s$^{-1}$\,pc$^{2}$)$^{-1}$; and another relation calibrated instead using $\alpha_{\rm CO}$=4.35 (H17$_{\alpha_{\rm CO}=4.35}$), which is the standard value for the Milky Way. These relations have the same slope and only differ in terms of normalisation. Molecular gas masses obtained with H17$_{\alpha_{\rm CO}=4.35}$ are thus shifted to lower values by $0.17\,$dex (Fig.~\ref{fig:append_B_lum_met}). This yields a quantitatively different evolution of the molecular gas mass fraction and depletion time with redshift than those inferred with H17 (Fig.~\ref{fig:append_B_H17_4p3}). However, the main conclusions of our analysis remains unchanged: the molecular gas fraction of MS galaxies still increases significantly with redshifts and decreases with stellar masses; and their depletion time still decrease more moderately with redshifts and remains mostly independent from their stellar mass.

Besides the H17 gas mass calibration, S16 also provides a metallicity-independent $L_{850}$-to-$M_{\rm mol}$ relation calibrated on a sample of 70 SFGs with both dust RJ and CO measurements. Then, assuming a constant mass-weighted dust temperature of 25\,K, a dust emissivity, $\beta$, of 1.8, and $\alpha_{\rm CO}$=6.5 $M_{\odot}\,$(K\,km\,s$^{-1}$\,pc$^{2}$)$^{-1}$, they calibrated a light-to-mass relation,
\begin{equation}
{\rm log}(M_{\rm mol})={\rm log}(L_{850})-19.83,
\end{equation}
where $M_{\rm mol}$ and $L_{850}$ are in units of $M_{\odot}$ and $\rm erg\, s^{-1}\, Hz^{-1}$, respectively. This relation yields molecular gas masses in very good agreement with those inferred here (Fig.~\ref{fig:append_B_lum_met}), with only a systematic offset of $+0.04\,$dex for S16. The redshift evolution of the molecular gas mass fraction and depletion time inferred using S16 are shown in Fig.~\ref{fig:append_B_S16}. Those agree qualitatively and quantitatively with the original findings of our study.

One can also measure the molecular gas mass of SFGs by applying first a standard $L_{850}$-to-$M_{\rm dust}$ relation and then a metallicity-dependent $M_{\rm dust}$-to-$M_{\rm mol}$ relation. To begin with, we thus convert our stacked rest-frame $850\,\mu$m luminosities into $M_{\rm dust}$ following \citet{2020ApJ...892...66M}, whom assumed a constant mass-weighted dust temperature of 25\,K, a dust emissivity of 1.8, and a photon cross-section to dust mass ratio at rest-frame $850\,\mu$m, $\kappa_{850}$, of 0.0431 m$^{2}$ kg$^{-1}$,
\begin{equation}
{\rm log}(M_{\rm dust})={\rm log}(L_{850})-21.86,
\label{eq:mdust}
\end{equation}
where $M_{\rm dust}$ is in unit of $M_{\odot}$. This dust mass can then be converted into molecular gas mass using the metallicity-dependent $M_{\rm dust}$-to-$M_{\rm mol}$ relation of, e.g., B18, which was calibrated using 78 local SFGs with known gas-phase metallicity,
\begin{equation}
{\rm log}(M_{\rm mol})={\rm log}(M_{\rm dust})+1.83+0.12\times((\rm 12+log(O/H))-8.67),
\end{equation}
where the gas-phase metallicity, 12+log(O/H), can be inferred using the redshift- and stellar mass-dependent relation given in \citet{2019ApJ...887..235L},
\begin{equation}
{\rm 12+log(O/H)}=
\begin{cases}
a\hspace{3.0cm}  {\rm if\,log}(M_{\star}/M_{\odot})\geq b(z),\\
a-0.087\times ({\rm log}(M_{\star}/M_{\odot})-b(z))^{2},\hspace{0.7cm} {\rm else},
\end{cases}
\end{equation}
where $a$=8.74 and $b(z)$=10.4+4.46$\times$log(1+$z$)-1.78$\times$(log(1+$z$))$^{2}$. Using this molecular gas mass calibration (i.e., B18) yields estimates which are $\sim$0.17\,dex lower than those from H17 (Fig.~\ref{fig:append_B_lum_met}). In addition to this global offset, this metallicity-dependent approach yields quantitatively different evolution of the molecular gas mass fraction and depletion time with redshift than those inferred with H17 (Fig.~\ref{fig:append_B_B18}). However, the main conclusions of our analysis remains unchanged: the molecular gas fraction of MS galaxies still increases significantly with redshifts and decreases with stellar masses; and their depletion time still decreases more moderately with redshifts and remains mostly independent from their stellar mass. 

Finally, we converted the dust masses from Eq.~\ref{eq:mdust} into molecular gas masses using instead the standard metallicity-dependent $M_{\rm dust}$-to-$M_{\rm mol}$ relation of \citet[][$\delta_{\rm GDR}$]{2011ApJ...737...12L}, which was calibrated using high-resolution observations of five local group galaxies,
\begin{equation}
{\rm log}(M_{\rm mol})={\rm log}(M_{\rm dust}) + 9.4 - 0.85\times(\rm 12+log(O/H)).
\end{equation}
We note that while this relation formally account for both molecular and atomic gas masses, we implicitly assume here that the gas in high-redshift ($z>0.5$) SFGs is dominated by their molecular phase \citet[see discussion in][]{2018ApJ...853..179T,2019ApJ...887..235L}. Overall, the molecular gas masses inferred using the $\delta_{\rm GDR}$ method are about $\sim$0.13\,dex higher than those obtained from H17 (Fig.~\ref{fig:append_B_lum_met}). As for B18, this gas mass calibration also yields slightly different, yet qualitatively consistent, molecular gas mass fraction and depletion time redshift evolution (Fig.~\ref{fig:append_B_L11}) than those inferred with H17.

To summarize, the gas mass calibrations are: $M_{\rm mol}^{\rm H17_{\alpha_{\rm CO}=4.35}}$ $\approx$ $M_{\rm mol}^{\rm B18}$ < $M_{\rm mol}^{\rm H17}$ $\approx$ $M_{\rm mol}^{\rm S16}$ < $M_{\rm mol}^{\delta_{\rm GDR}}$. The redshift evolution of the molecular gas fraction and depletion time obtained from S16 agree qualitatively and quantitatively with those from H17, while those measured using H17$_{\alpha_{\rm CO}=4.35}$, B18 and $\delta_{\rm GDR}$ agree only qualitatively with those from H17 (see Tab.~\ref{tab:appendix_B_best_fit}). However, because the main conclusion of our paper are not qualitatively affected by the particular choice of a given relation and because the H17 approach yield measurements which are bracket by others, we decided to use H17 in our paper.\\

The choice of using the $L_{850}$-to-$M_{\rm mol}$ relation from H17 could seem at odds since this relation was calibrated using $\alpha_{\rm CO}$=6.5, while our local reference, i.e., \citet[][]{2017ApJS..233...22S}, converted their CO measurements into molecular gas masses using $\alpha_{\rm CO}\sim$4 (at the high stellar masses of our study). Despite this apparent inconsistency, the redshift evolution of the molecular gas content of massive SFGs inferred by combining this local reference with our high-redshift H17 measurements is in much better agreement with \citet{2020ARA&A..58..157T} than when combining this local reference with our high-redshift $H17_{\alpha_{\rm CO}=4.35}$ measurements (see Fig.~\ref{fig:gas} and~\ref{fig:append_B_lum_met}). Because at high stellar masses results from \citet{2020ARA&A..58..157T} can be considered as the reference (as they are based in a fairly complete sample of massive SFGs and a thorough cross-calibration of the CO- and dust-based methods), we decided to use in the core of our paper the $L_{850}$-to-$M_{\rm mol}$ relation from H17. The agreement between these high-redshift H17 measurements and those from \citet{2020ARA&A..58..157T} is explained by the fact that using $\alpha_{\rm CO}$=6.5 instead of 4.3 to calibrate the local $L_{850}$-to-$M_{\rm mol}$ relation corrects indirectly (and to first order) for the fact that at a given stellar mass, high-redshift galaxies have lower metallicities than local galaxies, and thus have a higher gas-to-dust ratio \citep[e.g.,][]{2011ApJ...737...12L} and consequently should have lower $L_{850}$-to-$M_{\rm mol}$ ratio. This seems also confirmed by the good agreement at high stellar masses between our H17 measurements and those from the metallicity-dependent $\delta_{\rm GDR}$ method.

\begin{table*}
\centering
\caption{Best-fit coefficients for the molecular gas fraction (Eq.~\ref{eq:gas fraction}) and molecular gas depletion time (Eq.~\ref{eq:depletion}) functions.}
\label{tab:appendix_B_best_fit}
\begin{tabular}{lcccc}
\hline
\multicolumn{5}{c}{log$_{10}\ \mu_{\rm mol}$}\\ 
\multicolumn{5}{c}{with $a=0.4195$ and $ak=0.1195$ from \citet[][]{2019ApJ...887..235L}}\\
 & $b$ & $c$ & $ck$ & $d$ \\
\hline
H17 & -0.468$^{+0.070}_{-0.070}$ & -0.122$^{+0.008}_{-0.008}$ & 0.572$^{+0.059}_{-0.060}$ & 0.002$^{+0.011}_{-0.011}$\\
H17$_{\alpha_{\rm CO}=4.35}$ & -0.463$^{+0.070}_{-0.071}$ & -0.102$^{+0.008}_{-0.008}$ & 0.311$^{+0.061}_{-0.058}$ & 0.000$^{+0.010}_{-0.010}$ \\
B18 & -0.353$^{+0.069}_{-0.069}$ & -0.099$^{+0.008}_{-0.008}$ & 0.266$^{+0.057}_{-0.057}$ & -0.008$^{+0.010}_{-0.010}$ \\
S16 & -0.396$^{+0.070}_{-0.070}$ & -0.123$^{+0.008}_{-0.008}$ & 0.584$^{+0.059}_{-0.058}$ & -0.004$^{+0.011}_{-0.011}$ \\
L11 & -0.679$^{+0.068}_{-0.067}$ & -0.152$^{+0.008}_{-0.008}$ & 0.948$^{+0.057}_{-0.057}$ & 0.017$^{+0.010}_{-0.010}$ \\
\hline
\multicolumn{5}{c}{}\\
\hline
\multicolumn{5}{c}{log$_{10}\ \tau_{\rm mol}$ [Gy$^{-1}$]}\\ 
\multicolumn{5}{c}{with $a=-0.5724$ and $ak=0.1120$ from \citet[][]{2019ApJ...887..235L}}\\
 & $b$ & $c$ & $ck$ & $d$ \\
\hline
H17 & 0.055$^{+0.069}_{-0.071}$ & 0.049$^{+0.008}_{-0.008}$ & -0.643$^{+0.056}_{-0.057}$ & 0.016$^{+0.010}_{-0.010}$ \\
H17$_{\alpha_{\rm CO}=4.35}$ & 0.054$^{+0.071}_{-0.069}$ & 0.069$^{+0.008}_{-0.008}$ & -0.899$^{+0.055}_{-0.057}$ & 0.014$^{+0.010}_{-0.011}$ \\
B18 & 0.168$^{+0.069}_{-0.067}$ & 0.072$^{+0.008}_{-0.008}$ & -0.947$^{+0.057}_{-0.057}$ & 0.006$^{+0.010}_{-0.010}$ \\
S16 & -0.125$^{+0.070}_{-0.067}$ & 0.047$^{+0.008}_{-0.008}$ & -0.628$^{+0.058}_{-0.058}$ & 0.010$^{+0.010}_{-0.010}$ \\
L11 & -0.157$^{+0.068}_{-0.070}$ & 0.019$^{+0.008}_{-0.008}$ & -0.265$^{+0.059}_{-0.057}$ & 0.031$^{+0.010}_{-0.011}$ \\
\hline
\end{tabular}
\end{table*}

\end{appendix}
\end{document}